%% file: main_final.tex
\documentclass[a4paper,twocolumn,11pt,accepted=2025-01-13]{quantumarticle}
\pdfoutput=1

\usepackage[utf8]{inputenc}
\usepackage[english]{babel}
\usepackage[T1]{fontenc}
\usepackage{amsmath}
\usepackage{hyperref}

\input{preamble}
\usepackage[square,numbers,sort&compress]{natbib}
\usepackage{tikz}

\newcommand{\stylecolor}{quantumviolet}

\usepackage{bbold}
\usepackage{booktabs}
\usepackage{multirow}
\usepackage{enumitem}
\usepackage{booktabs}
\usepackage{fancyhdr, amssymb, graphicx}
\usetikzlibrary{shadows}
\usepackage[explicit]{titlesec}
\usepackage[numbib,nottoc]{tocbibind}

\begin{document}

\title{Designing open quantum systems with known steady states: Davies generators and beyond}

\author{Jinkang Guo}
\email{jinkang.guo@colorado.edu}
\affiliation{Department of Physics and Center for Theory of Quantum Matter, University of Colorado, Boulder, CO 80309, USA}
\orcid{0000-0001-6458-383X}
\author{Oliver Hart}
\orcid{0000-0002-5391-7483}
\affiliation{Department of Physics and Center for Theory of Quantum Matter, University of Colorado, Boulder, CO 80309, USA}
\author{Chi-Fang  Chen}
\affiliation{Institute for Quantum Information and Matter, California Institute of Technology, Pasadena, CA, 91125 USA}
\orcid{0000-0001-5589-7896}
\author{Aaron J.~Friedman}
\affiliation{Department of Physics and Center for Theory of Quantum Matter, University of Colorado, Boulder, CO 80309, USA}
\orcid{0000-0003-2159-6475}
\author{Andrew Lucas}
\email{andrew.j.lucas@colorado.edu}
\affiliation{Department of Physics and Center for Theory of Quantum Matter, University of Colorado, Boulder, CO 80309, USA}
\orcid{0000-0003-3098-5683}

\maketitle

\begin{abstract} \onecolumn
  \small
We provide a systematic framework for constructing generic models of nonequilibrium quantum dynamics with a 
target stationary (mixed) state. Our framework identifies (almost) all combinations of Hamiltonian and dissipative dynamics that relax to a steady state of interest,  generalizing the Davies' generator for dissipative relaxation at finite temperature to nonequilibrium dynamics targeting arbitrary stationary states. 
We focus on Gibbs states of stabilizer Hamiltonians, identifying local Lindbladians compatible therewith by constraining the rates of dissipative and unitary processes. Moreover, given terms in the Lindbladian not compatible with the target state, our formalism identifies the operations -- including syndrome measurements and local feedback -- one must apply to correct these errors.  Our methods also reveal new models of quantum dynamics: for example, we provide a ``measurement-induced phase transition'' in which measurable two-point functions exhibit critical (power-law) scaling with distance at a critical ratio of the transverse field and rate of measurement and feedback. Time-reversal symmetry -- defined naturally within our formalism -- can be broken both in effectively classical and intrinsically quantum ways. Our framework provides a systematic starting point for exploring the landscape of dynamical universality classes in open quantum systems, as well as identifying new protocols for quantum error correction.
\end{abstract}

\tableofcontents

\section{Introduction and summary of results}

\subsection{Introduction}

Over the past century, the laws of equilibrium statistical mechanics have been increasingly understood and organized via a Wilsonian renormalization group~\cite{WilsonI, WilsonII}.   However, beyond the familiar setting of equilibrium, new phenomena can arise -- e.g., spontaneous symmetry breaking in 
models of flocks of birds in two spatial dimensions, which is \emph{not} possible in equilibrium 
due to the Mermin-Wagner theorem. 
Such classical systems fall under the umbrella of \emph{active matter}~\cite{RamaswamyAnnuRev, Marchetti2013, Ramaswamy2017} -- 
i.e., systems whose constituent particles are ``self propelled,'' in that they contain internal sources of energy and entropy) -- which has led to a significant body of research into nonequilibrium classical 
phenomena.

Quantum systems may also be driven away from thermal equilibrium.  Indeed, one may directly try to add quantum fluctuations to a theory of classical active matter, such as flocking~\cite{Adachi2022, khasseh2023active, yamagishi2023defining}. However, this is certainly not the only setting in which nonequilibrium quantum systems may arise. For example, it is well established that systems may tend towards entangled states in a \emph{driven-dissipative system}~\cite{CiracDQC}, described in terms of open quantum dynamics where a coherently driven system is coupled to an environment or bath.  Perhaps the most important example of a uniquely quantum system driven out of thermal equilibrium is a  quantum computer.  If a quantum computer is built out of, e.g., surface-code qubits~\cite{bravyi1998,Fowler_2012}, it will fail to retain quantum information if it thermalizes -- an analogue of the ``Mermin-Wagner theorem'' forbids this in two spatial dimensions~\cite{hastingsfiniteT}.  Therefore, the long-time storage of quantum information in surface codes \emph{requires} ``activity'' -- i.e., being persistently driven out of equilibrium.\footnote{This analogy is imperfect since, after all, the \emph{decoders} that work for the surface code collect global classical information before decoding.  Moreover, we are no longer guaranteed that the stationary state does not have complex long-range correlations (which would invalidate any Mermin-Wagner-like result).  Still, this analogy serves as a useful inspiration for developing the theory in this work.}

The purpose of this paper is to give a \emph{systematic framework} for discovering and, just as importantly, organizing our understanding of these ``active'' quantum systems.  Inspired by recent work \cite{ACM} in \emph{classical statistical physics}, which provides a framework for classifying systems based on their steady-state probability distribution, here we provide an exhaustive classification of the most general local quantum many-body system that protects a target stationary (mixed) state $\sigma$. Our work also details how such Lindblad dynamics can be simulated in actual experiment, using the most local operators possible. When $\log \sigma$ is a sum of commuting operators (i.e., $\sigma$ is a ``stabilizer state"), we can find (almost all) \emph{local} many-body dynamics that protects the desired state.  An immediate -- and sometimes useful -- byproduct of this framework is a definition of time-reversal symmetry for such open quantum systems; this definition has independently been identified in Ref.~\citenum{HiddenTRS}.

Solving this technical problem is rather useful, as it immediately provides us with new (and unifying!) insight into a diverse array of problems from different subfields within physics.

In condensed matter and quantum statistical mechanics, there has been intense recent interest in discovering uniquely quantum nonequilibrium phases in \emph{monitored quantum systems}, in which unitary dynamics is interrupted by measurements~\cite{NahumMeasInduced, FisherQuantumZeno, SmithEntanglementDynamics,  FisherMeasurementDriven, ChoiQEC, Ludwig2020}.  In order to avoid a \emph{postselection problem} and realize
novel phases in the thermodynamic limit of any experiment, it is crucial to perform \emph{quantum error correction} (i.e., active feedback) based on the measurement outcomes to drive the system towards a deterministic state whose properties can be measured in experiment~\cite{RoySteering, herasymenko2022measurementdriven, iadecola2022dynamical, buchhold2022revealing, chertkov2022characterizing, milekhin2023measurementinduced, FriedmanMIPT, lee2022decoding}.  However, it is not a priori obvious whether such a phase realizes a \emph{uniquely quantum} state of matter, or whether it is essentially a classical phase that arises out of a microscopically quantum dynamical system.  Although we do not attempt to answer such a large open question definitively  herein, we believe that the framework we provide in this work is a well defined starting point for addressing such questions systematically.

In quantum information science, it is often desirable to protect entangled states against decoherence or other deleterious environmental effects.  For example, one may wish to prepare an entangled GHZ state \cite{GHZ89}, which is capable of performing quantum-enhanced sensing \cite{sensing_rmp}.  Alternatively, as highlighted previously, one may want to protect a quantum error-correcting code \cite{Calderbank_1996, Steane_1996}.  Usually, one devises some protocol that relies on few-qubit measurements and operations in order to protect such a code state, and then numerically simulates whether or not the protocol can protect against errors.  The formalism that we describe here is very well suited for discovering fault-tolerant quantum error-correcting protocols, and gives a systematic way of building all possible driven-dissipative systems that protect a target state $\sigma$ for long times; see Ref.~\citenum{viola2} for related ideas on short-time preparation.

Lastly, one anticipated application of future quantum computers is to simulate properties of quantum systems arising in physics, material science, and chemistry~\cite{feynman1982SimQPhysWithComputers}. In many perceivable cases (such as correlation functions or transport properties), preparing ground states or thermal states has been identified as a key algorithmic subroutine. Nevertheless, the complexity of practically relevant low-energy states has remained a debated topic and, thus far, there has not been a consensus on the ``go-to'' state-preparation algorithm~\cite{lee2023evaluating} (see, e.g.,~Table 1 of Ref.~\citenum{kastoryano2023quantum} for a catalog). More recent work introduced a new algorithmic family of \textit{quantum Gibbs samplers}~\cite{temme2011quantum, Shtanko2021AlgorithmsforGibbs, rall2023thermal, kastoryano2023quantum, Chen:2023zpu} that attempt to model and simulate the thermalization process in nature. The challenge is that, for a noncommuting Hamiltonian, there is conflict between the energy uncertainty, locality, and quantum detailed balance; the conventional Davies' generator is inefficient in noncommuting many-body systems, and only recently has this been reconciled~\cite{Chen:2023zpu, Rouze:2024ufx, Ding:2024mxo}. Although we have mostly focused on commuting Hamiltonians, the formalism described in this paper should, in principle, give an ``exhaustive'' classification of all possible ways to find noncommutative Gibbs samplers.  Since in classical statistical physics, the convergence rate of such samplers is often increased by breaking time-reversal symmetry~\cite{Bernard2009,Suwa2010}, it is natural to expect that a similar result also holds in quantum systems.  Our framework explains how to look for these T-broken samplers that still protect the same state.

\subsection{Summary of results}

\noindent 
Throughout the manuscript, we make use of the following fundamental assumptions:
\begin{enumerate}[label={\sf\color{\stylecolor}A\arabic*}]
    \item \label{asn:finite-dimensional} We restrict to \textsl{finite-dimensional} quantum systems;
    \item Time evolution of the system's state (density matrix) is captured by a \textsl{time-independent Lindbladian};
    \item \label{asn:positive} All stationary states (denoted by $\sigma$) that we consider are \textsl{positive definite}.
\end{enumerate}

\noindent
Under these global assumptions, we obtain the following key results:
\begin{itemize}
    \item We introduce the notions of time reversal T~\eqref{eq:T} and generalized time reversal gT~\eqref{eq:gT} for any open quantum system with  stationary state $\sigma$ in Secs.~\ref{subsec: T} and \ref{subsec:gT}. Systems that are invariant (symmetric) under gT have ``hidden time-reversal symmetry''~\cite{HiddenTRS, HiddenTRS2, HiddenTRS3, HiddenTRS4, HiddenTRS5}, which reduces to ``quantum detailed balance''~\eqref{eq:QDB} in special cases discussed after Eq.~\eqref{eq:gT}.
    \item Specializing to stationary states of the form $\sigma = \ee^{-\Phi}$, where $\Phi$ is a sum of commuting, strictly local operators, we provide in Sec.~\ref{sec:stationarity} a systematic construction of almost all possible local many-body open quantum dynamics (Lindbladians) for which $\sigma$ \eqref{eqn:ansatz} is a steady state (alternative dynamics are discussed in Appendix~\ref{sec:T-odd}). Our formalism can be used to generate both T-even and T-odd dynamics, with Davies' generator as a special case. Importantly, our formalism can be used for the systematic identification of open quantum dynamics \emph{beyond} Davies' generator (i.e.,~by allowing for nontrivial Hamiltonian terms and including T-odd contributions). 

    \item Further specializing to the case where $\Phi$ is a sum of commuting {Pauli strings}, the stationary state $\sigma = \ee^{-\Phi}$ can be targeted by adding terms to the Lindbladian $\Liouvillian$ that can be interpreted as measurements and outcome-dependent feedback (see  Sec.~\ref{sec:stabilizer-steady-states}). We also show how to ``correct'' for arbitrary, unwanted jump operators and/or Hamiltonian terms that may be present in Lindbladians describing real experiments.  Hence, our formalism provides a natural connection between protecting stabilizer states and quantum error correction.

    \item We explicitly show how to construct both T-even and T-odd nontrivial dynamics that protect certain stabilizer states and how to correct different kinds of errors. Our constructions include an experimentally detectable measurement-induced phase transition and biased quantum walks that do not have any classical analogues, which we discuss in Sec.~\ref{sec:examples}.  Many of the Lindbladians that we identify have novel features, such as non-Pauli feedback or new, T-breaking (i.e., biased) quantum random walks. These examples highlight how our framework extends beyond both Davies' generator and the stabilizer-based quantum error correction.
\end{itemize}


\section{Reversibility and effective theories of open systems}
\label{sec:classical and quantum}

We now review some preliminary facts about dissipative (open) systems in both the classical and quantum settings. 
As previously emphasized, in both cases, we organize our approach around the identification of a target stationary state, as in the related work \cite{ACM} for classical systems. This section serves to explain some of our physical motivations underlying this work, though readers primarily interested in the problem of engineering open systems with known steady states may skip to the formalism in Sec.~\ref{sec:theory}.

Because we assume that the open dynamics of interest are time-translation invariant, such a stationary state always exists. The existence of this stationary state implies a \emph{reversibility} transformation, which we associate with microscopic time reversal. Precise definitions of reversibility can be found in later subsections. One commonly defines a system to be in equilibrium if the dynamics is reversible; since our framework is organized around a known steady state, it is straightforward to distinguish equilibrium versus nonequilibrium phenomena.  Generalizing this reversibility symmetry to include additional transformations (e.g., spatial inversion) is straightforward. Given a target stationary state and any symmetries (especially reversibility) that we wish to impose, we can work out the most general possible local unitary and dissipative dynamics compatible therewith.  This approach embodies the spirit of Wilson's effective (field) theory \cite{WilsonI, WilsonII, ACM}, and is central to our framework.

\subsection{Classical systems}
\label{sec:ACM review}

We begin by reviewing Wilsonian approach \cite{WilsonI, WilsonII} to effective theories of dissipative \emph{classical} systems that relax to a known (or target) stationary state $\sigma$~\cite{ACM}. 
For simplicity, we focus on open classical systems captured by discrete state spaces (e.g., a collection of $N$ Ising spins). The corresponding dynamics are realized by a continuous-time Markov process, captured by a master equation \cite{Doi1, Doi2, Peliti1, Peliti2, KadanoffBook} -- i.e., a discrete analogue of the Fokker-Planck equation \cite{KadanoffBook, Langevin, Fokker, Planck}. Denoting by $W_{ba}$ the rate at which the configuration $a$ transitions to the configuration $b \neq a$, we have that
\begin{equation}
    W_{aa} = -\sum_{b , b \neq a} W_{ba} \, , ~~
\end{equation}
is the rate at which the system remains in state $a$. The system evolves via the classical \emph{master equation},
\begin{equation}
    \label{eqn:Fokker Planck}
    \pd{t} p_a(t)  = \sum\limits_b  W_{ab} \, p_b (t)  \, ,~~
\end{equation}
where $p_a(t)$ is the probability to find the system in configuration $a$ at time $t$.  The stationary state $\sigma$ is a probability distribution (with probability $\sigma_a$ for configuration $a$) such that
\begin{equation}
    \label{eq:classical stationary}
    \pd{t} \sigma = W \sigma = 0
    \, , ~~
\end{equation}
where $W$ is the transition (or ``rate'') matrix and $\sigma$ is a vector whose components $\sigma_a$ denote the probability of realizing the configuration $a$ in the stationary state $\sigma$. We also define the probability for the system to evolve from the initial state $a$ to the final state $b$ in time $t$ under $W$ as
\begin{equation}
\label{eq:classical transition probability}
    \text{Pr} \left[ a (0) \to b (t) ; W \right] = \matel*{b}{\hspace{0.4mm} \ee^{t W}}{a} \, , ~~
\end{equation}
where we used the fact that $W$ is time independent. This quantity is also known as the ``propagator'' \cite{ACM}.

Importantly, the existence of a stationary distribution $\sigma$ \eqref{eq:classical stationary} -- which need not correspond to thermal equilibrium -- implies the \emph{global balance} condition \cite{Kolmogorov-backward, KellyStochastic, ACM} 
\begin{equation}
    \label{eqn:Pin-Pout}
    \sum_{b,b \neq a} \left[ \sigma_b W_{ab} - W_{ba}\sigma_a \right] = 0
    \, ,~~
\end{equation}
for any configuration $a$. In other words, in the stationary state $\sigma$, the total probability to transition \emph{into} configuration $a$ is equal to the total probability to transition \emph{out of} configuration $a$. Using the global balance condition \eqref{eqn:Pin-Pout}, we identify the \emph{time-reversed} transition matrix
\begin{equation}
    \label{eqn:classical reversibility}
    \widetilde{W}_{ba} = \sigma_b W_{ab} / \sigma_a ~~~ \Longleftrightarrow ~~~ \widetilde{W} = \hat\sigma^{-1} \, W^T \, \hat\sigma \, ,~~
\end{equation}
where we have defined $\hat\sigma_{ab} = \sigma_a  \delta_{ab}$ as the ``stationary operator,'' the generators $W$ and $\widetilde{W}$ share the \emph{same} stationary distribution $\sigma$ \eqref{eq:classical stationary}, and, crucially, if the former generates the sequence of configurations $a,b,\ldots,y,z$ then the latter realizes the \emph{reversed} sequence of configurations $z,y,\ldots,b,a$ \cite{ACM, Kolmogorov-backward, KellyStochastic}. 

Moreover, noting that $\matel*{b}{W}{a} = \matel*{a}{W^T}{b}$, we have that
\begin{equation}
\label{eq:classical reversed transition probability}
    \text{Pr} \left[ a(0) \to b(t) ; W \right] = \matel*{a}{\ee^{t W^T}}{b} = \matel*{a}{\hat{\sigma}^{-1} \ee^{t \widetilde{W}} \hat{\sigma}}{b} = \ee^{\Phi(a) - \Phi(b)} \, \text{Pr} \left[ b(0) \to a(t); \widetilde{W} \right] \, ,~~
\end{equation}
since the $\hat{\sigma}$ acts as $\ee^{\Phi(b)}$ on the configuration $b$, so that the probability to go from $a \to b$ in time $t$ under $W$ is related to the probability to go from $b \to a$ in time $t$ under the time-reversed generator $\widetilde{W}$ \eqref{eqn:classical reversibility}, up to the ratio $\sigma_b / \sigma_a$ of the probabilities for those configurations in the stationary distribution $\sigma$~\cite{ACM}.

Accordingly, we associate the $\Ints_2$ ``reversibility'' transformation,
\begin{equation}
    \label{eqn:classical T}
    \text{T} : W \mapsto \widetilde{W} \, , ~~
\end{equation}
with \emph{time reversal}~\cite{ACM}. For discrete state spaces, T~\eqref{eqn:classical T} is the \emph{only} notion of time reversal, as it maps the Markov generator $W$ to its time-reversed partner $\widetilde{W}$. However, for continuous state spaces described by the Fokker-Planck equation, one often combines T~\eqref{eqn:classical T} with a ``microscopic'' $\Ints_2$ transformation on certain variables (e.g., the momentum transforms as $p \to -p$). In general, there are multiple notions of time reversal in open systems -- both classical and quantum. As a reminder, the existence of the stationary state $\sigma$ implies the transformation T~\eqref{eqn:classical T} and time-reversed generator $\widetilde{W}$ \eqref{eqn:classical reversibility} \cite{ACM, Kolmogorov-backward, KellyStochastic}.

We say that a system exhibits ``equilibrium'' dynamics when $W$ \eqref{eqn:Fokker Planck} is even under T -- i.e., $W = \widetilde{W}$. One often states that \eqref{eqn:classical reversibility} implies that T-even systems obey \emph{detailed balance}:
\begin{equation}
    \label{eqn:classical DB}
    \sigma_b W_{ab} = \sigma_a W_{ba}  ~~
\end{equation}
if and only if $W = \widetilde{W}$. 
We emphasize, however, that detailed balance is not necessary for global balance. There are many stochastic dynamical systems that break time-reversal symmetry T -- i.e., for which $W \neq \widetilde{W}$, while maintaining the same stationary state $\sigma$. In many physical cases of interest, one can identify an extra $\Ints_2$ transformation g (e.g., parity, charge conjugation, etc.) such that the product of g and T is a symmetry of the dynamics. We refer to this combined $\Ints_2$ symmetry gT as generalized time reversal, and in the classical setting, it is quite instructive to classify dynamics according to whether they respect, explicitly (or spontaneously) break T and/or gT~\cite{ACM}. Finally, we comment that it is possible to enforce generic (e.g., continuous) symmetries -- in both a weak and strong sense -- on the generator W~\cite{ACM}.


\subsection{Quantum systems}
\label{subsec: T}

We now consider the quantum analogues to the discussion of open classical systems in Sec.~\ref{sec:ACM review}. As before, we require continuous time-translation symmetry. We also take the bath to be Markovian (i.e., memoryless), as is standard in the literature on open quantum systems~\cite{lindblad1973entropy, Lindblad76, Gorini76, Gardiner2004, NielsenChuang2010, brasil2013simple, Daley2014, GarrahanLectures}. Although we only explicitly consider finite-dimensional quantum systems throughout, we see no \emph{conceptual} barrier to extending the framework to infinite-dimensional quantum systems, such as bosonic modes. A somewhat similar philosophy is presented in the context of noninteracting systems in Ref.~\citenum{Thompson:2023osk}.

The quantum analogue of the probability distribution $p (t)$ is the reduced density matrix $\rho (t)$, which captures the (generically mixed) state of the quantum system at time $t$ (e.g., after tracing over environmental degrees of freedom). The set of allowed updates to a density matrix $\rho$ correspond to completely positive and trace-preserving (CPTP) maps~\cite{KrausBook}. The quantum analogue of the master equation \eqref{eqn:Fokker Planck} for classical systems is the Lindblad master equation~\cite{lindblad1973entropy, Lindblad76, Gorini76, Gardiner2004, NielsenChuang2010, brasil2013simple, Daley2014, GarrahanLectures}, which captures generic CPTP maps.\footnote{One often interprets this as having ``integrated out''~\cite{lindblad1973entropy, Lindblad76, Gorini76, KrausBook, Gardiner2004, NielsenChuang2010, brasil2013simple, Daley2014, GarrahanLectures} the environment. However, from the perspective of effective theory, it is more natural to build the dissipative effective theory directly.} Such dynamics are generated by a  ``Lindbladian'' (or ``Liouvillian'') $\Liouvillian$ of the general form
\begin{equation}    
    \pd{t} \rho = \Liouvillian ( \rho ) = -\ii [ H , \rho ] + \sum\limits_{i,j} \gamma_{ij} \left( A^{\vpd}_i \rho A_j^\dagger - \frac{1}{2} \
    \{ A_j^\dagger A^{\vpd}_i , \rho \} \right) \, ,
    \label{eqn:master}
\end{equation}
where $\gamma_{ij}$ is a positive-semidefinite matrix, the ``jump operators'' $\{A_i\}$ form a complete basis for the operators acting on the system's Hilbert space, and the system Hamiltonian $H$ is Hermitian, and may differ from the na\"ive Hamiltonian $H_0$ for the system in isolation (i.e., integrating over the bath degrees of freedom to recover $H$ may ``renormalize'' terms in $H_0$ or generate new ones). 

We also comment that the choice of $H$ and the dissipative part $\gamma$ is not unique. However, if we require that $H$ and the jump operators $\{A_i\}$ are all traceless, then the Lindbladian \eqref{eqn:master} is unique, up to a change of basis on the jump operators $A_i$.
Due to time-translation symmetry of $\Liouvillian$ \eqref{eqn:master}, an initial density matrix $\rho$ at time $t=0$ evolves to the state $\rho(t) = \ee^{t \Liouvillian} \rho$ at time $t$.

As in the classical setting, we seek Lindbladians $\Liouvillian$ that protect a target stationary density matrix $\sigma$, which we assume is mixed. The stationarity condition \eqref{eq:classical stationary} for the quantum case corresponds to
\begin{equation}
    \label{eqn:stationarity}
    \Liouvillian (\sigma) = 0 \, ,~~
\end{equation}
where we find it convenient to write the stationary state $\sigma$ in the particular form
\begin{equation}
    \label{eqn:sigma phi}
    \sigma = \ee^{-\Phi} \, , ~~
\end{equation}
where we stress the following points about the stationary state $\sigma$ and the corresponding $\Phi$:
\begin{enumerate}
    \item Assumption~\ref{asn:positive}: We assume that $\sigma$ \eqref{eqn:sigma phi} is full rank -- and thus, \emph{invertible}. However, our results also extend to pure states upon writing $\Phi = \beta H_\text{eff}$ and taking the limit $\beta \to \infty$.\footnote{Lindbladians 
that capture relaxation to \emph{entangled} dark states are useful in designing state-preparation protocols~\cite{diehl2008quantum, Kraus2008entangled, CiracDQC}. As the $\beta \to \infty$ limit is singular, it does not necessarily provide \emph{all} such local dynamics that protect a dark state. If $\Liouvillian$ has a pure stationary state $\BKop{\psi}{\psi}$, we can add arbitrary dynamics to $\Liouvillian$ so long as it leaves $\ket{\psi}$ unchanged.  However if $\ket{\psi}$ is the ground state of Hamiltonian $H$ and $H$ has many eigenvalues, dynamics that protects $\Phi = \beta H$ for any $\beta$ forbids adding generic excited-state transitions, which are allowed if the only goal is to have a dark steady state.}
    %
    We only require that $\sigma > 0$ is positive definite, which is guaranteed when $\Phi = \Phi^\dagger$ is self-adjoint and bounded. 
    \item For convenience of presentation and without loss of generality, we neglect the overall normalization of $\sigma$ \eqref{eqn:sigma phi}, which is unimportant to the linear functions of $\sigma$ that we consider herein.
    \item Most importantly, the operator $\Phi$ is generically \emph{unrelated} to the Hamiltonian $H$ that generates the unitary part of the time evolution captured by $\Liouvillian$ \eqref{eqn:master}.
\end{enumerate}

In Sec.~\ref{sec:stationarity}, we build generic Lindbladians  $\Liouvillian$ that preserve a target stationary state $\sigma$ \eqref{eqn:sigma phi}. This is a departure from the standard approach in the literature, in which one first postulates the form of $H$, $A_i$, and $\gamma_{ij}$ based on \emph{microscopic},  phenomenological assumptions about locality, symmetries, and the dominant dynamical processes present in real experiments on a given system. As Ref.~\citenum{ACM} argues in the context of classical systems, it is often more instructive to take the ``inverse perspective'': rather than try to deduce $\sigma$ from $\Liouvillian$ \eqref{eqn:master}, \emph{we instead identify all Lindbladians $\Liouvillian$ compatible with a particular choice of $\sigma$} \eqref{eqn:sigma phi}.

As in the classical case, we also define a reversibility transformation \eqref{eqn:classical T} that we associate with a time-reversal transformation with respect to a stationary density matrix $\sigma$ of interest. Before making this transformation precise, we first define several inner products, along with the \emph{adjoint} Lindbladian $\Liouvillian^\dagger$. 

First, consider the standard ``Frobenius'' operator inner product, defined by
\begin{align}
    \label{eq:Frobenius inner}
    \expval{A,B} \equiv \frac{1}{D} \tr ( A^\dagger \hspace{0.4mm} B ) \, ,~~
\end{align}
where $D = \tr ( \ident )$ is the dimension of the underlying Hilbert space $\mathcal{H}$. When $D$ is finite, the space $\text{End}(\mathcal{H})$ of operators on $\mathcal{H}$ is itself a Hilbert space with dimension $D^2$, since all operators on $\Comps^D$ are bounded and trace class. When $\mathcal{H}$ corresponds to a system of $N$ qubits, the Pauli group -- i.e., the set of all Kronecker products of Pauli operators over $N$ qubits -- forms an orthonormal basis with respect to Eq.~\eqref{eq:Frobenius inner}. 

Importantly, the Frobenius inner product \eqref{eq:Frobenius inner} defines the \emph{adjoint Lindbladian} $\Liouvillian^\dagger$ via
\begin{align}
    \label{eq:adjoint Lindbladian}
    \expval{ A , \Liouvillian B} = \frac{1}{D} \tr[ A^\dagger \Liouvillian (B) ] = \frac{1}{D}  \tr[ B  \Liouvillian^\dagger (A^\dagger)]  = \tr [ B^\dagger \Liouvillian^\dagger (A) ]^* = \expval{ B, \Liouvillian^\dagger A }^*  = \expval{\Liouvillian^\dagger A, B} \, ,~~
\end{align}
where $\expval{A,B}^* = \expval{B,A}$ by skew symmetry of $\expval{A,B}$~\eqref{eq:Frobenius inner} under complex conjugation. 

Physically, we interpret $\Liouvillian$ \eqref{eqn:master} as the generator of time evolution of \emph{density matrices} $\rho$ in the Schr\"odinger picture, and the adjoint Lindbladian $\Liouvillian^\dagger$ \eqref{eq:adjoint Lindbladian} as the generator of time evolution of \emph{operators} in the Heisenberg picture. In particular, consider the time-dependent expectation value
\begin{equation}
\label{eq:adjoint expval}
    \expval{ O (t) } = \tr[ O \hspace{0.4mm} \rho (t) ]  = \tr [ O \hspace{0.4mm} \ee^{t  \Liouvillian} ( \rho ) ] \equiv \tr [ \rho \hspace{0.4mm} \ee^{t \Liouvillian^\dagger} ( O ) ]  = \tr [ \rho \hspace{0.4mm} O (t) ] \, ,~~
\end{equation}
where $\rho (t) = \ee^{t \Liouvillian} \rho$ and $O(t) = \ee^{t \Liouvillian^\dagger} O$ in the Schr\"odinger and Heisenberg pictures, respectively. Just as the Lindbladian $\Liouvillian$ \eqref{eqn:master} annihilates the stationary state $\sigma$ \eqref{eqn:stationarity}, the adjoint satisfies
\begin{equation}
    \label{eq:adjoint identity condition}
    \Liouvillian^\dagger (\ident ) = 0 \, ,~~
\end{equation}
as a result of $\Liouvillian$ \eqref{eqn:master} being trace preserving -- i.e., $1 = \tr [ \rho (t) ] = \tr [ \rho \hspace{0.4mm} \ee^{t \Liouvillian^\dagger}(\ident) ]$ for all times $t$.
 
Before considering the quantum analogue of the reversibility transformation \eqref{eqn:classical T}, we define another operator inner product. Physically, this inner product captures time-dependent correlation functions, i.e.,
\begin{align}
\label{eq:corr inner}
    \expval{A(t),B}_{\sigma} \equiv \tr \left[ A^\dagger (t) \hspace{0.4mm} \sigma^{1/2} B \sigma^{1/2} \right] = \tr \left[ \ee^{t \Liouvillian^\dagger} (A^\dagger) \hspace{0.4mm} \mathcal{T} (B) \right] \, ,~~
\end{align}
where we have implicitly defined the superoperator $\mathcal{T}$ via
\begin{equation}
\label{eqn:T-def}
    \mathcal{T}(\rho) = \sigma^{1/2} \rho \hspace{0.4mm}  \sigma^{1/2}
    \quad \text{and} \quad 
    \mathcal{T}^{-1}(\rho) = \sigma^{-1/2} \rho \hspace{0.4mm}  \sigma^{-1/2} \, , 
\end{equation}
where we have explicitly written the inverse $\mathcal{T}^{-1}$ for convenience. Applying the definition of the adjoint Lindbladian $\Liouvillian^\dagger$ \eqref{eq:adjoint Lindbladian} and other manipulations to the inner product \eqref{eq:corr inner} leads to
\begin{align}
    \expval{A(t),B}_{\sigma} &\equiv \tr \left[ \ee^{t \Liouvillian^\dagger} (A^\dagger) \hspace{0.4mm} \mathcal{T} (B) \right] = \tr \left[ A^\dagger \hspace{0.4mm} \ee^{t \Liouvillian} ( \mathcal{T} (B) ) \right] \notag \\
    &= \tr \left[ A^\dagger \hspace{0.4mm} \mathcal{T} \big(  \ee^{t \mathcal{T}^{-1} \Liouvillian \mathcal{T}} (B) \big)  \right] \notag \\
    &= \tr \left[ \ee^{t \mathcal{T}^{-1} \Liouvillian \mathcal{T}} ( B^\dagger ) \hspace{0.4mm} \mathcal{T} (A) \right]^* \notag \\
    &\equiv \tr \left[ \ee^{t \widetilde{\Liouvillian}^\dagger} (B^\dagger) \hspace{0.4mm} \mathcal{T} (A) \right]^* = \expval{A,\widetilde{B(t)}}_\sigma \, , ~~
    \label{eq:simple reversal}
\end{align}
where, in the first line, we used the definition of the adjoint \eqref{eq:adjoint Lindbladian} to move the time evolution from $A^\dagger$ to $B$; in the second line, we pulled the superoperator $\mathcal{T}$ through the exponential of $\Liouvillian$ \eqref{eqn:master}; in the third line, we used the facts that $\tr [ A \hspace{0.4mm} \mathcal{T} (B)] = \tr [A \hspace{0.4mm} \sigma^{1/2} B \hspace{0.4mm} \sigma^{1/2} ] = \tr [ \mathcal{T}(A) \hspace{0.4mm} B]$ and  $\tr [ O^\dagger ] = \tr [ O ]^*$; in the final line, we defined a ``reversed'' Lindbladian $\widetilde{\Liouvillian}$ as the adjoint with respect to the inner product \eqref{eq:corr inner}, i.e.,
\begin{align}
\label{eq:T}
    \text{T} \hspace{0.4mm} : \hspace{0.4mm} \Liouvillian \mapsto \widetilde{\Liouvillian} \equiv \mathcal{T} \hspace{0.4mm} \Liouvillian^\dagger \mathcal{T}^{-1} \quad  \, , ~~
\end{align}
so that $\widetilde{B(t)} \equiv \ee^{t \widetilde{\Liouvillian}^\dagger} (B) $ in Eq.~\eqref{eq:simple reversal}, and $\widetilde{\Liouvillian}$ is analogous to the time-reversed generator $\widetilde{W}$ in the classical case \eqref{eqn:classical T}. It is straightforward to verify that the transformation T \eqref{eq:T} is $\Ints_2$, as one expects of a time-reversal operation; accordingly, we identify $\widetilde{\Liouvillian}$ \eqref{eq:T} as the \emph{time-reversed partner} to $\Liouvillian$ \eqref{eqn:master}, where
\begin{equation}
    \label{eq:reversed Lindblad action}
    \widetilde{\Liouvillian}(\rho) = \sigma^{1/2} \Liouvillian^\dagger ( \sigma^{-1/2} \rho \hspace{0.4mm} \sigma^{-1/2} ) \hspace{0.4mm} \sigma^{1/2} \, ,~~
\end{equation}
and we note that $\widetilde{\Liouvillian}(\sigma) = \sigma^{1/2} \Liouvillian^\dagger (\ident ) \hspace{0.4mm} \sigma^{1/2} = 0$ by the trace-preserving condition \eqref{eq:adjoint identity condition}, so that the time-reversed Lindbladian $\widetilde{\Liouvillian}$ \eqref{eq:T} has the \emph{same} stationary state $\sigma$ \eqref{eqn:sigma phi} as the original Lindbladian $\Liouvillian$ \eqref{eqn:master}.

We also comment that the particular definition of the correlation-function inner product \eqref{eq:corr inner} is \emph{required} for the time-reversed Lindbladian $\widetilde{\Liouvillian}$ to be a valid CPTP map~\cite{FagnolaQMST, FagnolaSQMST}. More generally, one could instead define a family of correlation-function inner products \eqref{eq:corr inner} given by
\begin{equation}
\label{eq:s-dep inner}
    \expval{A(t),B}_{\sigma,s} \equiv \tr \left[ A^\dagger (t) \sigma^{s} B \hspace{0.4mm} \sigma^{1-s} \right] \, ,~~
\end{equation}
where the choices $s=0$ and $s=1/2$ are the most common in the literature. Although one can, in principle, define a time-reversal transformation T \eqref{eq:T} with respect to the $s$-dependent inner product \eqref{eq:s-dep inner}, it is only for the \emph{symmetric} choice $s=1/2$ that  $\widetilde{\Liouvillian}$ is a valid Lindbladian~\cite{Carlen2017Semigroups, FagnolaQMST, FagnolaSQMST}. For other choices of $s \neq 1/2$, $\widetilde{\Liouvillian}$ is not, in general, completely positive. We also note that the symmetric $s=1/2$ correlation function is common in the analysis of correlations, locality, and spectral properties in chaotic systems~\cite{Romero-Bermudez:2019vej,FriedmanMIPT, lowdensitybound}. A similar construction was discussed in Ref.~\citenum{kwon}.

Importantly, the $\Ints_2$ transformation T \eqref{eq:T} defines the notion of \emph{quantum detailed balance} \cite{FagnolaQMST, FagnolaSQMST,Parzygnat:2022ldx}. An open quantum system with a Lindbladian $\Liouvillian$ \eqref{eqn:master} is said to obey quantum detailed balance (QDB) if
\begin{equation}
    \label{eq:QDB}
    \Liouvillian (\rho) - \widetilde{\Liouvillian} (\rho) = - 2 \ii [ H_\sigma, \rho ] \, ,~~
\end{equation}
for any $\rho$, where $H_\sigma$ commutes with the stationary state $\sigma$ \eqref{eqn:sigma phi} \cite{FagnolaQMST, FagnolaSQMST}. Intuitively, open systems that exhibit thermal dynamics with $\Phi = \beta H$ \eqref{eqn:sigma phi} are expected to obey QDB \eqref{eq:QDB}. This is because the time-reversal operation T \eqref{eq:T} flips the sign of the Hamiltonian term in $\widetilde{\Liouvillian}$ compared to $\Liouvillian$ \eqref{eqn:master}, leading to Eq.~\eqref{eq:QDB} with $H_\sigma = H$. The dissipative dynamics of systems that relax to thermal equilibrium can then be captured by Davies' generator~\cite{Davies1974_I, Davies1976_II}, which we discuss in Sec.~\ref{sec:davies}. We emphasize that, since the operator $\Phi$ encoding the steady state is generically unrelated to the generator $H$ of unitary evolution, the framework detailed in Sec.~\ref{sec:theory} extends beyond thermal systems. While Eq.~\eqref{eq:QDB} extends to generic nonthermal stationary states $\sigma$ upon replacing $H$ with any $H_\sigma$ satisfying $[H_\sigma,\sigma]=0$, we find  QDB to be less useful as a classification criterion than invariance (or not) under T \eqref{eq:T}.

Separately, we say that a Lindbladian $\Liouvillian$ is T even if and only if%
\begin{equation}
\label{eq:KMS main}
    \Liouvillian = \widetilde{\Liouvillian} \quad \Longleftrightarrow \quad \expval{A(t),B}_\sigma = \expval{A,B(t)}_\sigma \, ,~~
\end{equation}
which differs slightly from the definition of quantum detailed balance, except when $H_\sigma = 0$. Instead, being T even \eqref{eq:KMS main} is related to the Kubo-Martin-Schwinger (KMS) invariance~\cite{Kubo_1966, martinschwinger, ArakiKMS, Liulec} of generic thermal systems. In fact, recent works extending the successes of thermal effective field theories (EFTs)  \cite{WilsonI, WilsonII} to hydrodynamic systems and even beyond equilibrium have been organized around KMS invariance~\cite{haehl2016fluid,eft1,eft2,jensen2018dissipative,guo22, Qi:2022vyu, Mullins:2023ott, ACM}.

In the context of open classical systems \cite{ACM}, the classification of dynamical generators $W$ \eqref{eqn:Fokker Planck} -- and even particular terms in the generator -- is crucial to the diagnosis of the possible \emph{phases of matter} associated with a stationary distribution $\Phi$. In the classical setting, all terms corresponding to Hamiltonian dynamics (i.e., leading to equations of motion characterized by Poisson brackets) \emph{and} all terms due to dissipation (i.e., coming from stochastic noise sources) are guaranteed to be even under classical T~\eqref{eqn:classical T}. Hence, (closed) Hamiltonian systems, those that relax to thermal stationary states $\sigma \propto \exp (- \beta H)$, and dissipative relaxation to thermal states are all T even. In the classical setting, these dynamics also obey KMS invariance and detailed balance. However, nonreciprocal (and even \emph{active}) dynamics require the presence of terms in $W$ \eqref{eqn:Fokker Planck} that are \emph{odd} under T~\eqref{eqn:classical T}. Physically, these terms do \emph{not} result from integrating out degrees of freedom that are in thermal equilibrium with the system itself. In the context of self-propelled particles (e.g., birds), these additional nonthermal degrees of freedom correspond to internal ``batteries,'' which act as local sources and sinks of energy and entropy, potentially leading to nonthermal dynamics and stationary states, and even violations \cite{TonerTu} of the Mermin-Wagner theorem~\cite{merminwagner}.

We expect such an analysis of open quantum systems -- described by a Lindbladian $\Liouvillian$ \eqref{eqn:master} -- to be similarly fruitful. As in the classical setting, there are numerous definitions of T \eqref{eq:T}, detailed balance \eqref{eq:QDB}, and KMS invariance \eqref{eq:KMS main}, which we discuss further in Sec.~\ref{subsec:gT}. In fact, there are arguably even \emph{more} definitions for quantum systems. We also note that there are more ways to break these notions of T (and also QDB and KMS) in quantum systems, as both the ``Hamiltonian'' term and the dissipative jump operators in the Lindbladian \eqref{eqn:master} can be T odd, in contrast to the classical case. Moreover, as we discuss in Sec.~\ref{subsec:Q break T}, one can break T in ways that have classical analogues, and also in ways that are unique to the quantum setting. We relegate a classification of the nonequilibrium phases of open quantum systems and their corresponding dynamics to future work, though we expect notions of T \eqref{eq:T} to play a crucial role.

\subsection{Incorporating symmetries}
\label{subsec:symmetries}

Before discussing generalizations of the time-reversal transformation T \eqref{eq:T}, we first briefly discuss the notions of weak versus strong symmetries in open systems and the action of symmetry transformations on the dynamical generator $\Liouvillian$ \eqref{eqn:master}. In the following discussion, we primarily highlight comparisons to the effective theories of open classical systems \cite{ACM} and the general action of symmetries on the Linbladian $\Liouvillian$ \eqref{eqn:master} in abstract terms. Although we expect that the existence -- and possibly, spontaneous breaking -- of one or more symmetries is important to constructing effective theories of open quantum systems, in the applications to quantum error correction that we consider herein, there is generally no symmetry restriction on the terms in $\Liouvillian$ \eqref{eqn:master}. Hence, we only briefly discuss the constraints imposed by symmetries on $\Liouvillian$ \eqref{eqn:master} when discussing applications to quantum error correction in Sec.~\ref{sec:application-to-QEC}.

In the context of open systems -- both classical and quantum -- there are two distinct notions of symmetries: weak and strong \cite{Buca2012weakstrong, Lieu2020, ACM}. In the classical setting discussed in Sec.~\ref{sec:ACM review}, symmetries are defined with respect to the Fokker-Planck generator $W$ \eqref{eqn:Fokker Planck}. A \emph{strong} symmetry of $W$ is one that holds on \emph{every} stochastic trajectory, captured by the condition $\ee^{-F} W \hspace{0.4mm} \ee^{F} = W$ for some conserved ``charge'' $F (\bvec{q})$, which is a function of the coordinates $\bvec{q} = \{q_a\}$.\footnote{The operator $W$ \eqref{eqn:Fokker Planck} may also depend on the coordinates $\{q_a\}$, and generically involves differential operators $\pd*{a} = \partial / \partial q_a$.} Conservation of $F$ (in the strong sense) is guaranteed provided that the operator $W$ \eqref{eqn:Fokker Planck} is invariant under shifting the differential operator according to $\pd*{a} \to \pd*{a} + [ \pd*{a} F]$ \cite{ACM}. Conversely, a \emph{weak} symmetry of $W$ is one that only holds \emph{on average}, meaning that $\pd{t} \expval{F(\bvec{q})} = 0$. This is guaranteed provided that $W^T F = 0$ \cite{ACM}.

In open \emph{quantum} systems, we again have both weak and strong notions of symmetries, which we now define explicitly. In particular, consider a symmetry group $G$, whose elements $g \in G$ have some \emph{unitary} representation $U(g)$ acting on the Hilbert space $\mathcal{H}$ of interest. For each element $g \in G$, we define the left (L) and right (R) action of $g$ on a density operator $\rho$ via the following pair of superoperators:
\begin{subequations}
    \begin{align}
        \mathcal{U}_{g,\text{L}}(\rho) &= U(g) \rho, \\
        \mathcal{U}_{g,\text{R}}(\rho) &= \rho \hspace{0.4mm} U^{-1} (g)  = \rho  \hspace{0.4mm} U (g^{-1}) \, , ~~
    \end{align}
\end{subequations}
where the subscript refers to the side of $\rho$ to which the unitary $U(g)$ (or its inverse) is applied. 

The group $G$ is a \emph{strong symmetry}  of the Lindbladian $\Liouvillian$ \eqref{eqn:master} if, for all $g \in G$, we have that
\begin{equation}
    \comm*{\Liouvillian}{\mathcal{U}_{g,\text{L}}} = \comm*{\Liouvillian}{\mathcal{U}_{g,\text{R}}} = 0 \, , ~~ \label{eq:strongsymmetry}
\end{equation}
meaning that the commutator of \emph{super}operators $\Liouvillian$ and ${\cal U}_{g,\text{L/R}}$ vanishes acting on any operator $\rho$. 

The group $G$ is a \emph{weak symmetry} of the Lindbladian $\Liouvillian$ \eqref{eqn:master} if we only have that
\begin{equation}
    \comm*{\Liouvillian}{\mathcal{U}_{g,\text{L}} \; \mathcal{U}_{g,\text{R}}} = 0 \, ,~~ \label{eq:weaksymmetry}
\end{equation}
meaning that only the \emph{combination} of left and right action of $g \in G$ commutes with $\Liouvillian$ \eqref{eqn:master}.

The distinction between weak and strong symmetries is important to, e.g., the application of our methods to quantum error correction \cite{Lieu2020,Lieu2023dissipative}, which we discuss in Sec.~\ref{sec:application-to-QEC}. Note that even a weak symmetry of the Lindbladian necessarily implies a symmetry of (at least one) stationary state $\sigma$, i.e.,
\begin{equation}
    \mathcal{U}_{g,\text{L}} \hspace{0.5mm} \mathcal{U}_{g,\text{R}} (\sigma) = U(g) \hspace{0.4mm} \sigma \hspace{0.4mm} U^{-1}(g) = \sigma \, , ~~
\end{equation}
though the converse is \emph{not} true: Symmetries of the stationary state $\sigma$ \eqref{eqn:sigma phi} do not imply weak or strong symmetries of $\Liouvillian$ \eqref{eqn:master}. This is particularly relevant to the discussion of quantum error correction in Sec.~\ref{sec:application-to-QEC}.

\subsection{Generalizations of time reversal}
\label{subsec:gT}

Even in the context of open classical systems, the definition of time reversal T \eqref{eq:T} is \emph{not} unique \cite{ACM}. For example, one may associate the na\"ive ``reversibility transformation'' T \eqref{eqn:classical T} with time reversal, as is common in systems with discrete state spaces. However, when working with continuous state spaces -- e.g., involving canonical positions $x_i$ and momenta $p_i$ -- it is common to combine T \eqref{eqn:classical T} with another ``microscopic'' $\Ints_2$ transformation $p_i \to - p_i$, which captures the fact that momentum coordinates are expected to be odd under time reversal. Additionally, there are numerous classical systems for which $W$ \eqref{eqn:Fokker Planck} is not symmetric under T, but is instead invariant under a \emph{generalized} time-reversal operation gT that combines the transformation T \eqref{eqn:classical T} with another $\Ints_2$ symmetry, such as a parity operation, spatial inversion, or swapping the roles of ``predator'' and ``prey'' in nonreciprocal Kuramoto models~\cite{ACM, kuramoto1975, Kuramoto_RMP}.

Indeed, alternative definitions of T to Eq.~\eqref{eq:T} exist in the context of open quantum systems. As in the classical case~\cite{ACM}, we expect that certain definitions of T may be more illuminating or analytically useful in the context of different physical systems.\footnote{We expect such details to be more important to classifying phases of open quantum systems than to the engineering of particular stationary states -- and correction of generic errors -- that we consider herein.} A particularly natural extension is to combine T \eqref{eq:T} with the \emph{microscopic} implementation of time reversal on a generic operator $O$, given by
\begin{equation}
\label{eq:Sigma micro T}
    \widetilde{O} = \mathcal{K} (O) = U K O K U^\dagger \, ,~~
\end{equation}
where we use a tilde to denote the time-reversed partner of a given operator $O$ (including density matrices), $K$ is the \emph{anti}unitary (and antilinear) operator that realizes complex conjugation, and $U$ is a unitary operator. The form of $U$ depends on the physical nature of the underlying degrees of freedom: intrinsic spins 1/2 have $U=Y$, so that $\mathcal{K} (\sigma^\nu) = - \sigma^\nu$ for any Pauli label $\nu=x,y,z$; other systems may have $U = \ident$. The form of $U$ is further constrained by the fact that, for the superoperator $\mathcal{K}$ \eqref{eq:Sigma micro T} to realize time reversal, it must be $\Ints_2$ valued -- i.e., an involution satisfying $\mathcal{K}^2 (O) = O$ for any operator $O$ -- to be $\Ints_2$ valued. 

For closed quantum systems, $\mathcal{K}$ \eqref{eq:Sigma micro T} provides the \emph{only} notion of time reversal, with
\begin{equation}
    \label{eq:T closed}
    \mathcal{K} [ \rho (t) ] = \mathcal{K} \left[ \ee^{-\ii t H}  \rho (0)  \ee^{\ii t H}  \right] = \ee^{\ii t H}  \rho (0)  \ee^{- \ii t H} = \rho (-t) \, ,~~ 
\end{equation}
assuming that the Hamiltonian $H$ and initial state $\rho(0)$ are T even (i.e., $ \widetilde{H} = \mathcal{K} (H) = H$, and likewise for $\rho$). However, this need not be the case in general. The transformation $\mathcal{K}(O)$ \eqref{eq:Sigma micro T} realizes time reversal for any operator $O$ to which it is applied, \`a la $p \to -p$ in the classical case; when applied to the unitary evolution operator, $\mathcal{K}$ sends $t \to -t$ (and may modify $H$ itself), as in Eq.~\eqref{eq:T closed}.

To realize a version of T \eqref{eq:T} incorporating the transformation $\mathcal{K}$ \eqref{eq:Sigma micro T}, we first define
\begin{align}
    \label{eq:corr inner T version}
    \expval{A(t),B}_{\sigma}^{K} &\equiv \tr \left[ \ee^{t \hspace{0.4mm} \widetilde{\Liouvillian}_K^\dagger} (\widetilde{B}^\dagger)  \widetilde{\mathcal{T}} ( \widetilde{A} ) \right]  = \tr \left[ \ee^{t \hspace{0.4mm} \mathcal{T}^{-1} \mathcal{K} \widetilde{\Liouvillian}_K^{\vpd} \mathcal{K} \mathcal{T}} (A^\dagger) \mathcal{T} (B) \right]\, ,~~
\end{align}
where $\widetilde{\Liouvillian}^{\,}_K$ is the as-yet undetermined analogue of $\widetilde{\Liouvillian}$ for the version of T \eqref{eq:T} that includes $\mathcal{K}$ \eqref{eq:Sigma micro T}, and all other tildes denote the application of $\mathcal{K}$ \eqref{eq:Sigma micro T}. In particular, we have that
\begin{align}
    \label{eq:super T tilde}
    \widetilde{\mathcal{T}} ( \widetilde{A} )  \equiv \widetilde{\sigma}^{1/2} \widetilde{A} \hspace{0.4mm} \widetilde{\sigma}^{1/2} = \mathcal{K} \left( \mathcal{T} (A ) \right) \, ,~~
\end{align}
and we recover an expression for the time-reversed Lindbladian $\widetilde{\Liouvillian}^{\,}_K$ \eqref{eq:corr inner T version} by demanding that
\begin{equation}
\label{eq: general T}
    \expval{A(t),B}^{\vpp}_{\sigma} = \expval{A(t),B}_{\sigma}^{K} \, ,~~
\end{equation}
and manipulating both sides leads to
\begin{align*}
    \tr \left[ \ee^{t \hspace{0.4mm}  \Liouvillian^\dagger} (A^\dagger) \mathcal{T} (B) \right] &= \tr \left[ \ee^{t \hspace{0.4mm} \widetilde{\Liouvillian}^{\dagger}_K} \left(\mathcal{K} (B^\dagger) \right) \hspace{0.4mm} \mathcal{K} \left( \mathcal{T} ( A ) \right) \right]  \\
    \tr \left[ \ee^{t \hspace{0.4mm}  \mathcal{T}^{-1} \Liouvillian \hspace{0.4mm} \mathcal{T}} (B) \mathcal{T} (A^\dagger) \right] &= \tr \left[ \ee^{t \hspace{0.4mm}  \mathcal{K} \widetilde{\Liouvillian}^{\dagger}_K \hspace{0.4mm}  \mathcal{K}} (B) \hspace{0.4mm}  \mathcal{T} ( A^\dagger ) \right]  \, , ~~
\end{align*}
since $\mathcal{K}$ \eqref{eq:Sigma micro T} is its own inverse and adjoint, and $\tr [ \mathcal{K} (A) \mathcal{K} (B) ] = \tr [A B]^* = \tr [ B^\dagger A^\dagger]$. We then find that
\begin{equation}
    \label{eq:gT}
    \text{gT} : \Liouvillian \mapsto \widetilde{\Liouvillian}^{\vps}_K \equiv \mathcal{K} \mathcal{T} \Liouvillian^\dagger \mathcal{T}^{-1} \mathcal{K} \, ,~~
\end{equation}
which is equivalent to the original transformation T \eqref{eq:T} up to sandwiching $\widetilde{\Liouvillian}$ with the superoperator $\mathcal{K}$ \eqref{eq:Sigma micro T} on both sides \cite{FagnolaQMST, FagnolaSQMST}. The notions of being ``T even'' and of quantum detailed balance are the same as before, and agree with the various notions of ``quantum detailed balance'' that appear in Refs.~\citenum{FagnolaQMST} and \citenum{FagnolaSQMST}, which only considers the T-even case in which $\mathcal{K}(\sigma)=\sigma$, simplifying matters substantially. In this special case, Eq.~\eqref{eq:gT} above follows automatically from the definition of T~\cite{FagnolaSQMST}.

In fact, this notion of gT \eqref{eq:gT} agrees with that of a recent series of papers on ``hidden time-reversal symmetry'' \cite{HiddenTRS, HiddenTRS2, HiddenTRS3, HiddenTRS4, HiddenTRS5}, which represent the dynamics generated by $\Liouvillian$ \eqref{eqn:master} on a doubled Hilbert space. There, the ``hidden'' notion of time reversal is intimately connected to the ability to compute the stationary state $\sigma$ for a given gT-even Lindbladian $\Liouvillian$. These papers present a distinct but complementary physical motivation for the construction of gT, which can play an important role in classifying \emph
{universal} dynamics and phase structure in open quantum systems -- i.e., identifying classes of Lindbladians $\Liouvillian$ whose behavior over large spatiotemporal scales is robust to microscopic details and fully characterized by only a handful of ``universal'' exponents.\footnote{See, e.g., Ref.~\citenum{ConstrainedRUC} for several examples of novel dynamical universality classes in \emph{isolated} quantum systems.} Compared to the standard approach -- which starts from a microscopic Lindbladian that may not have a known stationary state -- our framework allows for the direct classification of the phases of matter realized by a stationary state $\sigma$~\eqref{eqn:sigma phi}, as well as the ``universal'' dynamics that relax to that $\sigma$. Phases of matter, e.g., can be diagnosed by evaluating ``order parameters'' in the stationary state $\sigma$; more generally, universality can be diagnosed by considering correlation functions and linear response, either in the stationary state $\sigma$ or while the system relaxes to $\sigma$. 

Finally, we comment that other choices of T (and gT) may be identified, and may be more appropriate for particular open quantum systems. As in the classical case \cite{ACM}, it may be beneficial in certain contexts to combine T with another $\Ints_2$ symmetry (which may be a subgroup of a larger symmetry group) to obtain a gT symmetry. In practice, this would manifest in a modification of unitary $U$ in Eq.~\eqref{eq:Sigma micro T}. Note that one could, in principle, include only $U$ (and not $K$) in a given definition of T, so long as $U^2=\ident$. Because the appropriate choice of T is likely to depend on the particular system of interest, we relegate elsewhere further discussion of T, its variants, and their implications.


\section{Formalism}
\label{sec:theory}


\subsection{Lindbladians consistent with stationarity}
\label{sec:stationarity-eigenbasis}

We now discuss how the existence of a target stationary state $\sigma$ \eqref{eqn:sigma phi} constrains the form of the Lindbladian $\Liouvillian$~\eqref{eqn:master}. We first detail the implications of the time-reversal transformation T~\eqref{eq:T} on $\Liouvillian$~\eqref{eqn:master}, working in the eigenbasis of $\sigma$. We explicitly consider the canonical example of  Davies' generator, which describes relaxation to thermal states $\sigma = \exp (- \beta H)$. While this approach can be extended to nonequilibrium states $\sigma$, and allows one to recover the most general family of Lindbladians consistent with a given stationarity state $\sigma$, the corresponding dynamics are generically highly nonlocal. Deriving \emph{local} Lindbladians consistent with relaxation to arbitrary mixed states $\sigma$ is the subject of Sec.~\ref{sec:stationarity}.

\subsubsection{General construction}

Suppose that the full-rank stationary density matrix $\sigma$~\eqref{eqn:sigma phi} has eigenstates $\set{\ket{a}}$ and corresponding positive-semidefinite eigenvalues $\{ \sigma_a \}$, which may be arbitrarily degenerate. In other words, 
\begin{equation}
    \sigma = \sum_a \sigma_a \BKop{a}{a}
    \, ,~~
    \label{eqn:sigma-decomp}
\end{equation}
where the eigenvalues $0 \leq \sigma_a \leq 1$ are interpreted as a probability distribution over eigenstates $\ket{a}$ of $\sigma$. 
These conditions ensure that $\sigma$ is a valid density matrix. In the following discussion, we also require that $\sigma_a>0$, such that $\sigma$ is full rank [see assumption~\ref{asn:positive}] and thus invertible.
Since we work in the eigenbasis of $\sigma$, a natural set of jump operators are those that induce transitions between these eigenstates, such as $\BKop{a}{b}$. In this basis, we have that
\begin{equation} \label{eq:F}
    \Liouvillian (\rho) = -\ii [H, \rho] + \sum_{a a' b b'}  L^{aa'}_{bb'} \left( \BKop{a}{b} \rho \BKop{b'}{a'}  -  \kron{a,a'}  \frac{1}{2} \{ \BKop{b'}{b}, \rho \} \right) 
    \, ,~
\end{equation}
for some set of coefficients $L^{aa'}_{bb'}$
that define a positive-semidefinite matrix when $(ab)$ and $(a'b')$ are each treated as a single index. Specifically, the coefficient $L^{aa'}_{bb'}$ induces transitions between 
$\rho_{bb'}$ and $\rho_{aa'}$.
Substituting Eq.~\eqref{eqn:sigma-decomp} into Eq.~\eqref{eq:F}, we find that stationarity of $\sigma$ requires that the Hamiltonian $H$ and the coefficients $L^{aa'}_{bb'}$ be chosen in so as to satisfy
\begin{equation} 
\label{eq:stationarity}
    \matel{a}{\Liouvillian(\sigma)}{b} = -\ii (\sigma_b - \sigma_a) H_{ab} + \sum_c \left[\sigma_c L^{ab}_{cc} - \frac{1}{2} (\sigma_a + \sigma_b) L^{cc}_{ba} \right] \betterbe 0 \, ,~
\end{equation}
for all eigenstates $a$ and $b$ of $\sigma$. Note that, if we fix all the coefficients $L^{aa'}_{bb'}$,
the Hamiltonian matrix elements $H_{ab}$ corresponding to nondegenerate eigenvalues $\sigma_a \neq \sigma_b$ are uniquely determined by Eq.~\eqref{eq:stationarity}. We further observe that the diagonal elements $H_{aa}$ of the Hamiltonian do \emph{not} contribute to stationarity, and are thus arbitrary. On the other hand, to ensure stationarity, the coefficients $L^{aa'}_{bb'}$ must satisfy 
\begin{equation}
    \sum_c \sigma_c L^{aa}_{cc} - \sigma_a L^{cc}_{aa} \betterbe 0
    \, .~~
\end{equation}
We comment that the foregoing pair of equations are identical to the constraints required for stationarity of a \emph{classical} Markov process \eqref{eqn:Pin-Pout}, and can therefore be satisfied by finding a solution thereof.
Namely, $\{ \sigma_a \}$ is the equilibrium probability distribution, and the coefficients $L_{aa}^{cc}$ describe the rate of transitions between diagonal density matrix elements.
Finally, the matrix elements $H_{ab}$ corresponding to degenerate eigenvalues $\sigma_a = \sigma_b$ for $a \neq b$ once again do not affect the stationarity of $\sigma$ and can be chosen arbitrarily, but the constraints satisfied by the coefficients $L^{aa'}_{bb'}$ are significantly more involved.

We now elucidate the effect of the time-reversal transformation T~\eqref{eq:T}. The time-reversed Lindbladian $\widetilde{\mathcal{L}}$ is parameterized\footnote{
The jump operators $\ket{a}\bra{b}$ are not traceless. However, having fixed a basis, the time reversal transformation T is still uniquely determined, as is the reversed Lindbladian $\tilde{\Liouvillian}$.
} in terms of a time-reversed system Hamiltonian $\widetilde{H}$ and coefficients $\widetilde{L}^{aa'}_{bb'}$ according to
\begin{equation} \label{eq:widetilde L}
    \widetilde{\Liouvillian}(\rho) = -\ii [\widetilde{H}, \rho] + \sum_{a,a',b,b'}  \widetilde{L}^{aa'}_{bb'} \left( \BKop{a}{b} \rho \BKop{b'}{a'}  - \kron{aa'} \frac{1}{2} \{ \BKop{b'}{b} , \rho \} \right) 
    \, .~
\end{equation}
As in the classical case \cite{ACM}, the appropriate notion of generalized time reversal depends on the steady state. As a result, the Hamiltonian matrix elements and jump operator coefficients transform as
\begin{subequations}
\begin{align}
    \widetilde{H}_{ab} &= -\frac{1}{\sqrt{\sigma_a \sigma_b}} \left[ \frac{1}{2}(\sigma_a +\sigma_b) H_{ab} + \frac{\ii}{4}\left( \sigma_a - \sigma_b \right) \sum_c  L^{cc}_{ba}  \right] \\
    \widetilde{L}^{aa'}_{bb'} &= \sqrt{\frac{\sigma_a \sigma_{a'}}{\sigma_b \sigma_{b'}}}\, L^{b'b}_{a'a} \, . \label{eq:classicalTrule} 
 \end{align}
 \label{eq:T-transformations}%
\end{subequations}
Note that we used the stationarity conditions \eqref{eq:stationarity} to derive the relation above, and we recall that $\sigma$ is assumed to be invertible. As a result, the transformation T~\eqref{eq:T-transformations} can only be applied when $\Liouvillian(\sigma) = 0$.
For the diagonal matrix elements (i.e., $a'=a$ and $b'=b$), we have $\sigma_b \widetilde{L}^{aa}_{bb} = \sigma_a L^{bb}_{aa}$. Consequently, if the dissipative part of $\Liouvillian$ is T even, then the quantum detailed balance condition reduces to classical detailed balance~\eqref{eqn:classical DB} for the diagonal matrix elements. Moreover, if the dynamics is T even, then the Hamiltonian matrix elements $H_{ab}$ corresponding to degenerate eigenvalues $\sigma_a = \sigma_b$ are fixed to be zero. Hence, if $\Liouvillian$ protects $\sigma$ and is T even, then all the $H_{ab}$ are uniquely determined in terms of the coefficients $L^{aa'}_{bb'}$.


\subsubsection{Davies' generator as a special case}
\label{sec:davies}

In certain limits, a system interacting weakly with a Markovian bath can be described by an effective Lindbladian known as \emph{Davies' generator}~\cite{Davies1974_I,Davies1976_II}. We now show how Davies' generator naturally arises when considering dynamics that are even under T~\eqref{eq:T}. This serves as a useful point of reference for the more general framework presented in Sec.~\ref{sec:stationarity}.

Consider a target state generated by $\Phi = \beta H$ for some  Hamiltonian $H = \sum_a E_a \BKop{a}{a}$ and temperature $T = \beta^{-1}$. The goal is to identify a family of Lindbladians whose steady states are the Gibbs state $\sigma \propto \ee^{-\beta H}$. 
Consider the jump operators
\begin{align}
    A_\omega = \sum_{E_a-E_b = \omega} g_{ab}(\omega) \BKop{a}{b}
    \, ,~
    \label{eqn:davies-jumpop}
\end{align}
where the sum is over all states $a$ and $b$ satisfying $E_a = E_b + \omega$. As a result, the jump operator $A_\omega$ leads to transitions between eigenstates of $\sigma$ separated in energy (with respect to $H$) by $\omega$.
These operators satisfy
\begin{align}
    A_{-\omega} &= \sum_{E_a - E_b = \omega} g_{ba}(-\omega)  \ket{b}\bra{a} \\
    A^{\dagger}_{\omega} &= \sum_{E_a - E_b = \omega} g^*_{ab}(\omega) \ket{b}\bra{a}
    \, ,~
\end{align}
where the star denotes complex conjugation.
Comparing these two equations, we can ensure that $A_{-\omega} = A_{\omega}^\dagger$ if the coefficients $g_{ab}(\omega)$ are chosen such that they respect the constraint $g_{ab}^*(\omega) = g_{ba}(-\omega)$. Now we can construct dynamics generated by these jump operators with positive semidefinite rates $\gamma(\omega)$
\begin{equation}
    \Liouvillian(\rho) = \sum_{\omega \in B} \gamma(\omega) \left[ A^{\vpd}_\omega \rho A_\omega^\dagger - \frac{1}{2} \{ A^\dagger_\omega A^{\vpd}_\omega, \rho \} \right]
    \, ,
    \label{eqn:Davies}
\end{equation}
where the summation is over the Bohr frequencies, i.e., energy differences, $B = \{ E_a - E_b \, | \, E_a, E_b \in \spec(H) \}$. Note that the dynamics in Eq.~\eqref{eqn:Davies} is purely dissipative; since the Hamiltonian is diagonalized by the same eigenbasis as $\sigma$, we could choose to include unitary dynamics generated by $H$ in Eq.~\eqref{eqn:Davies} without affecting stationarity of $\sigma$.
To apply the time-reversal transformation~\eqref{eq:widetilde L}, we express the Lindbladian~\eqref{eqn:Davies} in the eigenbasis of $\sigma$. This leads to the coefficients
\begin{align}
    L^{aa'}_{bb'} = \delta(\omega_{ab}-\omega_{a'b'}) \gamma(\omega_{ab})  g_{ab}(\omega_{ab}) g^{*}_{a'b'}(\omega_{ab})
    \, ,
\end{align}
where $\delta(\omega)$ is the Kronecker delta. Using Eq.~\eqref{eq:widetilde L}, we require that the dynamics be even under T~\eqref{eq:T},
\begin{align}
    L^{aa'}_{bb'} \stackrel{!}{=}
    \widetilde{L}^{aa'}_{bb'} =
    \sqrt{\frac{\sigma_a \sigma_{a'}}{\sigma_b \sigma_{b'}}} L^{b'b}_{a'a} &= \ee^{-\beta\omega_{ab}} \delta(\omega_{ab}-\omega_{a'b'}) \gamma(-\omega_{ab})  g^{*}_{ba}(-\omega_{ab}) g_{b'a'}(-\omega_{ab}) \notag \\
    & = \ee^{-\beta\omega_{ab}} \delta(\omega_{ab}-\omega_{a'b'}) \gamma(-\omega_{ab})  g_{ab}(\omega_{ab}) g^{*}_{a'b'} (\omega_{ab})
    \, .
\end{align}
These equations can be satisfied if the decay rates are chosen to satisfy $\gamma(\omega) = \ee^{-\beta \omega} \gamma(-\omega)$,
frequently known as the Kubo-Martin-Schwinger (KMS) condition. In fact, Eq.~\eqref{eqn:Davies} supplemented by the KMS condition recovers the well known Davies' generator~\cite{Davies1974_I, Davies1976_II}.
Absent any unitary Hamiltonian contribution to the dynamics, being T even under~\eqref{eq:T} is equivalent to satisfying quantum detailed balance. Furthermore, using Eq.~\eqref{eq:stationarity}, one can easily verify  that Davies' generator annihilates the Gibbs states $\ee^{-\beta H}$, as required.


\subsection{Local dynamics compatible with stationarity for stabilizer \texorpdfstring{$\Phi$}{Phi}}
\label{sec:stationarity}


\subsubsection{A convenient operator basis}
\label{subsub:operator basis}

For many-body systems, a Lindbladian that takes the form of Eq.~\eqref{eq:F} is, in general, not very useful from the perspective of dissipative state preparation.
The reason is that the jump operators $\ket{a}\bra{b}$ are highly nonlocal, even for relatively simple many-body steady states $\sigma$.
In practice, efficient implementation of the dissipative dynamics requires \emph{local} Lindbladians -- i.e., those consisting of a local Hamiltonian $H$ and local jump operators.
We now show that, given a stationary state $\sigma = \ee^{-\Phi}$ in which $\Phi$ is a sum of commuting local operators \eqref{eqn: Phi}, we can
write down a simple ansatz that captures a large number of the possible (local) dynamics that protect  $\sigma$.
The remaining dynamics \emph{not} captured by this ansatz can instead be generated using the methods presented in Appendix~\ref{sec:T-odd}. Finally, a useful discussion of a restricted set of frustration-free Lindbladians with steady state $\sigma$ appears in Ref.~\citenum{viola1}.

The first step of the construction is to identify a convenient local basis for the jump operators that appear in the Lindbladian $\Liouvillian$~\eqref{eqn:master}.
To construct the most convenient basis, consider how the transformation $\Liouvillian \to \widetilde{\Liouvillian}$~\eqref{eq:T} affects the Lindbladian in a generic basis ${\{ B_i \}}$ for the jump operators. In particular, a Lindbladian of the form~\eqref{eqn:master} under T~\eqref{eq:T} is sent to
\begin{equation}    
    \widetilde{\mathcal{L}}(\rho) = \text{i}\left[ \mathcal{S}(H)\rho - \rho \mathcal{S}^{-1}(H) \right] + \sum_{ij}{\gamma}_{ij} \left[
    \mathcal{S}(B_j^{\dagger}) \rho \mathcal{S}^{-1}(B^{\vpd}_i) 
    - \frac{1}{2} \left(
        \mathcal{S}(B_j^\dagger) \mathcal{S}(B^{\vpd}_i)  \rho 
        + \rho \mathcal{S}^{-1}(B_j^\dagger) \mathcal{S}^{-1}(B^{\vpd}_i)   
        \right)
    \right]
    \, .
    \label{eqn:generic-transformed-L}
\end{equation}
In order to rewrite the reversed operator $\widetilde{\Liouvillian}$ in Lindblad form, we must complete the action of the superoperator $\mathcal{S}$ on the jump operators $B_i$,
where $\mathcal{S}$ is defined in terms of the steady state $\sigma$~\eqref{eqn:sigma phi} via
\begin{equation}
\label{eq:S}
    \mathcal{S}(A) = \sigma^{1/2} A \hspace{0.4mm} \sigma^{-1/2} 
    \, ,~~
\end{equation}
and we note the subtle difference between the superoperator $\mathcal{S}$ and the operator $\mathcal{T}$~\eqref{eqn:T-def} associated with time reversal.
Since the basis $\{ B_i\}$ is assumed to be complete and orthonormal, we can rewrite the action of the map $\mathcal{S}$~\eqref{eq:S} in terms of complex coefficients $s_{ij}$ satisfying
\begin{align}
    \mathcal{S}(B_i) 
    &= \sum_j s_{ij} B_j 
    \, ,
    \label{eqn:superop-eigenvalues}
\end{align}
where $s_{ij} = \tr [B_i^{\dagger} \mathcal{S}( B_j)]$, from which it follows that the matrix defined by $s_{ij}$ is Hermitian.
We can therefore diagonalize $s_{ij}$ to find a new operator eigenbasis $\{ A_i \}$ of the transformation $\mathcal{S}$~\eqref{eq:S}, i.e.,
\begin{equation}
    \label{eq:S eig}
    \mathcal{S}(A_i) = c_i A_i \, ,~~
\end{equation}
where, since $\sigma$ \eqref{eqn:sigma phi} is positive definite (Assumption~\ref{asn:positive}), the eigenvalues $c_i$ of the map~\eqref{eqn:superop-eigenvalues} are also positive. 

Recall that we have assumed $\Phi$ is a sum of strictly local commuting operators -- i.e., $\Phi = \sum_x \phi_x$, where $[\phi_x, \phi_y]=0$ and each $\phi_x$ is supported only in the vicinity of some site $x$. It follows that $\mathcal{S}(B_i)$~\eqref{eqn:superop-eigenvalues} depends only on the terms $\phi_x$ that do not commute with $B_i$. For any local operator $B_i$, $\mathcal{S}(B_i)$ is then a sum of products between $B_i$ and (a subset of) the terms $\phi_x$ that overlap with the chosen operator $B_i$, which are generated by nested commutators of the form $[\phi_x,[\dots,[\phi_y, B_i]\dots]]$. Since $B_i$ and $\phi_x$ are local, only a finite number of $\phi_x$ overlap with $B_i$. Similarly, the number of such products that are linearly independent is finite because we restrict to finite onsite Hilbert spaces (see Assumption~\ref{asn:finite-dimensional}). Furthermore, when $\mathcal{S}$~\eqref{eq:S} acts on these products of local operators, the result can be similarly decomposed onto the same set of linearly independent products. The action of $\mathcal{S}$~\eqref{eq:S} on this basis of tensor-product operators can therefore be represented by a  matrix, which can be diagonalized to obtain \emph{local} eigenoperators that satisfy Eq.~\eqref{eq:S eig}. Repeating this procedure for different local $B_i$ allows for the construction of a complete local basis $\{ A_i \}$ satisfying Eq.~\eqref{eq:S eig}. This procedure is carried out explicitly in Sec.~\ref{sec:qubit-jump-ops} for the special case in which the $\phi_x$ correspond to finite Pauli strings.

Note that the operators $\{ A_i^\dagger \}$ are also eigenoperators of $\mathcal{S}$~\eqref{eqn:superop-eigenvalues}, with ${\mathcal{S}(A_i^{\dagger}) = c_i^{-1} A_i^{\dagger}}$. As a consequence, for any given $i$, $A_i^\dagger$ is either equal to or orthogonal to $A_i$. We therefore introduce the permutation $\pi$ that describes the relationship between the $A^{\vpd}_i$ and the $A_i^\dagger$ operators, namely $A_{\pi(i)} = A_{i}^\dagger$. Hermitian jump operators are mapped to themselves, while other operators undergo a swap (transposition) with their corresponding Hermitian conjugate, which implies that the corresponding permutation is an involution (i.e., $\pi^2(i) = i$ for all $i$). Making use of this notation, if the unitary part of $\Liouvillian$ is described by the Hamiltonian $H = \sum_i h_i A_i$, the time-reversed Lindbladian $\widetilde{\Liouvillian}$~\eqref{eqn:generic-transformed-L} is written in the form of Eq.~\eqref{eqn:master} with
\begin{subequations}
\begin{align}  
    \widetilde{H} &= -\frac{1}{2} \sum_i  \left( c_i + \frac{1}{c_i} \right) h_i A_i - \frac{\ii}{4} \sum_{ij}  \gamma_{ij} \left( \frac{c_i}{c_j} - \frac{c_j}{c_i} \right) A_j^\dagger A_i^{\vpd}
    \, , \label{eqn:reversed-H-eigenbasis} \\
    \widetilde{\gamma}_{ij} &= \gamma_{\pi(j) \pi(i)} c_i c_j \label{eqn:reversed-gamma-eigenbasis}
    \, ,
\end{align}%
\label{eqn:eigenop-transformation}%
\end{subequations}
where, as in Eq.~\eqref{eq:T-transformations}, we made use of  stationarity of $\sigma$ to derive these transformations. Explicitly, stationarity of $\sigma$ enforces that the coefficients $h_i$ and $\gamma_{ij}$ satisfy
\begin{align} \label{stationary}
    \mathcal{T}^{-1} \Liouvillian(\sigma) =
    \ii\sum_i \left( c_i - \frac{1}{c_i} \right) h_i A_i +
    \sum_{ij} \gamma_{ij} \left[ \frac{1}{c_i c_j} A_i^{\vpd} A_j^{\dagger} - \frac{1}{2} \left(\frac{c_j}{c_i} + \frac{c_i}{c_j}\right) A_j^{\dagger} A_i^{\vpd} \right]  \stackrel{!}{=} 0
    \, .
\end{align}
Note that $\widetilde{H}$~\eqref{eqn:reversed-H-eigenbasis} is Hermitian, and that the transformed $\widetilde{\gamma}_{ij}$~\eqref{eqn:reversed-gamma-eigenbasis} remains Hermitian and positive semidefinite, such that the resulting reversed Lindbladian $\widetilde{\Liouvillian}$ is indeed CPTP, as required.
The relationship between the more general transformation~\eqref{eqn:eigenop-transformation} and the eigenbasis version presented in Eq.~\eqref{eq:widetilde L} can be understood by noting that the jump operators $\BKop{a}{b}$ are eigenoperators of $\mathcal{S}$~\eqref{eq:S} with $c_{ab} = \sqrt{\sigma_a / \sigma_b}$.
However, we seek a \emph{local} basis of jump operators that diagonalize $\mathcal{S}$. In cases where the eigenvalues $c_{ab}$ are degenerate, we have freedom in which linear combinations we take, and this flexibility can be utilized to construct a more local operator basis.


\subsubsection{Local jump operators}
\label{sec:local-jumps}

We now introduce the family of steady states with which we work. These states admit simple, strictly local eigenoperators of~$\mathcal{S}$~\eqref{eq:S}. Specifically, we work primarily with \emph{stabilizer} steady states, corresponding to finite-temperature stabilizer ``Hamiltonians.''
This choice not only allows us to make considerable analytical progress, but also gives us access to states that are of interest experimentally, often by virtue of their relevance to quantum error correction.
We consider steady states $\sigma = \ee^{-\Phi}$ with the stationary distribution
\begin{equation} \label{eqn: Phi}
    \Phi = -\sum_a \mu_a S_a    
    \, .
\end{equation}
The operators $\{ S_a \}$ are a set of mutually commuting Pauli strings, i.e., $S_a^2 = \ident$ and $[S_a, S_b] = 0$ for all $a$, $b$, and the $\mu_a$ are tunable chemical potentials. 
For our purposes, the stabilizer group $G$ is a subgroup of the Pauli group on $N$ qubits that defines a codespace, which is spanned by states satisfying $S \ket{\psi} = \ket{\psi}$ for all $S \in G$.
Importantly, we allow for any $S \in G$ to appear in our steady state distribution $\Phi$, as opposed to restricting our attention to a minimal generating set for $G$, although we will care most about cases where the $S_a$ in Eq.~\eqref{eqn: Phi} are local.

To construct strictly local jump operators, consider a Pauli string $P$ that is orthogonal to all $S_a$ in $\Phi$~\eqref{eqn: Phi} -- i.e., $\tr (P S_a)=0$ for all $a$. One useful property of Pauli strings is that  $P$ \emph{either} commutes or anticommutes with each $S_a$. We denote the set of labels $a$ for which $S_a$ anticommutes with $P$ by 
\begin{equation}
    \mathcal{A}_P \equiv \{ a \, | \, S_a P + P S_a = 0 \} 
    \, ,
    \label{eqn:anticommute-set}
\end{equation}
and if $P$ belongs to $G$, i.e. $\mathcal{A}_P=\emptyset$, then it commutes with the steady state, and is thus an eigenoperator with eigenvalue $c=1$.
By ``dressing'' nontrivial $P$ with a projection operator, we arrive at the desired strictly local jump operators (see Sec.~\ref{sec:stabilizer}):
\begin{equation}
    P(\mathbf{n}) = P \Pi_P(\mathbf{n}) \equiv P \left[ \prod_{a \in \mathcal{A}_P} \frac12 (\ident + n_a S_a) \right]   
    \, ,
    \label{eqn:stabilizer-eigenvector}
\end{equation}
where $\mathbf{n} \equiv \{n_a\}_{a \in \mathcal{A}_P}$, with $n_a \in \{-1, +1\}$, defines the projector $\Pi_P(\mathbf{n})$  onto the $n_a$ subspace of $S_a$ for all stabilizers that anticommute with $P$.
By construction, these jump operators are eigenoperators of Eq.~\eqref{eq:S} with the corresponding eigenvalues 
\begin{equation}
    c_P(\mathbf{n}) = \exp\left(-\frac12 \Delta \Phi_P\right) \text{~~where~~} \frac12 \Delta \Phi_P = \sum_{a \in \mathcal{A}_P} n_a \mu_a
    \, ,
\end{equation}
where, physically, $\Delta\Phi_P$ is the change in the $\Phi$ induced by $P$.
Note that any local operator $O$ can be written as a linear combination of \emph{finitely} many eigenoperators of the form of Eq.~\eqref{eqn:stabilizer-eigenvector}. This follows since any strictly local $O$ can be decomposed into a finite number of Pauli strings, each of which anticommutes with a finite number of the $S_a$ (which are also assumed local). 
We have therefore shown that, for stabilizer steady states defined by Eq.~\eqref{eqn: Phi}, we are able to identify a family of strictly local jump operators that diagonalize the superoperator $\mathcal{S}$~\eqref{eq:S}.


\subsubsection{Local dynamics}

Finally, we write down a simple ansatz for local dynamics generated by a Lindbladian $\Liouvillian$ whose steady state is $\sigma$.
We revert to the notation $\{ A_i \}$ for a generic jump operator basis, but it should be understood that the ansatz is most useful when the jump operators are local, e.g., belonging to the family identified in the previous subsection.
Consider the Lindbladian\footnote{This choice is motivated by the decomposition of the Fokker-Planck generator in classical nonequilibrium dynamics~\cite{ACM}.}
\begin{equation}
    \Liouvillian = -\sum_{ij} m_{ij} L_{i}^{\vpd} \mathcal{T} L_{j}^{\dagger} \mathcal{T}^{-1}
    \, .
    \label{eqn:ansatz}
\end{equation}
We now explore its properties and its behavior under the time-reversal transformation~\eqref{eq:T}.
First, observe that $\Liouvillian(\sigma)=0$ if the superoperators $L_i$ are trace preserving, such that $L_i^\dagger(\ident)=0$.
This condition also ensures that $\Liouvillian$ is trace preserving, since $L_i^\dagger(\ident)=0$ for all $i$ implies that $\Liouvillian^\dagger (\ident) = 0$.
A natural choice for the superoperator $L_i$ is therefore
\begin{equation}
    L_i(\, \cdot \,) = [A_i, \, \cdot \,]
    \, ,
    \label{eqn:L-op}
\end{equation}
where  $\{A_i\}$ satisfy $\mathcal{S}(A_i) = c_i A_i$, under $\mathcal{S}$~\eqref{eq:S}.
With this choice for $L_i$, one may verify that Eq.~\eqref{eqn:ansatz} takes the form of Eq.~\eqref{eqn:master}, with unitary and dissipative parts parameterized by
\begin{subequations} \label{coefromm}
\begin{align} 
H &= \frac{\ii}{2} \sum_{ij} \left( m_{ij} c_j - m_{\pi(j)\pi(i)} c_i \right) A_j^{\dagger} A_i^{\vpd} \, ,\\
\gamma_{ij} &= m_{ij} c_j + m_{\pi(j)\pi(i)} c_i \, , \label{gammafromm} 
\end{align}%
\end{subequations}
respectively. Hence, the condition $m_{ij} = m_{\pi(i) \pi(j)}^*$ is required to make $\gamma_{ij}$ and $H$ Hermitian. We separately enforce positivity of $\gamma_{ij}$ by making the diagonal elements $m_{ii}$~\eqref{eqn:ansatz} sufficiently large and positive.
Henceforth, we restrict to $m_{ij}$ satisfying these criteria. 

We now consider how the ansatz Lindbladian~\eqref{eqn:ansatz} transforms under T.
The transformation rules laid out in Eq.~\eqref{eqn:eigenop-transformation} may be applied directly to Eq.~\eqref{coefromm} to find that
\begin{subequations} \label{eq: coe from m after T}
    \begin{align}
        \widetilde{H} &= \frac{\ii}{2} \sum_{ij} \left( m_{ji}^* c_j - m_{ij} c_i \right) A_j^{\dagger} A_i^{\vpd} \, ,  \\
        \widetilde{\gamma}_{ij} &= m_{ji}^* c_j + m_{ij} c_i 
    \end{align}%
\end{subequations}
Compared to Eq.~\eqref{coefromm}, we observe that time reversal T~\eqref{eq:T} is implemented by sending the matrix $m$~\eqref{eqn:ansatz} to $m^\dagger$. As a result, the Hermitian  and anti-Hermitian parts of the matrix $m$~\eqref{eqn:ansatz} correspond, respectively, to T-even and T-odd dynamics.

Observe that the Linbladian~\eqref{eqn:ansatz} contains Davies' generator (Sec.~\ref{sec:davies}) as a special case. In particular, the operators $A_\omega$ defined in Eq.~\eqref{eqn:davies-jumpop} are eigenoperators of $\mathcal{S}$ satisfying $\mathcal{S}(A_\omega) = \mathrm{e}^{-\beta \omega /2} A_\omega$. We may therefore use them in Eq.~\eqref{eqn:L-op}, in conjunction with a real, nonnegative, diagonal $m$ matrix (i.e., $m_{ij} = m_{i}\delta_{ij}$), which leads to a diagonal $\gamma_{ij}$ and a vanishing Hamiltonian. 
Hermiticity of $\gamma_{ij}$ in Eq.~\eqref{gammafromm} is then guaranteed by choosing $m$ such that $m_i = m_{\pi(i)}$, which is equivalent to $m_\omega = m_{-\omega}$.
From Eq.~\eqref{gammafromm}, we see that $\gamma(\omega) = 2 m_\omega c_\omega$, which automatically satisfies the KMS condition $\gamma(\omega) = \ee^{-\beta \omega} \gamma(-\omega)$. Therefore, the ansatz Lindbladian~\eqref{eqn:ansatz} captures not only Davies' generator, but more general T-even dynamics, as well as some of the possible T-odd dynamics. 

However, the ansatz Lindbladian~\eqref{eqn:ansatz} does not capture the most generic T-odd contributions to the dynamics. We now discuss the additional types of terms needed to identify the most general possible (local) $\Liouvillian$ compatible with stationarity.  Using the definition of the permutation $\pi$, we deduce that the $\gamma_{ij}$ matrices produced by Eq.~\eqref{eqn:ansatz} satisfy
\begin{equation} \label{gammafromm2}
    \begin{pmatrix}
        \gamma_{ij} \\
        \gamma_{\pi(j)\pi(i)}
    \end{pmatrix} 
    =
    \begin{pmatrix}
        c_j & c_i \\ 
        c_j^{-1} & c_i^{-1}
    \end{pmatrix}
    \begin{pmatrix} 
        m_{ij} \\
        m_{ji}^{*} 
    \end{pmatrix}
    \, .
\end{equation}
Hence, if the matrix in~\eqref{gammafromm2} is invertible, we can find $m_{ij}$ that generate the corresponding $\gamma_{ij}$.
If $c_i = c_j$, then the matrix has zero determinant, and there exist $\gamma_{ij}$ that cannot be generated by $m_{ij}$. This can be seen more transparently from Eq.~\eqref{gammafromm}: if the eigenvalues are equal, then $\gamma_{ij} = c_i (m_{ij} + m_{ji}^*)$, which projects out the anti-Hermitian part of $m_{ij}$. As a result, the time-reversal transformation, which sends $m \to m^\dagger$, gives $\gamma_{ij} = \widetilde{\gamma}_{ij}$ for degenerate indices $(i, j)$.
This implies that only T-even dynamics can be generated for such pairs of indices.
By similar reasoning, if $c_i = c_j$, then the corresponding contribution to $H$ is of the form $\propto \ii c_i (m_{ij} - m_{ji}^*)$, which is again even under the time-reversal transformation. Since $A_j^\dagger A_i^{\vpd}$ is an eigenoperator of $\mathcal{S}$ with eigenvalue $c_i / c_j$, we observe that the ansatz also fails to capture T-odd contributions to Hamiltonian dynamics corresponding to operators $A_k$ with eigenvalue $c=1$.\footnote{If we fix $\gamma_{ij}$, the Hamiltonian terms with $c_k \neq 1$ are uniquely determined by stationarity of $\sigma$. Consequently, the only freedom we have when constructing dynamics that protect $\sigma$ is varying the coefficients of terms in $H$ that correspond to jump operators with eigenvalue $c_k=1$.}
However, this omission can easily be remedied: since the Hamiltonian terms $\sum_k h_k A_k$ with $c_k = 1$ do not contribute to stationarity from Eq.~\eqref{stationary}, they can be freely added to Eq.~\eqref{eqn:ansatz} without affecting the steady state.

To summarize, the ansatz Lindbladian~\eqref{eqn:ansatz} captures all T-even $\gamma_{ij}$, and all T-odd $\gamma_{ij}$ for nondegenerate indices $(i, j)$. The Hamiltonian contribution is essentially fixed by stationarity, up to the terms that correspond to operators $A_k$ with eigenvalue $c_k=1$, which can be varied freely without affecting stationarity. The remaining T-odd contributions to $\gamma_{ij}$ are discussed in Appendix~\ref{sec:T-odd}.
Specifically, we explain how to generate all one-dimensional, translation-invariant, local classical dynamics that
does not produce transitions between different symmetry-broken states,
and all local quantum dynamics for $\gamma_{ij}$ with ${c_i = c_j}$. The distinction between ``classical'' and ``quantum'' dynamics is made more precise in Sec.~\ref{sec:examples}.

\subsubsection{Weak and strong symmetries}

We next deduce the consequences of imposing a strong or weak symmetry on the form of $\mathcal{L}$, which we take to be of the ansatz form~\eqref{eqn:ansatz}.   It is natural to focus on steady states that are themselves symmetric: $\mathcal{U}_{g,\text{L}} \sigma =\mathcal{U}_{g,\text{R}} \sigma  = \sigma$.  This implies that 
\begin{equation}
    0 = [\mathcal{U}_{g,\text{L}},\mathcal{T}]= [\mathcal{U}_{g,\text{R}},\mathcal{T}] \, ,~~
\end{equation}
and a simple calculation shows that
\begin{equation} \label{eq:ULURg}
    \mathcal{U}_{g,\text{L}}\mathcal{U}_{g,\text{R}} L_i = L_{g\cdot i} \, \mathcal{U}_{g,\text{L}} \mathcal{U}_{g,\text{R}} \, ,~~
\end{equation}
where, using the decomposition \eqref{eqn:ansatz}, we have defined
\begin{equation}
    L_{g\cdot i} = [U(g) A_i U(g^{-1}), \, \cdot \, ] \,.~~
\end{equation}
Furthermore, from Eq.~\eqref{eq:ULURg}, we see that
\begin{equation}
   L_i \hspace{0.4mm} \mathcal{U}_{g^{-1},\text{L}} \mathcal{U}_{g^{-1},\text{R}}  = \mathcal{U}_{g^{-1},\text{L}}  \mathcal{U}_{g^{-1},\text{R}} L_{g^{-1}\cdot i} \, .~~
\end{equation} 
It is also useful to define a $g$-dependent matrix $a(g)$, such that
\begin{equation}
   U(g)A_iU(g^{-1}) = 
   \sum_j A_j \hspace{0.4mm} a_{ji} (g) \, ,~~
\end{equation}
satisfying $a(g)a(h) = a(gh)$. Combining the above formulas together, we see that a weak $G$ symmetry \eqref{eq:weaksymmetry} requires that $m_{ij}$ is invariant under the adjoint action of $a(g)$
\begin{equation}
   m_{ij} = \sum_{k,\ell} a_{ik} (g) \, m_{k \ell} \, a_{\ell j} (g^{-1}) \, ,~~
\end{equation}
for all group elements $g \in G$. Importantly, Schur's Lemma implies that the nonvanishing elements of $m_{ij}$ \eqref{eqn:ansatz} must contain jump operators in the same irreducible representations of $G$.  For example, with a single qubit, $\SU{2}$ invariance requires that $\Phi=0$ and the only allowed nontrivial jump operator is the fully depolarizing channel $\Phi_{\text{dp}} (\rho) = X\rho X + Y\rho Y + Z\rho Z - 3\rho$, since the full set of possible jump operators $(X,Y,Z)$ forms a three-dimensional representation of $\SU{2}$.

A strong symmetry \eqref{eq:strongsymmetry} is more constraining.  In particular, we note that if $\mathcal{U}_{g,\text{L}} \Liouvillian = \Liouvillian \hspace{0.4mm} \mathcal{U}_{g,\text{L}} $, then 
\begin{equation}
    \left(\mathcal{U}_{g,\text{L}} \Liouvillian \right)^\dagger(\ident) = \Liouvillian^\dagger[ U(g^{-1}) ] = 0
\end{equation}
for all $g\in G$.  This condition holds if and only if $[A_i,g]=0$ for all $m_{ij} \ne 0$ \eqref{eqn:ansatz}.  
Hence, with a strong symmetry, only singlet operators (in the trivial representation of $G$) can appear in the Lindbladian \eqref{eqn:ansatz}.

\subsubsection{Generalized time reversal}

Here we briefly discuss how the derivations and results of Sec.~\ref{subsub:operator basis} onward are modified upon replacing the transformation T \eqref{eq:T} with the \emph{generalized} transformation gT \eqref{eq:gT}. If we write $\widetilde{\Liouvillian}_K$ in the form
\begin{equation}    
    \widetilde{\Liouvillian} ( \rho ) = -\ii [ \widetilde{H} , \rho ] + \sum\limits_{i,j} \widetilde{\gamma}_{ij} \left( \widetilde{A}^{\vpd}_i \rho \widetilde{A}_j^\dagger - \frac{1}{2} \
    \{ \widetilde{A}_j^\dagger \widetilde{A}^{\vpd}_i , \rho \} \right) \, ,
    \label{eqn:master for general T}
\end{equation}
where $\widetilde{H} = \mathcal{K}(H)$ \eqref{eq:Sigma micro T} -- and likewise for $\mathcal{A}_j$ -- and we still require that the jump operators satisfy $\mathcal{S}(A_i) = c_i A_i$ \eqref{eq:S eig} [see also Eq.~\eqref{eq:S}], which translates to the condition that
\begin{equation}
\label{eq:S for general T}
    \widetilde{\mathcal{S}}(\widetilde{A}_i) = \widetilde{\sigma}^{1/2} \hspace{0.4mm} \widetilde{A}_i \widetilde{\sigma}^{-1/2} = c_i \widetilde{A}_i
\end{equation}
in the time-reversed language -- i.e., after transforming all terms under $\mathcal{K}$ \eqref{eq:Sigma micro T}.

As a result, Eq.~\eqref{eqn:eigenop-transformation} need only be modified slightly, according to
\begin{subequations}
\begin{align}
    \widetilde{H} &= \frac{1}{2} \sum_i  \left( c_i + \frac{1}{c_i} \right) h_i^* \widetilde{A}_i - \frac{\ii}{4} \sum_{ij}  \gamma_{ij}^* \left( \frac{c_i}{c_j} - \frac{c_j}{c_i} \right) \widetilde{A}_j^\dagger \widetilde{A}_i^{\vpd}
    \, , \label{eqn:reversed-H-eigenbasis for general T} \\
    \widetilde{\gamma}_{ij} &= \gamma_{\pi(j) \pi(i)}^* c_i c_j \label{eqn:reversed-gamma-eigenbasis for general T}
    \, ,
\end{align}%
\label{eqn:eigenop-transformation for general T}%
\end{subequations}
and using the ansatz decomposition~\eqref{eqn:ansatz}, applying time reversal leads to
\begin{equation}
    \widetilde{\Liouvillian}_K = -\sum_{ij} \widetilde{m}_{ij} \widetilde{L}_{i}^{\vpd} \widetilde{\mathcal{T}} \widetilde{L}_{j}^{\dagger} \widetilde{\mathcal{T}}^{-1} = -\sum_{ij} \widetilde{m}_{ij} \mathcal{K} L_{i}^{\vpd} \mathcal{T} L_{j}^{\dagger} \mathcal{T}^{-1} \mathcal{K}
    \, ,
    \label{eqn:ansatz after general T}
\end{equation}
where $\widetilde{m} = m^T$ and $\widetilde{L}_i = \mathcal{K} L_i \mathcal{K}$. The superoperator $L_i$~\eqref{eqn:ansatz} is the same, so that Eq.~\eqref{eq: coe from m after T} becomes
\begin{subequations} \label{eq: coe from m after general T}
    \begin{align}
        \widetilde{H} &= \frac{\ii}{2} \sum_{ij} \left( m_{ji} c_j - m_{ij}^* c_i \right) \widetilde{A}_j^{\dagger} \widetilde{A}_i^{\vpd} \, ,  \\
        \widetilde{\gamma}_{ij} &= m_{ji} c_j + m_{ij}^* c_i\, .
    \end{align}%
\end{subequations}
To generate T-even dynamics (with $\widetilde{\sigma} = \sigma$), we simply modify Eq.~\eqref{eqn:ansatz} to
\begin{equation}
    \Liouvillian = -\sum_{ij} m_{ij} L_{i}^{\vpd} \mathcal{T} \widetilde{L}_{j}^{\dagger} \mathcal{T}^{-1}
    \, ,
    \label{eqn:ansatz for general T}
\end{equation}
where  $m_{ij} = m_{\pi(i) \pi(j)}^*$~\eqref{eqn:ansatz} ensures that $\gamma$ and $H$ are Hermitian, and the time-reversal transformation sends $m \to m^T$, so that the symmetric part of $m$ is T even and the antisymmetric part is T odd.

Finally, the gT analogue of Eq.~\eqref{eqn:generic-transformed-L} is given by
\begin{align}    
    \widetilde{\Liouvillian}^{\vps}_K(\rho) &= -\ii \left[ \widetilde{\mathcal{S}}(\widetilde{H})\rho - \rho \widetilde{\mathcal{S}}^{-1}(\widetilde{H}) \right] \notag \\
    &~~+ \sum_{ij}{\gamma}_{ij}^* \left[
    \widetilde{\mathcal{S}}(\widetilde{B}_j^{\dagger}) \rho \widetilde{\mathcal{S}}^{-1}(\widetilde{B}^{\vpd}_i) 
    - \frac{1}{2} \left(
        \widetilde{\mathcal{S}}(\widetilde{B}_j^\dagger) \widetilde{\mathcal{S}}(\widetilde{B}^{\vpd}_i)  \rho 
        + \rho \widetilde{\mathcal{S}}^{-1}(\widetilde{B}_j^\dagger) \widetilde{\mathcal{S}}^{-1}(\widetilde{B}^{\vpd}_i)   
        \right)
    \right]
    \, ,
    \label{eqn:generic-transformed-L for general T}
\end{align}
where it is straightforward to check that $\widetilde{\Liouvillian}$ is a CPTP map satisfying Eq.~\eqref{eq: general T},  and $\widetilde{\Liouvillian}(\widetilde{\sigma})=0$. The derivation of the above follows those of Sec.~\ref{subsub:operator basis} up to the inclusion of tildes, which represent $\mathcal{K}$~\eqref{eq:Sigma micro T}.

\section{Steering towards stabilizer steady states}
\label{sec:stabilizer-steady-states}

As highlighted in Sec.~\ref{sec:stationarity}, our framework is particularly powerful at identifying dynamics that  protect a target state $\sigma = \ee^{-\Phi}$, where $\Phi$ is a sum of commuting, local terms. In this section, we classify exhaustively all possible Lindblad dynamics that flow towards such \emph{stabilizer steady states} $\sigma$; additionally, we provide \emph{physical interpretations} of such dynamics.  In many cases of interest, such interpretations suggest natural experimental protocols, even in ``digital'' quantum settings where discrete gates are more natural than continuous time evolution.

\subsection{Warm up: single qubit}
\label{sec:qubit-jump-ops}

As an elementary example, consider a single qubit, whose Hilbert space is spanned by the states $\ket{b} \in \{\ket{0}, \ket{1} \}$, satisfying $Z\ket{b} = (-1)^b\ket{b}$. Suppose that the stationary state is of the form $\sigma = \ee^{\mu Z}$ (i.e., $\Phi = -\mu Z$).
To find the appropriate jump operators that steer the qubit toward $\sigma$, we must identify the eigenoperators of the map $\mathcal{S}(\rho) = \sigma^{1/2} \rho \sigma^{-1/2}$~\eqref{eqn:superop-eigenvalues}.
First, observe that the projectors $\Pi^\pm \equiv (\ident \pm Z)/2$, being functions of $Z$ alone, commute with the steady state $\sigma$, and are therefore eigenoperators with unit eigenvalue $\mathcal{S}(\Pi^\pm) = \Pi^\pm$.
Additionally, the action of $\mathcal{S}$~\eqref{eq:S} on the operators $X$ and $Y$ remains closed. Explicitly, we have that
\begin{equation}
    \mathcal{S} 
    \begin{pmatrix}
        X  \\
        Y
    \end{pmatrix}
    = 
    \begin{pmatrix}
        \cosh \mu & -\ii \sinh \mu \\
        \ii \sinh \mu & \cosh\mu
    \end{pmatrix}
    \begin{pmatrix}
        X  \\
        Y
    \end{pmatrix}
    \, .
    \label{eqn:action-on-qubit}
\end{equation}
The eigenvectors are of the form $(X \pm \ii Y)/2$, with $\mu$-dependent eigenvalues $\ee^{\pm \mu}$, respectively. The eigenoperators of $\mathcal{S}$~\eqref{eq:S} are therefore (\emph{i}) projectors onto the computational basis states ($\Pi^\pm$) and (\emph{ii}) spin raising and lowering operators ($X^{\pm}$). Together, these eigenoperators form a basis for the operators on the single-qubit Hilbert space.
For $\mu \gg 1$, where $\sigma$ targets the ground state $\ket{0}$ (up to an exponentially small statistical admixture of $\ket{1}$), the eigenvalue of the spin raising operator $\ket{0}\bra{1}$ (i.e., $c=\ee^\mu$) is exponentially enhanced with respect to that of the lowering operator $\ket{1}\bra{0}$ (i.e., $c=\ee^{-\mu}$). 

To summarize, the complete basis of jump operators for the single-qubit system can be written
\begin{subequations}
\begin{align}
    \Pi^\pm \equiv \frac12(\ident \pm Z) \quad &\text{with} \quad c = 1 \, , \label{eqn:qubit-c=1} \\
    X^\mp \equiv \frac12 X (\ident \pm Z) \quad &\text{with} \quad c = \mathrm{e}^{\mp \mu} \, , \label{eqn:qubit-raise}
\end{align}%
\end{subequations}
when targeting a steady state of the form $\sigma = \ee^{\mu Z}$. Given this simple eigenbasis, we can utilize the ansatz form~\eqref{eqn:ansatz} to deduce minimal Lindbladians that flow to $\sigma$.
Note that the permutation $\pi$ describing the relationship between jump operators $\{ A_i \}$ and $\{ A_i^\dagger \}$ swaps the raising and lowering operators in Eq.~\eqref{eqn:qubit-raise} but acts trivially on the projectors in Eq.~\eqref{eqn:qubit-c=1}.
In particular, for diagonal $m$~\eqref{eqn:ansatz}, we can write
\begin{equation}
    \Liouvillian(\rho) = \gamma \Big( Z \rho Z - \frac12 \{ Z , \rho \} \Big) + \sum_{n = \pm} \Gamma \ee^{-n\mu} \left( X \Pi^n \rho \Pi^n X - \frac12 \{ \Pi^n , \rho \} \right) 
    \, .
    \label{eqn:T1T2-dynamics}
\end{equation}
The first term recovers familiar ``phase-damping'' dynamics~\cite{NielsenChuang2010}, which annihilates off-diagonal matrix elements of $\rho$ in the computational basis. The second term corresponds to generalized (i.e., finite-temperature) ``amplitude-damping'' dynamics, which captures spontaneous emission~\cite{NielsenChuang2010}, and is responsible for stabilizing the correct populations of the two levels at late times. In addition to these familiar contributions, one can consider matrices $m$~\eqref{eqn:ansatz} with off-diagonal contributions. An example of off-diagonal terms between the operators in Eq.~\eqref{eqn:qubit-raise} is captured by sending
\begin{equation}
    \Liouvillian \to \mathcal{L} + \Delta \cosh\mu X^+ \rho X^+ + \Delta^* \cosh \mu X^- \rho X^-
    \, ,
\end{equation}
which has the effect of modifying the transient relaxation dynamics without modifying the steady state (since $X^\pm \BKop{b}{b} X^\pm = 0$).  
There are also off-diagonal terms that give rise to a nontrivial Hamiltonian contribution, such as off-diagonal terms between $Z$ and $X^\pm$. These produce, e.g.,
\begin{subequations}
\begin{gather}
    H = -h \cosh\mu X \\
    \gamma_{ZX^\pm} = \pm \ii h \ee^{\pm \mu} \, , \quad \gamma_{X^\pm Z} = \gamma_{ZX^\pm}^*
\end{gather}
\end{subequations}
The dynamics due to the transverse magnetic field ($X$) are compensated for by the dissipative contribution, leaving the steady state unchanged. Notice that we also need to introduce diagonal dephasing terms $\gamma_{ZZ}$ and $\gamma_{X^{\pm}X^{\pm}}$ to ensure that $\gamma_{ij}$ as a whole is positive semidefinite.  We show in Sec.~\ref{sec:unitary-pauli-errors} that there exists a general correction procedure for compensating arbitrary Hamiltonian terms using jump operators.

\subsection{Generic stabilizer steady states}
\label{sec:stabilizer}

Now consider a system composed of $N$ qubits, and suppose that we target the ``stabilizer'' steady states introduced in Sec.~\ref{sec:local-jumps}. That is, the steady state is of the form $\sigma = \ee^{-\Phi}$, with $\Phi = -\sum_a \mu_a S_a$, with each $S_a$ a Pauli string.
To identify the eigenoperators of the map $\mathcal{S}$~\eqref{eq:S}, consider the action of $\mathcal{S}$ on some Pauli string $P$ that is orthogonal to all Pauli strings in the stabilizer group.
Such a string then either commutes or anticommutes with each $S_a$; as in Eq.~\eqref{eqn:anticommute-set} we denote the set of $a$ for which $S_a$ anticommutes with $P$ by $\mathcal{A}_P$.
Since all $S_a$ mutually commute, we have that
\begin{equation}
    \ee^{\Phi/2} = \prod_a \exp\left( \frac12 \mu_a S_a \right) \equiv \prod_a \ee^{ \Phi_a / 2}
    \, ,
\end{equation}
and we can consider conjugation by each $\ee^{\Phi_a/2}$ separately:
\begin{equation}
    \ee^{ \Phi_a / 2 } P \ee^{-  \Phi_a / 2 } = \cosh( \mu_a ) 
    P -\sinh( \mu_a ) P S_a 
    \, ,~
\end{equation}
so that the action of $\mathcal{S}_a(\, \cdot \,) = \ee^{ \Phi_a /2 } \, \cdot \, \ee^{-  \Phi_a /2 }$ on the Pauli strings $P$ and $P S_a$ remains closed, and the system of equations essentially reduces to the eigenvalue problem for the two-level system~\eqref{eqn:action-on-qubit}. Specifically, the generalization of the spin lowering and raising operators are $P \frac{1}{2} (\ident \pm S_a)$, with eigenvalues $\ee^{\mp\mu_a}$. This procedure of reducing to a $2 \! \times \! 2$ eigenvalue problem can be iterated for all elements of $\mathcal{A}_P$ to arrive at eigenoperators of the form
\begin{equation}
    P(\mathbf{n}) = P \Pi_P(\mathbf{n}) \equiv P \left[ \prod_{a \in \mathcal{A}_P} \frac{1}{2} (\ident + n_a S_a) \right]   
    \, ,
    \label{eqn:eigenop-generic}
\end{equation}
as stated previously~\eqref{eqn:stabilizer-eigenvector}. The operator $P(\mathbf{n})$ projects onto a state with definite stabilizer eigenvalues $\mathbf{n}$ (which one may regard as the post-measurement state if measurement outcomes $\mathbf{n}$ were obtained), then $P$ flips the eigenvalues of these stabilizers. Alternatively, $P(\mathbf{n})$~\eqref{eqn:eigenop-generic} can be regarded as a controlled $P$ operation. The corresponding eigenvalues are
\begin{equation}
    c_P (\mathbf{n}) = \exp\left(-\frac{1}{2} \Delta \Phi_P\right) \text{~~where~~} \frac{1}{2} \Delta \Phi_P = \sum_{a \in \mathcal{A}_P} n_a \mu_a
    \, .
\end{equation}
The generalization of the phase-damping contributions~\eqref{eqn:T1T2-dynamics} are elements of the stabilizer group. Since these operators commute with $\Phi$, they have eigenvalue $c=1$.

This procedure also allows us to decompose any strictly local operator $O$ in terms of a \emph{finite} number of eigenoperators, as described in Eq.~\eqref{eqn:eigenop-generic}. Without loss of generality, we write some local operator $O$ as $O = \sum_k c_k P_k$, where every $P_k$ is a Pauli string whose nonidentity operator content is supported only within a finite (fixed) spatial region, and $c_k = 2^{-N} \tr (O P_k)$. In the controlled-$P$ operation~\eqref{eqn:eigenop-generic}, each $P_k$ can be decomposed as $P_k =  \sum_{\mathbf{n}} P_k(\mathbf{n})$, where $\mathbf{n}$ runs over the measurement outcomes of the stabilizers $S_a$ that anticommute with $P_k$. If the stabilizers $S_a$ are also local, each $P_k$ only anticommutes with a finite number ($\abs{\mathcal{A}_{P_k}}$) of stabilizers, since their support must overlap in order to anticommute. Each $P_k$ can therefore be written in terms of $2^{\abs{\mathcal{A}_{P_k}}}$ jump operators, and hence, any strictly local operator $O$ can be decomposed in terms of a finite number of jump operators.

\subsection{Interpretation: Measurements and feedback}
\label{sec:trajectories}

\subsubsection{Projective measurement}

We now show how the dynamics we have derived can -- in certain cases -- be interpreted in terms of projective measurements of stabilizer operators and subsequent unitary feedback. 
For the purposes of this discussion, consider a Lindbladian that is diagonal in the jump operators derived in Secs.~\ref{sec:qubit-jump-ops} and \ref{sec:stabilizer}.
\begin{equation}
    \Liouvillian(\rho) = - \ii [H, \rho] + \sum_{i \geq 1} \gamma_i \left( L^{\vpd}_i \rho L_i^\dagger - \frac12  \{ L_i^\dagger L^{\vpd}_i, \rho \} \right)
    \, ,
    \label{eqn:diagonal-lindbladian}
\end{equation}
which describes time evolution of the state $\rho(t)$.
This scenario will be of interest for correcting a large family of Hamiltonian and dissipative errors, see Secs.~\ref{sec:unitary-pauli-errors} and \ref{sec:correcting-nonunitary}, respectively. The time evolution~\eqref{eqn:diagonal-lindbladian} can alternatively be interpreted in terms of Kraus operators $K_i$ that map the state $\rho(t) \to \rho(t+\thed t)$ over the time interval $\thed t$ via the operator sum decomposition $\rho(t+\thed t) = \sum_{i\geq 0} K_i \rho(t) K_i^\dagger$. The Kraus operators that achieve this decomposition are
\begin{subequations}
\begin{align}
    K_0 &= \ident - \left( \ii H + \frac{1}{2} \sum_{i \geq 1} \gamma_i L^\dagger_i L^{\vpd}_i \right) \thed t \, , \label{eqn:K0} \\
    K_i &= L_i \sqrt{\gamma_i \thed t} \, , \text{ for } i \geq 1 , \label{eqn:K-jump}
\end{align}%
\label{eqn:Lindblad-Kraus}%
\end{subequations}
where the Kraus operators satisfy the completeness relation $\sum_{i\geq 0} K_i^\dagger K^{\vpd}_i = \ident$, and $K^{\vpd}_0$ describes deterministic evolution according to the effective (non-Hermitian) Hamiltonian defined by the parenthetical terms in Eq.~\eqref{eqn:K0}, while the operators $K_i$ for $i\geq 1$ correspond to discontinuous ``jumps.''

For the single-qubit example of Sec.~\ref{sec:qubit-jump-ops}, the formalism presented in Sec.~\ref{sec:stationarity} gives rise to the dissipative contribution (i.e., absent any Hamiltonian evolution),
\begin{equation}
    \Liouvillian (\rho) \supset  \sum\limits_{n = \pm 1} \Gamma \ee^{-n \mu} \left( X \Pi^n \rho \Pi^n X - \frac{1}{2} \acomm*{\Pi^n}{\rho} \right) + \gamma_n \left( \Pi^n \rho \Pi^n - \frac{1}{2} \acomm*{ \Pi^n}{\rho} \right)
    \label{eqn:exp(Z)-dynamics}
\end{equation}
where $\Pi^n$ is the projection introduced in Eq.~\eqref{eqn:qubit-c=1}, 
and $\Gamma$ and $\gamma_n$ are undetermined (nonnegative) constants that follow from $m_{ij}$ via~Eq.~\eqref{gammafromm}. 
As discussed in Sec.~\ref{sec:qubit-jump-ops}, the first term corresponds to generalized amplitude damping~\cite{NielsenChuang2010}. The second term corresponds to measurements in the $Z$ basis (at least when $\gamma_n$ are independent of $n$).
To derive an exact equivalence between the Kraus operators~\eqref{eqn:Lindblad-Kraus} and measurements followed by unitary feedback, we are free to choose specific values for the $\gamma_n$ constants. Recall that the $\gamma_n$ coefficients can be varied freely without affecting the stationarity of $\sigma = \ee^{\mu Z}$ since the projectors $\Pi^n$ commute with the steady state.
Specifically, we take\footnote{Taking $\gamma_n = C - \Gamma \ee^{-n\mu}$ for some $C > \Gamma \ee^{\abs{\mu}}$ would also work, although the probability of applying feedback would be correspondingly diminished.} $\gamma_n = \Gamma(\ee^{\abs{\mu}} - \ee^{-n\mu})$, allowing us to write the contribution from Eq.~\eqref{eqn:K0} and Eq.~\eqref{eqn:K-jump} as
\begin{equation}
    \thed \rho \supset \sum_{n = \pm} \thed t \Gamma \ee^{\abs{\mu}} \tr[\Pi^n \rho \Pi^n]  \left\{ p_\text{f}(n) \frac{X\Pi^n \rho \Pi^n X}{\tr[\Pi^n \rho \Pi^n] } + [1-p_\text{f}(n)] \frac{\Pi^n \rho \Pi^n}{\tr[\Pi^n \rho \Pi^n] } \right\}
    - \rho \thed t \Gamma \ee^{\abs{\mu}}
    \, ,
\end{equation}
in a time interval $\thed t$,
where $p_{\text{f}}(n) = \ee^{-(1+n \sgn \mu)\abs{\mu}}$ is the probability that the unitary feedback $X$ is applied to the post-measurement state $\propto \Pi^n \rho \Pi^n$.
That is, during a time interval $\text{d}t$, there is a probability $\text{d}t \Gamma \ee^{\abs{\mu}}$ that the system is measured in the $Z$ basis. If the system is measured, $\tr[\Pi^n \rho \Pi^n]$ represents the Born probability for measurement outcome $n = \pm 1$. Finally, the operator $X$ is applied with the outcome-dependent probability $p_\text{f}(n)$. For our choice of $\gamma_n$, we have $p_\text{f}(n)=1$ for $n=-\sgn\mu$; for $\mu \gg 1$, the state $\ket{0}$ is being targeted, so feedback $X$ is applied with unit probability when the state $\ket{1}$ is obtained (and with exponentially small probability when $\ket{0}$ is obtained).
Note that the probability of applying feedback is precisely the acceptance probability of the Metropolis-Hastings algorithm \cite{Metropolis_1953}.

This interpretation can also be generalized to the case of generic stabilizer steady states studied in Sec.~\ref{sec:stabilizer}.
For the contribution from jump operators that correspond to the dressing of a particular Pauli string $P$~\eqref{eqn:stabilizer-eigenvector}, we write
\begin{equation}
    \Liouvillian(\rho) \supset \sum_{\mathbf{n}} \Gamma_{\mathbf{n}} c_P(\mathbf{n}) \left( P \Pi_P(\mathbf{n}) \rho \Pi_P(\mathbf{n}) P - \frac12 \{ \Pi_P(\mathbf{n}) , \rho \} \right)
    +
    \gamma_\mathbf{n} \left(  \Pi_P(\mathbf{n}) \rho \Pi_P(\mathbf{n}) - \frac12 \{ \Pi_P(\mathbf{n}), \rho \} \right)
    \label{eqn:drho-qubit}
\end{equation}
where the coefficients $\Gamma_\mathbf{n}$ satisfy $\Gamma_{\mathbf{n}} = \Gamma_{-\mathbf{n}}$ and are related to $m_{ij}$ via Eq.~\eqref{gammafromm}. Denoting the outcomes $\mathbf{n}$ for which the coefficient $\Gamma_\mathbf{n} c_P(\mathbf{n})$ is maximal by $\mathbf{n}_\star$, we can choose $\gamma_\mathbf{n} = \Gamma_{\mathbf{n}_\star} c_P(\mathbf{n}_\star) - \Gamma_\mathbf{n} c_P(\mathbf{n})$ to write
\begin{equation}
    \text{d}\rho \supset  \sum_{\mathbf{n}} 
        \text{d}t \Gamma_{\mathbf{n}_\star} c_P(\mathbf{n}_\star) \left\{
            p_\text{f}(\mathbf{n}) P\Pi_P(\mathbf{n}) \rho \Pi_P(\mathbf{n}) P + [1-p_\text{f}(\mathbf{n})]  \Pi_P(\mathbf{n}) \rho \Pi_P(\mathbf{n})
        \right\}
        - \rho \thed t \Gamma_{\mathbf{n}_\star} c_P(\mathbf{n}_\star) 
        \, ,
\end{equation}
where $p_\text{f}(\mathbf{n}) = \Gamma_\mathbf{n} c_P(\mathbf{n}) / [\Gamma_{\mathbf{n}_\star} c_P(\mathbf{n}_\star)]$.  The interpretation is analogous to that of Eq.~\eqref{eqn:drho-qubit}: During a time interval $\thed t$, there is a probability $\thed t \Gamma_{\mathbf{n}_\star} c_P(\mathbf{n}_\star)$ that all stabilizers that anticommute with $P$ are measured. If the system is measured, $\tr[\Pi_P(\mathbf{n}) \rho \Pi_P(\mathbf{n})]$ equals the Born probability for the set of outcomes $\mathbf{n}$. Finally, the Pauli string $P$ is applied with the outcome-dependent probability $p_\text{f}(\mathbf{n})$, where the feedback probability is maximal for the measurement outcomes $\mathbf{n}_\star$.

\subsubsection{Generalized measurement}
\label{sec:generalized-meas-interpretation}

In the most general setting, one also encounters corrections that cannot be implemented using only \emph{projective} measurements and unitary feedback.
Corrections requiring such an interpretation are discussed in Sec.~\ref{sec:coherent-nonunitary}.
However, we may always view Lindbladian time evolution as \emph{generalized} measurements with accompanying feedback. 
Consider a Lindbladian of the form
\begin{equation}
    \Liouvillian(\rho) = \sum_{\mathbf{n}} \Lambda_{\mathbf{n}} \left( 
        L^{\vpd}_\mathbf{n} \rho L_\mathbf{n}^\dagger - \frac{1}{2} \{ L_\mathbf{n}^\dagger L^{\vpd}_\mathbf{n}, \rho \}
    \right)
    \label{eqn:generic-dissipative}
\end{equation}
A concrete protocol for implementing~\eqref{eqn:generic-dissipative} is made clear by applying a singular value decomposition (SVD): $L_\mathbf{n} = U_{\mathbf{n}} \Sigma_\mathbf{n} V^\dagger_\mathbf{n} = U_{\mathbf{n}} E_{\mathbf{n}} = \sum_a \ket{u_{a\mathbf{n}}}\bra{v_{a\mathbf{n}}} \sum_b \sigma_{b\mathbf{n}} \ket{v_{\mathbf{n}b}} \bra{v_{\mathbf{n}b}}$, where $\ket{u_{\mathbf{n}a}}$ and $\ket{v_{\mathbf{n}a}}$ are the left and right singular vectors, respectively, and $\sigma_{b\mathbf{n}} \geq 0$ are the singular values.
To interpret this situation physically, we return to the Kraus representation of Eq.~\eqref{eqn:Lindblad-Kraus} with Kraus operators $K_i$. We interpret the infinitesimal time evolution as a positive operator-valued measurement (POVM)~\cite{NielsenChuang2010}. The state $\ket{\psi}$ of the system is sent to $ \propto K_i\ket{\psi}$ with probability $\braket{\psi | K_i^\dagger K^{\vphantom{\dagger}}_i | \psi}$. This can be achieved using only unitary operations and protective measurements by considering an ancilla that contains as many states as there are measurement outcomes $\mathbf{n}$ plus the default state $\ket{0}$. The unitary $U$ on the enlarged Hilbert space takes $U \ket{\psi}\ket{0} = \sum_{i \geq 0} K_i \ket{\psi}\ket{i}$. A subsequent projective measurement of the ancilla returns the correct states of the system with the appropriate probabilities.
The Kraus operators in Eq.~\eqref{eqn:K-jump} are then just $E_\mathbf{n} \sqrt{g c_\mathbf{n}^2 \thed t}$, and, if a nondefault measurement outcome $\mathbf{n}$ is obtained when measuring the ancilla, subsequent unitary feedback $U_\mathbf{n}$ is applied to the system.
Note that, if the $E_\mathbf{n}$ operators are just projectors, then the protocol can be replaced by a projective measurement of the system. 

In this way, all of our correction procedures may be implemented by utilizing \emph{generalized} measurement (optionally followed by unitary feedback). However, we stress that there can be other physical interpretations for a given $\mathcal{L}$, which may be more convenient for designing protocols for particular experimental systems.

\subsection{Correcting for errors}

Next, we show how our formalism can be used to correct for erroneous terms in the Hamiltonian (``Hamiltonian'' or ``unitary''  errors) and in the jump operators (``incoherent'' errors) that, absent any corrective terms, would violate stationarity of the desired steady state $\sigma$. Namely, for stabilizer steady states, we explicitly construct the jump operators and/or Hamiltonian terms that can be added to the Lindbladian to maintain stationarity of $\sigma$ in the presence of these erroneous terms. 
In this way, we provide simple probabilistic protocols that are able to correct for both Hamiltonian errors and incoherent errors in a Lindbladian that protects an arbitrary stabilizer steady state.

Consider the scenario introduced in Sec.~\ref{sec:stabilizer}: The desired steady state is of the form $\sigma = \ee^{-\Phi}$ with $\Phi = - \sum_a \mu_a S_a$, where the $S_a$ are mutually commuting Pauli strings. We will first consider the case in which the Hamiltonian contains a term $\propto P$, where $P$ is a Pauli string that does not commute with all the $S_a$, thereby violating the stationarity of $\sigma$ without additional corrective terms. Second, we consider incoherent Pauli errors arising from multiplication by some Pauli string at some rate. Finally, we look at the most general case in which the incoherent non-Pauli errors correspond to generic linear combinations of Pauli strings.

\subsubsection{Hamiltonian errors}
\label{sec:unitary-pauli-errors}

Suppose that we have some Lindbladian $\cal{L}$ that protects $\sigma$ (i.e., $\Liouvillian[ \sigma ] = 0$), which may be obtained using the methods presented in Sec.~\ref{sec:theory}.
This Lindbladian is then modified by adding a term
\begin{equation}
    H \to H - g P 
    \, ,
    \label{eqn:pauli-perturbation}
\end{equation}
in the Hamiltonian, where $P$ is a Pauli string that does not commute with all the stabilizers $S_a$.
If $P$ were to commute with \emph{all} $S_a$, then it could be added to the Hamiltonian $H$ freely without affecting the stationarity of $\sigma$. 
Since any Hamiltonian can be decomposed in terms of a sum of Pauli strings, the following discussion is able to correct for \emph{arbitrary} errors in the unitary evolution (each term in the sum can be treated separately in the manner described below).  Note that, obviously, we do not consider the trivial error correcting scheme of just ``modifying $H \to H+gP$'' to cancel the unwanted offset.

The first step towards correcting for the presence of $P$ is to decompose $P$ in terms of eigenoperators~\eqref{eqn:stabilizer-eigenvector} of $\mathcal{S}$~\eqref{eq:S}. This is achieved by resolving the identity
\begin{equation}
    P =  \sum_{\mathbf{n}} P\Pi_P(\mathbf{n}) = \sum_{\mathbf{n}} B^\dagger_{1\mathbf{n}} B^{\vpd}_{2\mathbf{n}} =  \frac12 \sum_{\mathbf{n}} \left( B^\dagger_{1\mathbf{n}} B^{\vpd}_{2\mathbf{n}} +  B^\dagger_{2\mathbf{n}} B^{\vpd}_{1\mathbf{n}} \right)
    = \frac12 \sum_{\mathbf{n}} \sum_{\alpha\beta} B^\dagger_{\alpha\mathbf{n}} \sigma^x_{\alpha\beta} B^{\vpd}_{\beta\mathbf{n}}
    \, ,
    \label{eqn:decompose-perturbation}
\end{equation}
where $\sigma^x$ is the $x$ Pauli matrix.
In the first equality, we write the identity as a sum over projectors $\Pi_P(\mathbf{n})$ onto measurement outcomes $\mathbf{n}$ that correspond to measuring all stabilizers $S_a$ that anticommute with $P$ (denoted by $a \in \mathcal{A}_P$).
In the second equality, each term in the summation over $\mathbf{n}$
has been written in terms of the jump operators $ B^\dagger_{1\mathbf{n}} = B^{\vpd}_{1-\mathbf{n}} = P\Pi_P(\mathbf{n})$ and $B^{\vpd}_{2\mathbf{n}} = B^\dagger_{2\mathbf{n}} = \Pi_P(\mathbf{n})$. 
In this section, it turns out to be more notationally convenient to work with the operators $B_{1\mathbf{n}}^\dagger$ and $B^{\vpd}_{2\mathbf{n}}$ in place of the operators $A^{\vpd}_{1\mathbf{n}} = P \Pi_P(\mathbf{n})$ and $A_{2\mathbf{n}} = \Pi_P(\mathbf{n})$ used elsewhere. The operators $B^{\vpd}_{1\mathbf{n}}$ and $B^{\vpd}_{2\mathbf{n}}$ 
are eigenoperators of the superoperator $\mathcal{S}$ with eigenvalues $d_{1\mathbf{n}} = \exp(\sum_{a \in \mathcal{A}_P} n_a \mu_a )$ and $d_{2\mathbf{n}} = 1$, respectively.
For a Hamiltonian parameterized by $H = \sum_{ij} h_{ij} A_j^\dagger A^{\vpd}_i$, with $\{ A^{\vpd}_i \}$ eigenoperators of $\mathcal{S}$ with eigenvalues $\{c_i\}$, stationarity of $\sigma$ is then implemented by enforcing:
\begin{equation}
    \mathcal{L}[\sigma] =
    \sqrt{\sigma}  
    \sum_{ij} \left\{
            -\ii \left( \frac{c_j}{c_i} - \frac{c_i}{c_j} \right) h_{ij}  + \gamma_{\pi(j)\pi(i)} c_i c_j  - \frac{1}{2} \gamma_{ij} \left(\frac{c_j}{c_i} + \frac{c_i}{c_j}\right)
    \right\} A_j^\dagger A^{\vpd}_i \sqrt{\sigma}
    \stackrel{!}{=} 0
    \, ,
    \label{eqn:stationary-general}
\end{equation}
interpreted as a constraint on the dissipative part of the dynamics $\gamma_{ij}$.
Hence, if the Hamiltonian is modified according to Eq.~\eqref{eqn:pauli-perturbation}, its effect can, in principle, be compensated for by adjusting either $\gamma_{ij}$ or $\gamma_{\pi(j)\pi(i)}$. Compensating for the change using the $\gamma_{\pi(j)\pi(i)}$ coefficients, we arrive at
\begin{equation}
    \delta\gamma_{\pi(\beta\mathbf{n}) \pi(\alpha\mathbf{n})} = \frac{\ii}{d_{\alpha\mathbf{n}} d_{\beta\mathbf{n}} } \left( \frac{d_{\beta\mathbf{n}}}{d_{\alpha\mathbf{n}}} - \frac{d_{\alpha\mathbf{n}}}{d_{\beta\mathbf{n}}} \right) \delta h_{\alpha\beta}(\mathbf{n})
    \, ,
    \label{eqn:compensate-unitary}
\end{equation}
for the stationarity of $\sigma$ to be maintained, where $\delta h_{\alpha\beta}(\mathbf{n})$ corresponds to the change in $h_{ij}$ induced by Eq.~\eqref{eqn:pauli-perturbation}.
Note that the indices $i$ and $j$ in Eq.~\eqref{eqn:stationary-general} run over \emph{all} jump operators, whereas $\alpha, \beta \in \{ 1, 2 \}$, and $\mathbf{n}$ capture the jump operators that are ``perturbed'' according to Eq.~\eqref{eqn:decompose-perturbation}, and that we are working in a basis defined by the $B_{\alpha \mathbf{n}}$ operators. The nonzero matrix elements -- all of which are off diagonal -- are
\begin{equation}
    \delta\gamma_{\pi(1\mathbf{n}) \pi(2\mathbf{n})} = \delta\gamma_{\pi(2\mathbf{n}) \pi(1\mathbf{n})}^* = \frac{\ii}{2 } g d_{1\mathbf{n}}^{-1} \left( d_{1\mathbf{n}}^{-1} - d_{1\mathbf{n}} \right) \equiv \frac{\ii}{2} s_\mathbf{n} \Lambda_\mathbf{n} 
    \, ,
    \label{eqn:compensate-nonzero}
\end{equation}
where $\Lambda_{\mathbf{n}}/2$ equals the modulus of~\eqref{eqn:compensate-nonzero}, and $s_\mathbf{n}$ contains the phase, $s_\mathbf{n}^2 = 1$.
While the modification~\eqref{eqn:compensate-unitary} will preserve stationarity of $\sigma$, it must be the case that $\mathcal{L}$ remains a valid Lindbladian, i.e., the matrix $\gamma_{ij}$ must remain both Hermitian and positive semidefinite. Hermiticity is inherited from $\delta h_{\alpha\beta}(\mathbf{n})$ in Eq.~\eqref{eqn:compensate-unitary}, while positivity can be ensured by additionally modifying the diagonal elements $\gamma_{\pi(\alpha\mathbf{n})\pi(\alpha\mathbf{n})}$ in such a way as to maintain protection of $\sigma$. Note that the diagonal terms have not already been modified by Eq.~\eqref{eqn:compensate-unitary}, since $\sigma^x_{\alpha\beta}$ is purely off-diagonal.
Taking
\begin{equation}
    \delta\gamma_{\pi(1\mathbf{n}) \pi(1\mathbf{n})} = \delta\gamma_{\pi(2\mathbf{n}) \pi(2\mathbf{n})} = \frac{1}{2  } d_{1\mathbf{n}}^{-1} \abs{ g(d_{1\mathbf{n}}^{-1} - d_{1\mathbf{n}}) } = \frac12 \Lambda_\mathbf{n}
    \, ,
    \label{eqn:unitary-error-diagonal}
\end{equation}
we observe that (i) positivity of $\gamma_{ij}$ is enforced, and (ii) the coefficients automatically satisfy $d_{1\mathbf{n}}^2 \delta\gamma_{\pi(1\mathbf{n}) \pi(1\mathbf{n})}  = \delta\gamma_{(1\mathbf{n}) (1\mathbf{n})}$, which implies that the modification of the diagonal elements will not affect the stationarity of $\sigma$.
Hence, the corrective part of $\Liouvillian$ may be written
\begin{equation}
    \Liouvillian(\rho) \supset 
    \sum_{\mathbf{n}} \Lambda_{\mathbf{n}} \left(   L^{\vpd}_\mathbf{n} \rho L^\dagger_\mathbf{n} - \frac12 \{ L_\mathbf{n}^\dagger L^{\vpd}_\mathbf{n}, \rho \} \right)
    \, ,
    \label{eqn:lindblad-modification-unitary}
\end{equation}
where we defined the diagonal jump operators $L_\mathbf{n} = \frac{1}{\sqrt{2}}(B_{1\mathbf{n}}^\dagger - \ii s_\mathbf{n} B_{2\mathbf{n}}^\dagger) = \frac{1}{\sqrt{2}}(P-\ii s_\mathbf{n} \ident)\Pi_{P}(\mathbf{n}) $, which may be written as $L_\mathbf{n} = \ee^{-\ii \pi s_\mathbf{n} P/4} P\Pi_{P}(\mathbf{n})$ to make the the projection and unitary components manifest. 
Utilizing the interpretation of Eq.~\eqref{eqn:lindblad-modification-unitary} from Sec.~\ref{sec:trajectories}, we find that the following protocol will correct for the presence of a Pauli string $P$ in the Hamiltonian. Let $\mathbf{n}_\star$ be the measurement outcomes for which $\Lambda_\mathbf{n}$ is maximized.
In time interval $\delta t$
\begin{enumerate}
    \item with probability $\delta t \Lambda_{\mathbf{n}_\star}$ measure the stabilizers $S_a$ that anticommute with the perturbation $P$,
    \item if the system was measured during the time interval, apply unitary feedback  $U_\mathbf{n} = \frac{1}{\sqrt{2}}(P-\ii s_\mathbf{n} \ident)$ with probability $\Lambda_\mathbf{n}/\Lambda_{\mathbf{n}_\star}$.
\end{enumerate}
If the stabilizers that anticommute with $P$ are measured at a rate $\tilde{\Lambda}$ that exceeds $\Lambda_{\mathbf{n}_\star}$, the probability of applying unitary feedback must correspondingly be reduced to $\Lambda_{\mathbf{n}}/\tilde{\Lambda}$ in order to protect $\sigma$.
We emphasize that the corrective procedure described above is sufficient to remove Hamiltonian perturbations of the form of Eq.~\eqref{eqn:pauli-perturbation}, but it is not unique. For instance, making the diagonal entries unequal in Eq.~\eqref{eqn:unitary-error-diagonal} (while maintaining positivity of $\gamma$) may lead to a different unitary feedback operator  $U_\mathbf{n} \propto ( P -\ii \alpha s_{\mathbf{n}} \mathds{1})$ for $\alpha \in \mathbb{R}$.
This freedom may lead to improved error thresholds in the context of measurement errors (see Sec.~\ref{section: measurement error}). 

The minimal dynamics we have described herein is not T even. While it is possible to write down local dynamics that is T even, the construction is not particularly illuminating, so we have chosen to omit it.

\subsubsection{Incoherent Pauli errors}
\label{sec:correcting-nonunitary}

Consider now the case where there is an erroneous term in the \emph{dissipative} part of the Lindbladian. Such terms may arise when considering bit-flip or phase-flip errors in quantum error-correcting codes. Specifically, we take the incoherent Pauli error to be of the form
\begin{equation}
    \Liouvillian(\rho) \to \Liouvillian(\rho) + g\left( P\rho P - \rho \right)
    \, ,
    \label{eqn:nonunitary-deform}
\end{equation}
for some Pauli string $P$ that does not commute with all the stabilizers $S_a$ defining the steady state, and for some $g>0$.
That is, at some rate, the system is subjected to ``$P$ errors,'' corresponding to the multiplication of the state by the operator $P$.
To correct for such errors, we again decompose the Pauli string $P$ into eigenoperators of $\mathcal{S}$ by resolving the identity, $P = \sum_\mathbf{n} P \Pi_P(\mathbf{n}) = \sum_\mathbf{n} A_\mathbf{n}$:
\begin{equation}
    g\left( P\rho P - \rho \right) = 
    g\sum_{\mathbf{m},\mathbf{n}} \left( A^{\vpd}_{\mathbf{m}} \rho A_{\mathbf{n}}^\dagger - \frac12 \{ A_{\mathbf{n}}^\dagger A^{\vpd}_{\mathbf{m}}, \rho \} \right)
    \, ,
    \label{eqn:nonunitary-error-expanded}
\end{equation}
where $A_{\mathbf{n}} = P \Pi_P(\mathbf{n})$ (note that we have dropped the `$1$' label with respect to Sec.~\ref{sec:unitary-pauli-errors} for simplicity of notation). Note that, since $A_\mathbf{m}^\dagger A_\mathbf{n} = \delta_{\mathbf{m}\mathbf{n}} \Pi_P(\mathbf{n})$, only the diagonal terms in Eq.~\eqref{eqn:nonunitary-error-expanded} 
contribute to stationarity and need to be compensated for.
The dissipative part of $\Liouvillian$ can be used to compensate for the diagonal terms by taking 
\begin{equation}
    \delta \gamma_{\mathbf{n}\mathbf{n}} = g c_\mathbf{n}^2
    \, .
    \label{eqn:nonunitary-correction}
\end{equation}   
Since $\delta\gamma_{\pi(\mathbf{n}) \pi(\mathbf{n})} = \delta\gamma_{-\mathbf{n} -\mathbf{n}} = gc_\mathbf{n}^{-2}$, we obtain $(g+gc_\mathbf{n}^{-2})c_\mathbf{n}^{2} - (g+gc_\mathbf{n}^{2}) = 0$ for the diagonal contribution to Eq.~\eqref{eqn:stationary-general}, as required. Hence, $P$ errors may be corrected using the following protocol. Let $\mathbf{n}_\star$ be the set of measurement outcomes for which $c_\mathbf{n}^2$ is maximal. Then, in time interval $\delta t$,
\begin{enumerate}
    \item with probability $\delta t  g c_{\mathbf{n}_\star}^2$, measure the stabilizers that satisfy $\{ S_a, P \} = 0$,
    \item if the system was measured, apply unitary feedback $P$ with probability $(c_{\mathbf{n}} / c_{\mathbf{n}_\star})^2$.
\end{enumerate}
Again, one can trade off the rate at which the anticommuting stabilizers are measured with the probability of unitary feedback $P$ being applied.
While this procedure appears similar in spirit to error-mitigation techniques such as probabilistic error cancellation (PEC)~\cite{TemmeZNE}, we emphasize that the correction protocol genuinely (re-)steers the system into the stationary state $\sigma$ with no classical postprocessing of the data, as opposed to reproducing its correlations on average once the results have been reweighted according to some quasiprobability distribution.

This scheme is extremely similar to the standard quantum error-correcting scheme involving measurement and feedback (see Sec.~\ref{sec:application-to-QEC}).

Note that Eq.~\eqref{eqn:nonunitary-correction} is a particularly simple choice, but it is not the only way to correct for the error while maintaining stationarity. More precisely, any $\delta\gamma$ that satisfies the stationarity condition
\begin{equation}
    (g+\delta\gamma_{\pi(\mathbf{n})\pi(\mathbf{n})}) c_{\pi(\mathbf{n})}^{-1} \stackrel{!}{=} (g + \delta\gamma_{\mathbf{n}\mathbf{n}})c_{\mathbf{n}}^{-1}
\end{equation}
will suffice. Sending $\mathbf{n} \to \pi(\mathbf{n})$ reveals that this set of equations can be highly underdetermined. Another particularly convenient solution is to set $\delta\gamma_{\mathbf{n}\mathbf{n}}=0$ for all $\mathbf{n}$ such that $c_\mathbf{n} \leq 1$. Then, for all remaining $\mathbf{n}$,
\begin{equation}
    \delta\gamma_{\mathbf{n}\mathbf{n}} = g(c_\mathbf{n}^2-1) \text{ for } c_\mathbf{n} > 1
    \, .
\end{equation}
This redundancy is analogous to the different update rules that satisfy detailed balance in Markov-chain Monte Carlo, such as Metropolis-Hastings, Glauber, and heatbath dynamics.

\subsubsection{Incoherent, non-Pauli errors}
\label{sec:coherent-nonunitary}

Finally, we consider the most general class of incoherent errors,
namely those in which the error in Eq.~\eqref{eqn:nonunitary-deform} is generalized from a single Pauli string $P$ to some generic linear combination of Pauli strings, $P \to \sum_q a_q P_q$, with complex coefficients $a_q$. As before, these Pauli strings can always be written in terms of the eigenoperators $A_i$ by decomposing the identity as $\ident = \sum_\mathbf{n} \Pi(\mathbf{n})$, i.e., we can write $P_q = \sum_{\mathbf{n}} P_q \Pi(\mathbf{n}) \equiv \sum_{\mathbf{n}} A_{q\mathbf{n}}$. Note that the various Pauli strings may commute with different numbers of stabilizers; we take $\mathbf{n}$ to be the measurement outcomes for the \emph{union} of all stabilizers that anticommute with $\{ P_s \}$. The perturbation to $\Liouvillian$ can then be written
\begin{equation}
    \Liouvillian(\rho) \to \Liouvillian(\rho) + g \sum_{qr,\mathbf{m}\mathbf{n}} a_q\bar{a}_r \left(  A^{\vpd}_{q\mathbf{m}}\rho A^\dagger_{r\mathbf{n}} - \frac12 \left\{ A^\dagger_{r\mathbf{n}}  A^{\vpd}_{q\mathbf{m}} , \rho \right\} \right)
    \, .
    \label{eqn:nonunitary-deform-general}
\end{equation}
The effects of the diagonal contributions ($q=r$) can be removed using the results of the previous subsection using only stabilizer measurements and unitary feedback.
Here, we remove the effects of the \emph{off-diagonal} terms -- when possible -- by modifying the Hamiltonian.
Specifically, if the eigenvalues $c_{r\mathbf{n}}$ and $c_{q\mathbf{m}}$ are \emph{nondegenerate}, we are able to modify the Hamiltonian according to
\begin{equation}
     \delta h_{{q\mathbf{m}},r\mathbf{n}} = \ii g \left[
    \frac{1}{2} a_q\bar{a}_r \left(\frac{c^2_{r\mathbf{n}} + c^2_{q\mathbf{m}} }{c^2_{r\mathbf{n}} - c^2_{q\mathbf{m}}} \right)
    - a_r\bar{a}_q \frac{c^2_{q\mathbf{m}} c^2_{r\mathbf{n}} }{c^2_{r\mathbf{n}} - c^2_{q\mathbf{m}}} 
    \right]
    \, .
    \label{eqn:hamiltonian-correction}
\end{equation}
The term in the square brackets is anti-Hermitian, leading to the Hermiticity of the matrix $\delta h$. The case of degenerate eigenvalues will be dealt with shortly.
Recall that the Hamiltonian defined by the matrix $h_{ij}$ is $H = \sum_{ij} h_{ij} A_{j}^\dagger A^{\vpd}_i$, and that the operators $A_{j}^\dagger A^{\vpd}_i$ are not linearly independent from the operators $A_i$, which form a complete basis.
Indeed, we have $A^\dagger_{r\mathbf{n}}  A_{q\mathbf{m}} = \Pi(\mathbf{n}) P_r P_q \Pi(\mathbf{m})$, which is only nonzero for measurement outcomes that satisfy $\varphi_r(\mathbf{n}) = \varphi_q(\mathbf{m})$, where the function $\varphi_q$ flips the sign of measurement outcomes of stabilizers that anticommute with $P_q$, i.e., $P_q \Pi(\mathbf{n}) P_q =  \Pi(\varphi_q(\mathbf{n}))$. For such $\mathbf{m}$, $\mathbf{n}$, the jump operators satisfy $A^\dagger_{r\mathbf{n}}  A_{q\mathbf{m}} = P_rP_q \Pi(\mathbf{m})$, which is just another jump operator that diagonalizes $\mathcal{S}$. Furthermore, since $P_q$ and $P_r$ either commute or anticommute, the operator $A^\dagger_{q\mathbf{n}}  A_{r\mathbf{m}}$ also contributes to the coefficient for the jump operator $P_rP_q \Pi(\mathbf{m})$ in the Hamiltonian $H$. 

Next, consider what happens if the two (or more) jump operators have identical eigenvalues. In this case, the Hamiltonian cannot be used to compensate for such terms, since the contribution from $h_{ij}$ is projected out in the stationarity condition~\eqref{eqn:stationary-general}. Hence, we must instead modify $\gamma_{ij}$ to mitigate the effects of these terms.
Consider the case in which both $P_q$ and $P_r$ anticommute with the \emph{same} set of stabilizers.
Consequently, $\varphi_q(\mathbf{n}) = \varphi_r(\mathbf{n})$ for all $\mathbf{n}$, and therefore only measurement outcomes satisfying $\mathbf{m} = \mathbf{n}$ contribute nontrivially to the stationarity condition.
Restricting to measurements on the stabilizers that anticommute with $\{ P_q \}$ belonging to the degenerate block, stationarity is preserved if
\begin{equation}
    \gamma^{rq}_{-\mathbf{n}, -\mathbf{n}} c_\mathbf{n}^2 - \gamma^{qr}_{\mathbf{n}, \mathbf{n}} \pm \left[ \gamma^{qr}_{-\mathbf{n}, -\mathbf{n}} c_\mathbf{n}^2 - \gamma^{rq}_{\mathbf{n}, \mathbf{n}} \right] = 0
    \, ,
    \label{eqn:stationarity-degen}
\end{equation}
for all $q$ and $r$ belonging to the degenerate block. The $\pm$ sign follows from whether $P_q$ and $P_r$ commute ($+$) or anticommute ($-$).
It will be most convenient to choose $\gamma$ such that the two terms (i.e., inside and outside of the square brackets) both vanish separately.
This occurs if we modify $\gamma \to \gamma + \delta\gamma$ such that
\begin{equation}
    \delta \gamma^{qr}_{\mathbf{n} , \mathbf{n}} = \bar{a}_q a_r g  c_\mathbf{n}^2
    \, .
    \label{eqn:degenerate-correction}
\end{equation}
The diagonal elements are clearly positive, and Eq.~\eqref{eqn:degenerate-correction} generically produces a positive semidefinite matrix. Furthermore,
note that the correction to the diagonal elements matches the correction~\eqref{eqn:nonunitary-correction} obtained  previously.
Note that Eq.~\eqref{eqn:stationarity-degen} is highly underdetermined and, hence, we emphasize that Eq.~\eqref{eqn:degenerate-correction} is merely a particularly simple choice for the correction.
Since the correction $\delta \gamma$~\eqref{eqn:degenerate-correction} factorizes, we immediately identify that the additional jump operators required to correct the erroneous terms in Eq.~\eqref{eqn:nonunitary-deform-general} are $L_\mathbf{n} \propto \sum_q \bar{a}_q A_{q\mathbf{n}}$. This correction can be implemented microscopically using the interpretation given in Sec.~\ref{sec:generalized-meas-interpretation}.

Finally, we consider the most challenging case to correct: when two (or more) of the constituent jump operators have degenerate eigenvalues, but anticommute with \emph{different} stabilizers. Specifically, consider two jump operators $A_{q\mathbf{n}}$ and $A_{r\mathbf{n}}$ that anticommute with different stabilizers; the two sets could be completely disjoint, or have some (but not full) overlap.
While only $\mathbf{m}$, $\mathbf{n}$ satisfying $\varphi_q(\mathbf{n}) = \varphi_r(\mathbf{n})$ contribute to stationarity [see the discussion below Eq.~\eqref{eqn:hamiltonian-correction}], we find it more convenient to satisfy the equations
\begin{equation}
    \gamma^{rq}_{\varphi_r(\mathbf{n}) \varphi_q(\mathbf{m})} c_{q\mathbf{m}} c_{r\mathbf{n}} - \gamma_{\mathbf{m} \mathbf{n}}^{qr} \stackrel{!}{=} 0
\end{equation}
for \emph{all} $q\mathbf{m}$ and $r\mathbf{n}$ satisfying the degeneracy condition $c_{q\mathbf{m}} = c_{r\mathbf{n}}$. These equations can easily be satisfied by modifying $\gamma^{qr}_{\mathbf{m} \mathbf{n}}$ according to
\begin{equation}
    \delta \gamma^{qr}_{\mathbf{m} \mathbf{n}} =  g\bar{a}_q a_r c_{q\mathbf{m}} c_{r\mathbf{n}}    
    \, , \quad \forall \: (q\mathbf{m}), (r\mathbf{n}) \text{ s.t.~} c_{q\mathbf{m}} = c_{r\mathbf{n}}
    \, .
    \label{eqn:degenerate-correction-generic}
\end{equation}
Now consider all $(q\mathbf{m}), (r\mathbf{n})$ such that $c_{q\mathbf{m}} = c_{r\mathbf{n}} = c$ for some $c$. Also denote the set of jump operator indices $q\mathbf{m}$ that contribute to this degenerate block by $q\mathbf{m} \in c$. Hence, for each degenerate block, we have the diagonal jump operator $L_c = \sum_{q\mathbf{m} \in c} \bar{a}_q A_{q\mathbf{m}}$, which appears in the Lindbladian with the rate $gc^2$. Essentially, the condition $c_{q\mathbf{m}} = c_{r\mathbf{n}}$ breaks up the jump operators $A_{q\mathbf{m}}$ into equivalence classes where $q\mathbf{m} \sim r\mathbf{n}$ if the eigenvalues satisfy $c_{q\mathbf{m}} = c_{r\mathbf{n}}$. The jump operators that we add to restore stationarity then correspond to a linear combination of all jump operators in an equivalence class. 
The interpretation of these jump operators is analogous to the simpler case of degeneracy considered above: For each degenerate block, the jump operator $L_c$ can be decomposed using an SVD. This provides us with a Kraus operator for each degenerate block, and a generalized measurement in which the system is coupled to an ancillary degree of freedom that is subsequently measured can correspondingly be constructed.

\begin{figure}
    \centering
    \includegraphics[width=0.55\linewidth]{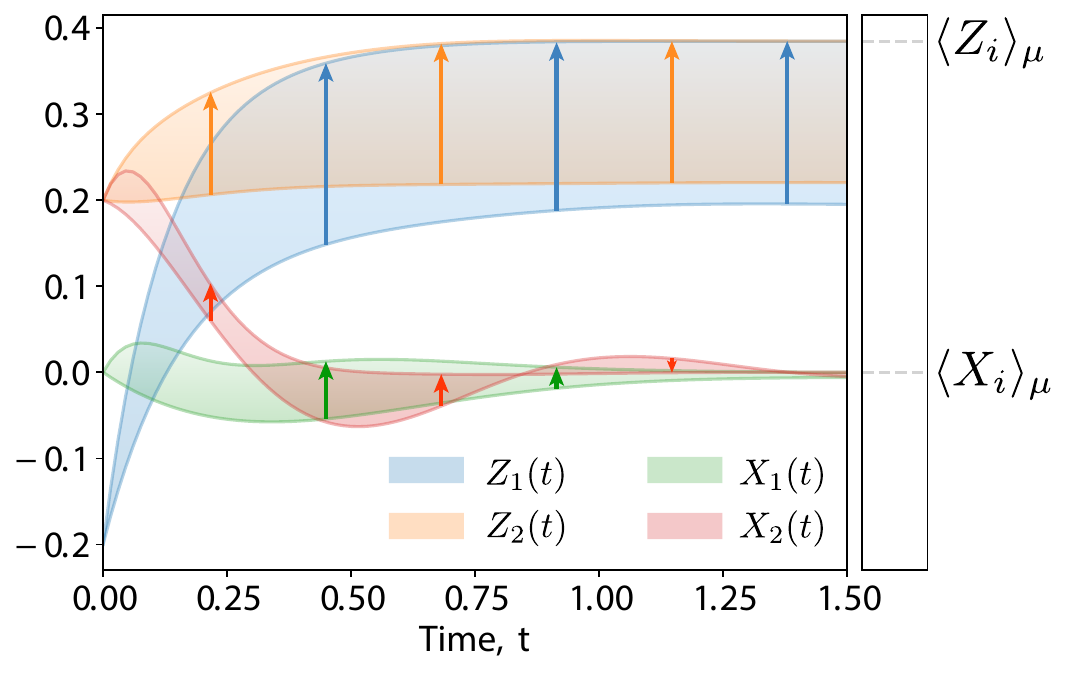}
    \caption{Implementing the correction procedure for Rydberg errors in a two-qubit system. The unperturbed system is driven towards the steady state $\Phi = -\mu \sum_i Z_i$ by the dynamics~\eqref{eqn:exp(Z)-dynamics} on each qubit with $\Gamma=1$, $\gamma_n=0$, and $\mu = \log(3/2)$, in addition to coherent dynamics specified by the Hamiltonian $H_0 = Z_1 - 3Z_2$ (which commutes with $\Phi$). The system is then perturbed by Rydberg errors~\eqref{eqn:Rydberg-error-L} with $g=5/2$, and corrected using the procedure outlined in Sec.~\ref{subsubsec:rydberg}. We illustrate the expectation values of single-qubit observables $X_i$ and $Z_i$ both in the presence and absence of the correction procedure (with the arrows pointing from uncorrected to corrected). The expectation values of $X_i$ and $Z_i$ in the target state $\ee^{-\Phi}$ are illustrated on the right-hand side, showing that the correction procedure leads to the required expectation values at late times.}
    \label{fig:rydberg-correction}
\end{figure}

\subsubsection{``Rydberg" errors}
\label{subsubsec:rydberg}

Given the abstract nature of the correction procedure for general ``degenerate'' errors, we provide here an explicit example of ``Rydberg errors.''  This example is motivated by the prospect of neutral-atom quantum computation~ \cite{Saffman2010, Bluvstein_2022, Lis_2023, Bluvstein_2023}; in such a platform, a two-qubit gate arises from a Rydberg blockade whereby two nearby neutral atoms interact via a Hamiltonian of the form
\begin{equation}
    H_{\text{int},12} = V_0 \frac{(1-X_1)(1-X_2)}{4}
    \, ,
    \label{eqn:CZ-generator}
\end{equation}
where $V_0$ is some constant.   The atoms are moved by trapping them in optical tweezer arrays, with the light beams readily adjusted by moving mirrors.  To apply the correct two-qubit gate, therefore, one needs to bring the atoms together for a specific length of time, and any uncertainty in the time in which the atoms are nearby causes a correlated one- and two-qubit error.
Averaging over this uncertainty leads to
\begin{equation}
    \int \thed \theta \, w(\theta)  \ee^{-\ii \theta H} \rho \ee^{\ii \theta H} = (1-p) \rho + p  L^{\vpd}_\text{R} \rho L_\text{R}^\dagger
    \, ,
\end{equation}
where $w(\theta) = w(-\theta)$ is the probability density function for the random variable $\theta$,
the jump operator $L_\text{R} = \frac12 (1+X_1 +X_2 - X_1 X_2)$ is just the \textsf{CZ} gate written in the $X$ basis, and $p = \expval{ \sin^2(\theta/2)}_{w(\theta)}$. If $p = g \delta t $, then we obtain Lindbladian time evolution with
\begin{equation}
    \mathcal{L}(\rho) \supset g \left[ L^{\vpd}_\text{R} \rho L_\text{R}^\dagger - \frac12 \{ L_\text{R}^\dagger L^{\vpd}_\text{R} , \rho \} \right]
    \, ,
    \label{eqn:Rydberg-error-L}
\end{equation}
Suppose, for concreteness, that we wish to stabilize the simple paramagnetic stationary state with $\Phi = -\mu \sum_i Z_i$, and
let $\{ P_q \} = \{ \ident, X_1, X_2 , X_1 X_2 \}$ for $q=0, \dots, 3$ label the Pauli strings that contribute to the jump operator $L_\text{R}$. The operator basis $A_{q\mathbf{m}}$ is then $A_{q\mathbf{m}} = P_q \Pi(\mathbf{m})$, where $\mathbf{m} = \{ m_1, m_2 \}$ is the set of measurement outcomes for the operators $Z_1$ and $Z_2$. The operators $A_{q\mathbf{m}}$ exhibit degeneracy of their eigenvalues for certain measurement outcomes, which complicates the correction procedure and means that we must consider the most general case presented in Sec.~\ref{sec:coherent-nonunitary}. Specifically, the eigenvalues satisfy
\begin{align}
    c_{0\mathbf{m}} &= c_{3\mathbf{n}} \text{ for } n_1 + n_2 = 0 \, , \\
    c_{1\mathbf{m}} &= c_{2\mathbf{n}} \text{ for } m_1 - n_2 = 0
    \, .
\end{align}
The diagonal jump operators that we need to add to $\mathcal{L}$ in order to correct the error in Eq.~\eqref{eqn:Rydberg-error-L} are $[\ident -  X_1 X_2 \sum_{n_1+n_2=0}\Pi(\mathbf{n})]/2$ corresponding to the $c=1$ block, and $[X_1 \Pi(n_1=\pm 1) + X_2 \Pi(n_2=\pm 1)]/2$ corresponding to the $c=\ee^{\mp\mu}$ block. The nondegenerate diagonal elements corresponding to $c=\ee^{\pm 2\mu}$ are corrected for using simple projective measurements and feedback as described in Sec.~\ref{sec:correcting-nonunitary}.
The nondegenerate off-diagonal terms are compensated for using a Hamiltonian correction
\begin{equation}
     \delta h_{{q\mathbf{m}},r\mathbf{n}} =  
    \frac{\ii g}{2} \left(\frac{c^2_{r\mathbf{n}} + c^2_{q\mathbf{m}} - 2 c^2_{q\mathbf{m}} c^2_{r\mathbf{n}} }{c^2_{r\mathbf{n}} - c^2_{q\mathbf{m}}} \right) a_q a_r
    \, ,
\end{equation}
which multiplies the operator $A_{r\mathbf{n}}^\dagger A_{q\mathbf{m}}$. Since the operators $P_q$ for $q=1,2,3$ commute with $P_0$, the correction terms $\delta h_{{0\mathbf{m}},1\mathbf{n}} = \delta h_{{0\mathbf{m}},2\mathbf{n}} = \delta h_{{0\mathbf{m}},3\mathbf{n}} = 0 $, as well as their Hermitian conjugates, may be taken to be zero. Other (nondegenerate) matrix elements, however, do produce nonzero contributions. The correction procedure for Rydberg errors of the form of Eq.~\eqref{eqn:Rydberg-error-L} for a target steady state with $\Phi = -\mu \sum_i Z_i$ is illustrated in Fig.~\ref{fig:rydberg-correction}.

\subsubsection{Measurement errors} \label{section: measurement error}

When using measurements and feedback to correct 
errors, we can also account for the possibility of ``readout errors'' in the measurement itself -- i.e., that the eigenvalue recorded by the apparatus differs from the true measurement outcome. It turns out, such errors are correctable; one need only modify the projection operators: 
\begin{equation}
\label{eq: measurement error rate}
    \Pi_P(\mathbf{n})\, \rho \, \Pi_P(\mathbf{n}) \to \sum_{\mathbf{n}'} p(\mathbf{n}' | \mathbf{n})\, \Pi_P(\mathbf{n}')\, \rho\, \Pi_P(\mathbf{n}'),
\end{equation}   
where $p(\mathbf{n}' | \mathbf{n})$ is the probability of getting states in subspace $\mathbf{n}'$ instead of the desired $\mathbf{n}$ subspace.  

When correcting incoherent Pauli errors, if we 
evolve under $\Liouvillian_{P(\mathbf{n})}(\rho) \equiv P \Pi_P(\mathbf{n})\, \rho\, \Pi_P(\mathbf{n}) P - \frac{1}{2} \{\Pi_P(\mathbf{n}), \rho\}$, we end up getting $\sum_{\mathbf{n}'} p(\mathbf{n}' | \mathbf{n}) \mathcal{L}_{P(\mathbf{n}')}$. From \eqref{eqn:stationary-general}, we know that the dynamics $\mathcal{L}_{P(\mathbf{n})} + c^2_{\mathbf{n}} \Liouvillian_{P(\mathbf{-n})}$ can keep $\sigma$ stationary . From the perspective of stationarity, we can say $\Liouvillian_{P(\mathbf{n})}$ and $c^2_{\mathbf{n}} \Liouvillian_{P(\mathbf{-n})}$ effectively cancel each other out. Similarly, for Hamiltonian errors we find that applying $U_\mathbf{n} = \frac{1}{\sqrt{2}}(P-\ii s_\mathbf{n} \ident)$ after the projection $\Pi_P(\mathbf{n})$ can cancel out the effect of applying $U_\mathbf{n}$ after the projection $\Pi_P(\mathbf{-n})$. Therefore, 
to implement the dynamics generated by Eq.~\eqref{eqn:lindblad-modification-unitary} or Eq.~\eqref{eqn:nonunitary-correction}, as long as the probability of measurement readout errors is below a certain threshold, one simply modifies the rates of the measurements and feedback such that with the measurement errors and the cancellation of some part of the errors, the resulting dynamics can be the same as in Eq.~\eqref{eqn:lindblad-modification-unitary} or Eq.~\eqref{eqn:nonunitary-correction}.  

As a simple example, consider a pair of qubits where $\Phi = -\mu Z_1 Z_2 $, and suppose that the measurement of a singular stabilizer $Z_1 Z_2$ returns the wrong outcome with probability $q$. We then have that  $p(+|+) = p(-|-) = 1-q$ and $p(+|-) = p(-|+) = q$, where $n=\pm 1$ labels the two outcomes. Now if we try to add the following dynamics by measurement and feedback,
 \begin{align} \label{eq: measurement error}
    \mathcal{L}(\rho) \to \mathcal{L}(\rho) +  X_1 \frac{1}{2}(\ident - Z_1 Z_2)\, \rho \,\frac{1}{2}(\ident - Z_1 Z_2) X_1 -\frac{1}{2} \left\{\frac{1}{2}(\ident - Z_1 Z_2),\rho \right\} ,
\end{align}
leads to the Lindbladian
\begin{align} \label{eq: measurement error 2}
    \mathcal{L}(\rho) \to\mathcal{L}(\rho) &+ p(-|-) \left[X_1 \frac{1}{2}(\ident - Z_1 Z_2)\, \rho \,\frac{1}{2}(\ident - Z_1 Z_2) X_1 -\frac{1}{2} \left\{\frac{1}{2}(\ident - Z_1 Z_2),\rho \right\} \right] \notag \\
    &+ p(+|-) \left[X_1 \frac{1}{2}(\ident + Z_1 Z_2)\, \rho \,\frac{1}{2}(\ident + Z_1 Z_2) X_1 -\frac{1}{2} \left\{\frac{1}{2}(\ident + Z_1 Z_2),\rho \right\} \right].
\end{align}
Effectively reproducing Eq.~\eqref{eq: measurement error} requires that $q/(1-q) < \ee^{-2\mu}$, as we now explain.  From the perspective of the stationary state, the two terms in Eq.~\eqref{eq: measurement error 2} should cancel out. If $p(-|-) > \ee^{2\mu}\, p(+|-)$, after the cancellation, only the first term appears, with coefficient $p(-|-) - \ee^{2\mu}\, p(+|-)$. If instead, $p(-|-) < \ee^{2\mu}\, p(+|-)$, then only the second term appears, with coefficient $p(+|-) - \ee^{-2\mu}\, p(-|-)$. Therefore, recovering Eq.~\eqref{eq: measurement error} requires that $p(-|-) > \ee^{2\mu}\, p(+|-)$, which means $q/(1-q) < \ee^{-2\mu}$. Similar analysis can be applied to more complicated systems.

\subsection{Application to error correction}
\label{sec:application-to-QEC}
The formalism above has a natural application to the theory of quantum error correction.   In this paper, we will discuss this application in rather abstract terms, and will discuss specific applications in other papers.

The most common kind of quantum error correcting code is a Calderbank-Shor-Steane (CSS) code \cite{Calderbank_1996,Steane_1996}.  In such a code, the physical Hilbert space has $n$ qubits, and stores $k$ logical qubits.  More precisely, we define $k$ logical $X$ operators $X_{\text{L},1},\ldots, X_{\text{L},k}$ and $k$ logical $Z$ operators $Z_{\text{L},1},\ldots, Z_{\text{L},k}$, such that each $X_{\text{L},i}$ is a product of physical $X$ Paulis, and each $Z_{\text{L},i}$ is a product of physical $Z$ Paulis.  These logical operators obey
\begin{equation}
    X_{\text{L},i} Z_{\text{L},j} = (1-2 \kron{i,j}) Z_{\text{L},j} \; X_{\text{L},i} \, .~
\end{equation}
The products of arbitrary logical Paulis generate a group $\SU{2^k}$, corresponding to logical gates on the code.  
If, by some miracle of nature, our open dynamics has a strong $\SU{2^k}$ symmetry, then it perfectly protects a logical qubit stored in the system.  

The theory of error correction has been developed to protect information in systems where such a strong symmetry does not exist. Indeed, we will use the remaining $n-k$ ``effective qubits" in the system to detect errors as follows:  pick a set of $n-k$ commuting Pauli strings $\lbrace S_a\rbrace$, which are either products of physical $X$ or $Z$ Paulis. These are called the stabilizers of the code.  We choose these stabilizers so that 
\begin{equation}
    [S_a, X_{\text{L},i}]= [S_a, Z_{\text{L},i}] = 0 \,,~~
\end{equation}
while, ideally, single-qubit Paulis all anticommute with at least one $S_a$.   Quantum error correction then typically proceeds by measuring stabilizers $S_a$, attempting to locate the physical errors that occurred based on the measurement outcomes, and applying the error a second time to undo it.  Note that during this measurement process, the wavefunction collapse effectively converts generic errors into either $X$ or $Z$ type errors, which we attempt to correct.

The crudest possible kind of quantum ``error-correcting protocol'' corresponds to a classical ``Gibbs sampler.''  The goal is to drive the system toward the stationary state generated by
\begin{equation}
    \Phi = -\sum_a \mu_a S_a ,
\end{equation}
corresponding to a sum of local stabilizers.  Using the protocol of Sec.~\ref{sec:correcting-nonunitary} corresponds to measuring syndromes and introducing local errors in such a way as to drive the system towards the steady state $\ee^{-\Phi}$.  Such a protocol is rather similar in spirit to the typical error correction scheme, which also proceeds by measuring stabilizers; however, the vast majority of error correction schemes studied in the literature then rely on ``active" decoding, where \emph{global} information about the measurement outcomes is used to infer the locations of errors.  In contrast, the scheme based on Sec.~\ref{sec:correcting-nonunitary} will only apply feedback on the system based on \emph{local} measurement outcomes, and thus represents a passive ``decoding" scheme.

The passive decoder obtained in our framework will drive the system to the steady state $\mathrm{e}^{-\Phi}$; however, for many of the simplest error-correcting codes such as the surface code \cite{bravyi1998,Fowler_2012}, such a steady state is \emph{not} useful.  The reason is that logical errors can easily proliferate in this thermal ensemble, somewhat analogously to how a domain wall can propagate easily in the one-dimensional Ising model (thus preventing any finite-temperature phase transition to an ordered state).   What is instead often desired is a \emph{thermal phase transition} where, upon making the chemical potentials $\{\mu_a\}$ large enough, the steady state $\ee^{-\Phi}$ condenses onto a small fraction of Hilbert space (with overwhelming probability in the thermodynamic limit, the system is found ``close'' to a logical state, and any residual errors are easily decoded).   Such a phase transition would be to a topologically ordered phase: the simplest known example corresponds to the four-dimensional toric code \cite{Dennis_2002,alicki2008thermal}.  

There are then two natural ways to use the framework described herein to engineer passive decoders capable of protecting information.  (\emph{1}) We can choose $\Phi$ to be a more complicated sum over \emph{products} of stabilizers, such that $\Phi$ exhibits a finite temperature phase transition.  (\emph{2}) More directly, we observe that the crucial property of the phase transition is not, per se, the existence of a thermodynamic ordered phase itself, but rather that the \emph{mixing time} in which a logical qubit can be corrupted is long.  We can instead aim to  build open quantum dynamics with slow mixing times directly.  Ordinarily, this also involves looking for phase transitions, but the two phenomena can be distinct \cite{Hong:2024vlr}.

Lastly, a crucial aspect of quantum error correction in experiments is the imperfections in the measurement and feedback used to detect and correct errors.  We already illustrated how to incorporate imperfect stabilizer measurement in Sec.~\ref{section: measurement error}.  Again, the simplest surface code is quite vulnerable to such measurement errors, and a ``spacetime history'' of stabilizer measurement outcomes is needed for the accurate detection and correction of errors errors \cite{Fowler_2012}. There is, therefore, significant interest in models that can achieve \emph{single-shot error correction}, where measuring stabilizers once is sufficient to correct for any errors \cite{Bombin_2015}.  A passive decoder with a slow mixing time is even more desirable than single-shot error correction; not only will it accurately protect against all kinds of errors, but it also is amenable to implementation via  ``measurement-free'' quantum error correction \cite{Ahn_2002}, as discussed in Ref.~\citenum{Hong:2024vlr}.

\section{Quantum error correction and the repetition code}
\label{sec:examples}
The manipulations up to this point have all been rather formal. We now present an explicit illustration of (\emph{i}) how to construct nontrivial dynamics that protect a particular stationary state $\sigma$ in the presence of both unitary dynamics and measurement and feedback, and (\emph{ii}) how to correct for errors -- both Hamiltonian errors and incoherent errors -- that occur at a known rate in the familiar context of the repetition code. 

Specifically, we consider a system composed of spin-$1/2$ degrees of freedom arranged on a square lattice in two spatial dimensions.
Suppose that we wish to protect the steady state $\sigma = \ee^{-\Phi}$ where $\Phi$ takes the form of Ising interactions between neighboring vertices of the square lattice
\begin{equation}
    \Phi = - \mu \sum_{\expval{x,y}} Z_x Z_y
    \, ,~~
    \label{eqn:ising-on-graph}
\end{equation}
where the sum runs over neighboring sites $x$ and $y$. Note that $\mu$ thus plays the role of an inverse temperature for the discussion that follows.
Using the procedure outlined in Sec.~\ref{sec:stabilizer}, we are able to find a convenient basis for jump operators. This basis corresponds to the eigenoperators of the map $\mathcal{S}$~\eqref{eqn:superop-eigenvalues}, using which we can construct a family of dynamics that protects $\sigma$ defined by Eq.~\eqref{eqn:ising-on-graph}.
The most local possible nontrivial choices of these jump operators can be found by ``dressing'' the single-site operator $X_x$. Specifically, from Sec.~\ref{sec:stabilizer}, we deduce that jump operators take the form of \emph{conditional spin flips}:
\begin{equation}
    A_x(\mathbf{n}) = X_x \Pi_x(\mathbf{n}) = X_x \prod_{ y : \langle xy \rangle} \frac12 (1 + n_{xy} Z_x Z_y) 
    \, ,
    \label{eqn:Ai-graph-rep-code}
\end{equation}
where the product is over the edges emanating from vertex $x$, since only the stabilizers on edges touching $x$ anticommute with $X_x$. $n_{xy}=\pm 1$ corresponds to the tentative measurement outcomes that would occur if $Z_xZ_y$ were measured -- namely, Eq.~\eqref{eqn:Ai-graph-rep-code} corresponds to an operator that applies $X_x$ after projecting onto certain stabilizer eigenvalues. 
The operator $A_x(\mathbf{n})$~\eqref{eqn:Ai-graph-rep-code} is an eigenvector of $\mathcal{S}$~\eqref{eq:S} with eigenvalue 
\begin{equation}
    c_x(\mathbf{n}) = \exp\left[-\mu \sum_{y : \langle xy \rangle} n_{xy}\right] =  \exp\left[-2\mu \left(2-|\mathbf{n}|\right)\right] = \exp\left(-\frac12 \Delta \Phi_x\right),
\end{equation}
where we defined $|\mathbf{n}|$ as the number of $-1$ in the stabilizer eigenvalues $\mathbf{n}$ and $\Delta\Phi_x$ is the change in $\Phi$~\eqref{eqn:ising-on-graph} induced by flipping the spin on site $x$.
If $|\mathbf{n}| = 2$, then flipping the spin on site $x$ does not change the number of antiferromagnetic bonds (i.e., leaving $\Phi$ unchanged).
This is reflected in the fact that $c_x = 1$ for such configurations of $\{\mathbf{n}\}$; diagonal terms composed of such operators, which locally rearrange domain wall configurations when acting on computational basis states, can be added freely to $\mathcal{L}$ without affecting stationarity of $\sigma$.
On the other hand, if $|\mathbf{n}| = 0$ or 4 (all bonds are either ferromagnetic or antiferromagnetic), then $c_i$ will lead to an exponential suppression or enhancement of the rate at which such processes occur for $\mu \gg 1$. Configurations with an unequal mixture of ferromagnetic and antiferromagnetic bonds will also be suppressed or enhanced, but to a lesser degree.
Considering jump operators that flip spins belonging to some connected cluster of sites on the lattice leads to similar conclusions: only the edges $\langle xy \rangle$ at the \emph{boundary} of the cluster are (\emph{i}) projected out in the generalization of Eq.~\eqref{eqn:Ai-graph-rep-code}, and (\emph{ii}) contribute to the eigenvalues via $n_{xy}$, since $\Phi$ is locally unchanged in the interior of the cluster.

\subsection{Correcting for errors}

\subsubsection{Incoherent Pauli errors} \label{sec:nonunitary error correction}

To gain some intuition, we now outline the simplest class of (T-even) dynamics compatible with stationary of $\sigma$ that can be deduced from the framework presented in Sec.~\ref{sec:stationarity}.
Recall that all T-even dynamics derive from a Hermitian matrix $m$, which gives rise to $\gamma_{xy} = m_{xy} (c_x + c_y)$. The simplest dynamics that protects $\sigma$ follows from taking $m_{xy}$ to be a real, diagonal matrix with nonnegative entries. In this case, the Hamiltonian term vanishes in $\mathcal{L}$, and the dissipative part of $\mathcal{L}$ has a simple interpretation.
Indexing sites with $x$ and measurement outcomes for the stabilizers along edges emanating from $x$ by $\mathbf{n}$ 
\begin{equation}
    \mathcal{L} = \sum_{x, \mathbf{n}} \gamma_x(\mathbf{n})  \mathcal{D}\left[A_{x}(\mathbf{n})\right]
    \, ,
    \label{eqn:rep-code-diag-dynamics}
\end{equation}
where we introduced the ``dissipator'' $\mathcal{D}$ via
\begin{align}
    \mathcal{D}\left[A\right](\rho) \equiv A \rho A^{\dagger}-\frac{1}{2}\{A^{\dagger} A, \rho \}
    \, ,
\end{align}
we observe that at position $x$, spin flips conditioned on stabilizer eigenvalues $\mathbf{n}$ occur at rate $\gamma_x(\mathbf{n})$. While there is much freedom in how the diagonal matrix elements $m_x(\mathbf{n})$ are chosen, the constraint $m_{\pi(x)\pi(y)} = m^*_{xy}$ -- required to ensure Hermiticity of $\gamma_{xy}$ -- reduces to $m_x(\mathbf{n}) = m_x(-\mathbf{n})$. This constraint guarantees that a conditional spin flip and the reversed process occur at the rates appropriate to stabilize $\sigma$: $\gamma_{x}(\mathbf{n})/c_x(\mathbf{n}) = \gamma_{x}(-\mathbf{n})/c_x(-\mathbf{n})$. In the basis of stabilizer eigenstates, the dynamics \eqref{eqn:rep-code-diag-dynamics} can be mapped to classical dynamics and the constraints of the ratio between $\gamma_{x}(\mathbf{n})$ and $\gamma_{x}(\mathbf{-n})$ is equivalent to the classical detailed balance condition~\eqref{eqn:classical DB}.

\begin{figure}[t]
    \centering
    \includegraphics[width=\textwidth]{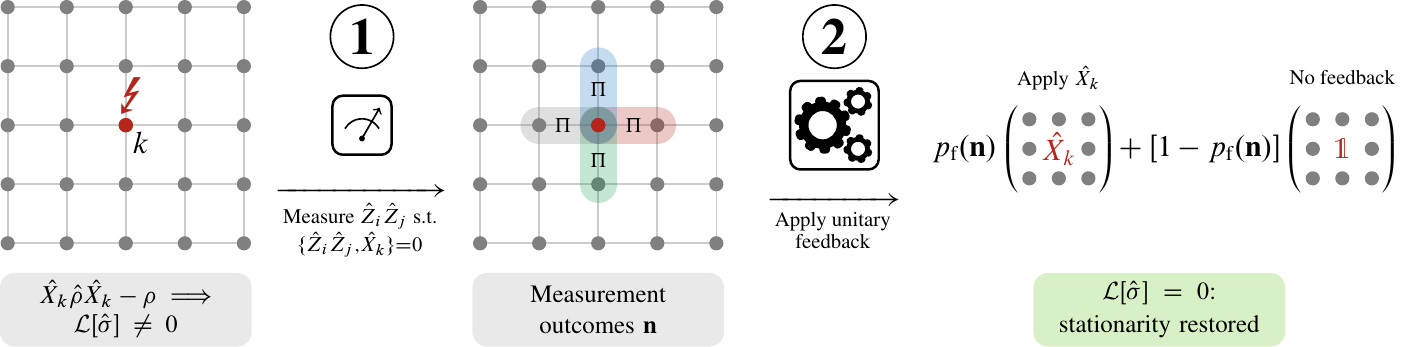}
    \caption{Schematic illustration of the measurement and feedback protocol for correcting incoherent errors in the Lindbladian that take the form of a Pauli $X$ operator. The desired steady state is $\sigma = \ee^{-\Phi}$, where $\Phi$ corresponds to a 2D classical Ising model. In the presence of stray transverse fields, this state is no longer the steady state of the Lindbladian, $\mathcal{L}[\sigma] \neq 0$. By performing measurements of the stabilizers and probabilistic unitary feedback, the desired steady state $\sigma$ can be restabilized. The measurements occur at a rate that is proportional to the strength of the stray fields.}
    \label{fig:enter-label}
\end{figure}

One of the simplest possible solutions to the above conditions on $\gamma_{xy}$ is given by 
\begin{equation} \label{eqn:coefficient-simple-T}
\gamma_x(\mathbf{n}) 
=  \gamma\left(|\mathbf{n}|\right) 
= 
\begin{cases}
    \gamma_4 \, \ee^{\mp 4\mu} &\text{ if } 2-|\mathbf{n}| = \pm 2  , \\
    \gamma_2 \, \ee^{\mp 2\mu} &\text{ if } 2-|\mathbf{n}| = \pm 1  , \\
    0 &\text{ otherwise,}
\end{cases}
\end{equation}
where $\gamma_2$ and $\gamma_4$ are nonnegative constants.  Such dynamics corresponds to a canonical (continuous-time) Gibbs sampler, along the lines of, e.g., the classic Metropolis algorithm \cite{Metropolis_1953}.

A slightly more interesting scenario arises if we consider that the system is instead subject to incoherent $X$ errors, which occur at some rate $g_0$. At the level of the Lindbladian description, $\mathcal{L}$ in Eq.~\eqref{eqn:rep-code-diag-dynamics} is modified to 
\begin{equation}
    \mathcal{L}(\rho) \to \mathcal{L}(\rho) +  g_0 \sum_x \left( X_x \rho X_x - \rho \right)
    \, .
    \label{eqn:rep-code-X-error}
\end{equation}
If this perturbation is not corrected, the stationarity of $\sigma$ is broken. To maintain the stationarity of $\sigma$, we can perform measurements and feedback designed to cancel the effects of the perturbation exactly~\eqref{eqn:rep-code-X-error}. In other words, we can add some terms to Eq.~\eqref{eqn:rep-code-X-error} corresponding to measurements and feedback, so that the dynamics instead realize Eq.~\eqref{eqn:rep-code-diag-dynamics} with coefficients given by Eq.~\eqref{eqn:coefficient-simple-T}. Using the results of Sec.~\ref{sec:correcting-nonunitary}, we can perform the following protocol. 
For each site $x$, in a time step $\delta t$, measure the stabilizers on edges connecting to site $x$ with probability $g_0\left(\ee^{8\mu}-1\right) \delta t$. After we get the measurement result $n_{xy}$ for each neighboring site $y$, apply $X_x$ with probability $\Lambda_\mathbf{n}$, where $\Lambda_\mathbf{n}|_{|\mathbf{n}| = 3} = \left(\ee^{4\mu}-1\right)/\left(\ee^{8\mu}-1\right)$, $\Lambda_\mathbf{n}|_{|\mathbf{n}| = 4} = 1$ and all other $\Lambda_\mathbf{n}=0$. The resulting dynamics is 
\begin{align} \label{eq: correct nonunitary error}
    \mathcal{L} \to \mathcal{L} +  g_0 \sum_x \mathcal{D}\left[X_x\right]  
    + \sum_{x, \mathbf{n}} g_0\left(\ee^{8\mu}-1\right) \left( \Lambda_\mathbf{n}\, \mathcal{D}\left[X_x \Pi_x(\mathbf{n})\right]   +\left(1-\Lambda_\mathbf{n}\right) \mathcal{D}\left[\Pi_x(\mathbf{n})\right] \right)
    \, ,
\end{align}
using the projector $\Pi_x(\mathbf{n})$~\eqref{eqn:Ai-graph-rep-code}. 
This amounts to measurements that are Poisson distributed in time, with a probability of applying feedback that depends on the outcomes of the measurements.  The method is summarized in Table~\ref{table:error correction rate} and Fig.~\ref{fig:enter-label}.  Mathematically,  we have simply adjusted the choice of $\gamma_{2,4}$ in the Gibbs sampler~\eqref{eqn:coefficient-simple-T} to be compatible with the presence of the unwanted error~\eqref{eqn:rep-code-X-error}.  This choice is not unique; here we have made the choice that leads to the smallest possible $\gamma_{2,4}$.

\begingroup
\renewcommand*{\arraystretch}{1.4}
\begin{table}[t]
\centering
\begin{tabular}{@{}ccccc@{}}
\toprule
\multirow{2}{*}{Measurement result} & \multicolumn{2}{c}{incoherent Pauli errors}        & \multicolumn{2}{c}{Hamiltonian errors}                                               \\ \cmidrule(l){2-5} 
                                    & Probability                            & Feedback & Probability                             & Feedback                                   \\ \midrule
$|\mathbf{n}| = 0$                   & $0$                             &     N/A     & $\ee^{-8\mu}$                    & $\frac{1}{\sqrt{2}}(\ii X_x - \mathds{1})$ \\
$|\mathbf{n}| = 1$                   & $0$                             &    N/A      & $(1-\ee^{-4\mu})/(\ee^{8\mu}-1)$ & $\frac{1}{\sqrt{2}}(\ii X_x - \mathds{1})$ \\
$|\mathbf{n}| = 2$                   & $0$                             &   N/A       & $0$                              &           N/A \\
$|\mathbf{n}| = 3$                  & $(\ee^{4\mu}-1)/(\ee^{8\mu}-1)$ & $X_x$    & $(\ee^{4\mu}-1)/(\ee^{8\mu}-1)$  & $\frac{1}{\sqrt{2}}(\ii X_x + \mathds{1})$ \\
$|\mathbf{n}| = 4$                  & $1$                             & $X_x$    & $1$                              & $\frac{1}{\sqrt{2}}(\ii X_x + \mathds{1})$ \\ \bottomrule
\end{tabular}
\caption{Summary of protocols for correcting incoherent Pauli errors and Hamiltonian errors, discussed in Secs.~\ref{sec:nonunitary error correction} and \ref{sec:unitary error correction}, respectively: for each site $x$, in a time step $\delta t$, measure the 4 stabilizers that touch $x$ with probability $g_0\left(\mathrm{e}^{8\mu}-1\right) \delta t$ or $h_0\left(\mathrm{e}^{8\mu}-1\right) \delta t$. Next, apply feedback based on the measurement results as described in the table. The probability of applying feedback and the operators that need to be applied only depends on the number of $-1$ in the measurement results of the stabilizers $|\mathbf{n}|$.}
\label{table:error correction rate}
\end{table}
\endgroup

If we think of the repetition code as storing a logical qubit, notice that a single $Z$ is a logical operator; therefore, in the presence of single-$Z$ errors, no quantum error correcting code exists.  After all, in our formalism, adding $Z$ errors does not modify the steady state at all!  
As a reminder, protecting a steady state $\sigma$ is \emph{not} equivalent to protecting quantum information, as discussed in Sec. \ref{sec:application-to-QEC}.  To build a quantum error-correcting code using this framework, one must also look for dynamics that ``slowly mixes'' between different sectors of logical operators.

\subsubsection{Hamiltonian errors} \label{sec:unitary error correction}

If the Hamiltonian of the system is subjected to an $X$ field
\begin{equation}
    H \to H - h_0 \sum_x X_x
    \, ,
\end{equation}
an analogous sequence of measurements and unitary feedback can be applied to maintain stationarity of $\sigma$.
We assume that $h_0>0$. According to Sec. \ref{sec:unitary-pauli-errors}, we can easily get the modifications we need to make $\sigma = \ee^{-\Phi}$ stationary:
  \begin{align} \label{eq: example 2d}
     \mathcal{L}(\rho) \to \mathcal{L}(\rho) +\ii h_0 \sum_x \comm{X_x}{\rho} 
     + h_0\left(\ee^{8\mu}-1\right) \sum_{\mathbf{x},|\mathbf{n}| \neq 2} \left( \Lambda_\mathbf{n} \, \mathcal{D}\left[U_\mathbf{x}(\mathbf{n}) \Pi_\mathbf{x}(\mathbf{n})\right](\rho) + \left(1-\Lambda_\mathbf{n}\right) \mathcal{D}\left[ \Pi_\mathbf{x}(\mathbf{n})\right] (\rho) \right) ,
 \end{align}
 where 
 \begin{subequations}
 \begin{align}
 &\Lambda_\mathbf{n} =  \left|1-\ee^{-4\mu \left(2-|\mathbf{n}|\right)}\right| / \left(\ee^{8\mu}-1\right), \label{eq: lambda_n}\\
 &U_\mathbf{x}(\mathbf{n}) = \frac{1}{\sqrt{2}} \left(X_{\mathbf{x}} - \sgn\left(2-|\mathbf{n}|\right) \ii \ident\right).
 \end{align}
 \end{subequations}

The dissipative part can be generated by measurements and feedback: in time interval $\delta t$, measure the stabilizers with probability $h_0\left(\ee^{8\mu}-1\right)\delta t$, then apply $U_x(\mathbf{n})$ with probability $\Lambda_\mathbf{n}$ based on the measurement result $\mathbf{n}$. The protocol is also summarized in Table \ref{table:error correction rate}.

\begin{figure}[t]
\centering
\includegraphics[scale=0.66]{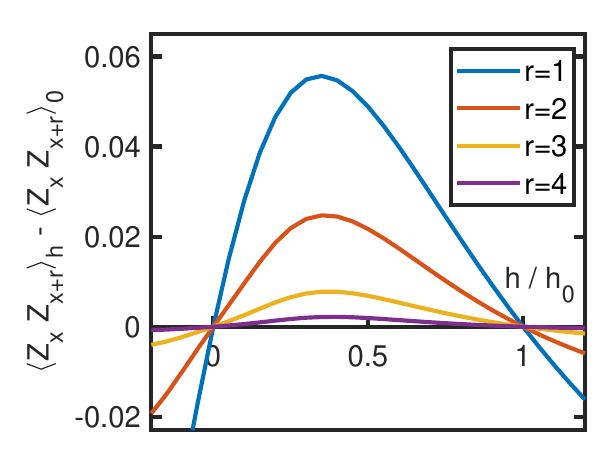}
\caption{We numerically simulate dynamics analogous to Eq.~\eqref{eq: example 2d} for 1D systems of size $L=11$ sites and $\mu = \frac{1}{4}\log{2}$ and change the Hamiltonian error to $h \sum_x X_x$. The change of the correlation function $\langle Z_x Z_{x+r} \rangle$ for different $h/h_0$ with respect to its value at $h=h_0$, denoted $\langle Z_x Z_{x+r} \rangle_0$. The absolute values of the correlator $\langle Z_x Z_{x+r} \rangle_0$ for $r=1, 2, 3, 4$ are, respectively, $0.172, 0.030, 0.005, 0.001$.}
\label{fig:2D Ising}
\end{figure}

One desirable property of our construction is that it, at least naively, gives rise to an \emph{experimentally detectable} measurement-induced phase transition.  To understand why, suppose that we study the dynamics on a square lattice at $\mu=\mu_{\mathrm{c}}$, where $\mu_{\mathrm{c}}$ is the (inverse) critical temperature of the 2D Ising model.   At $h=h_0$ the dynamics sample from the Ising critical point, and thus result in critical fluctuations; at $h=\infty$ we do not detect long-range order. At $h=0$, where we only apply measurement and feedback, notice that the gate $U_{x}$ is Clifford. One picture for the dynamics comes from considering an initial product state in the $Z$ basis; the ensuing time evolution alternates between measurements of syndromes of the form $Z_x Z_y$ and the unitary gates $U_{x}$. However, because the syndromes are measured between pairs of qubits \emph{before} any $U$s are applied, it is never possible for the set of qubits that are not in $Z$ eigenstates to be adjacent.  Therefore, the system remains in a product state, albeit not necessarily in the $Z$ basis.   As a consequence, all $ZZ$ correlation functions are effectively modeled by approximating that the state stays in a product state for all times, but after syndromes are measured there is only a 50\% chance of modifying the state (via Pauli $X$) on that site.   This adjustment does \emph{not} change expectation values of the (products of) $Z$s. Hence $\langle Z_xZ_y\rangle$ must be \emph{identical} at late times if $h=0$ and $h=h_0$.   For other values of $h$, the most plausible scenario is therefore one in which there is criticality or long-range order for $h\in [0,h_0]$, and not outside of this domain.  While we cannot exactly simulate this open system in 2D, 
direct application of time-evolution superoperator $\ee^{\Liouvillian t}$ on very small 1D systems suggests that this picture is correct: see Fig.~\ref{fig:2D Ising}.  We therefore conjecture that our construction leads to an \emph{experimentally observable measurement-induced phase transition} at $h=h_0$, where the experimentalist can simply measure the criticality in $\expval{ZZ}$ correlators to detect the transition. This statement holds assuming that the feedback rate is always the same; one can also simply tune through the thermal phase transition in the steady state, but in this situation the \emph{relative} rates of error correction depending on the number of flipped syndromes change in a ``complicated way,'' such that the phase diagram is not a simple function of $h/h_0$.

Lastly, we remark that, although the feedback scheme in this problem is Clifford, and (at discrete time steps) the continuous Hamiltonian evolution also generates Clifford gates, it does not appear to be the case that classical simulations of a circuit approximation to our model could accurately reproduce the dynamics.  Indeed, note that already when $h=0$, the Clifford feedback on its own prepares a critical state.  Introducing any additional $X$-type errors on top of this dynamics should lead to a short-range-correlated phase in classical simulations, and yet we see that the critical state is also robust at $h=h_0$.  The reason appears to be that the relative phases in the Clifford error correction tend to cancel the phases accumulated via continuous time evolution under the transverse field; this is precisely the kind of quantum effect that cannot be captured via Clifford simulation.

\subsubsection{Measurement errors} \label{sec:measurement error example}

In the previous example, where we corrected for Hamiltonian errors that took the form of a transverse field, we assumed that the syndromes can be measured perfectly.  If the syndrome measurements are imperfect (intuitively because the experimenter reads out an incorrect syndrome measurement outcome, with a \emph{known} error rate for such measurements), it is still possible to identify exactly  the location of a measurement-induced phase transition by specifying $\Phi$ to be the critical Ising model. Such a construction can be thought of as a toy model for fault-tolerant passive error correction using memoryless local decoding.

\begin{figure}[t]
\centering
\includegraphics[scale=0.63]{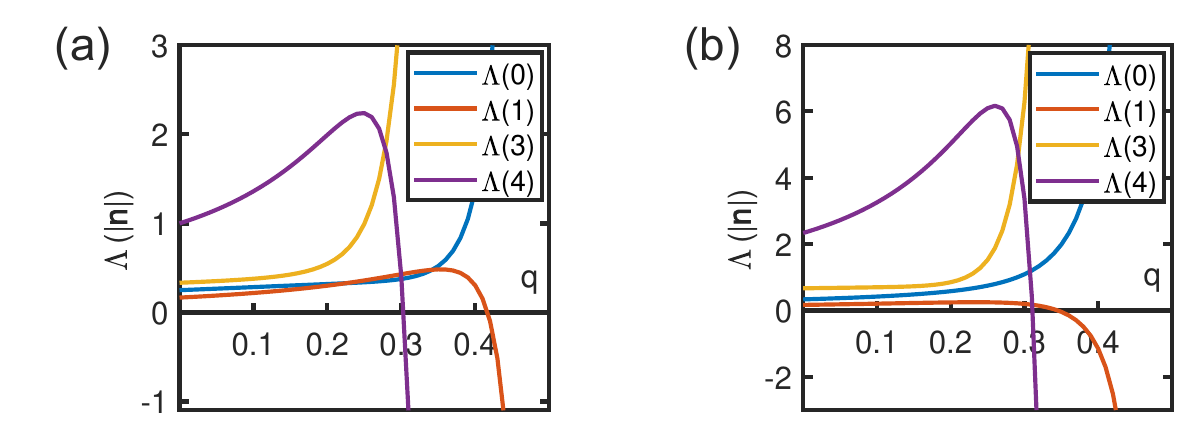}
\caption{ (a) The coefficients $\Lambda(|\mathbf{n}|)$ as functions of measurement error rate $q$ for Hamiltonian error correction~\eqref{eq: example 2d} for 2D Ising model. We choose $\mu = \frac{1}{4}\log{2}$. (b) The coefficients $\Lambda\left(|\mathbf{n}|\right)$ as functions of error rate $q$ for incoherent Pauli error correction~\eqref{eq: correct nonunitary error} for 2D Ising model. We choose $\gamma_4 = 2$, $\gamma_2 = \ee^{-2\mu}$ and $\mu = \frac{1}{4}\log{2}$.}
\label{fig:measurement error}
\end{figure}

Following Sec.~\ref{section: measurement error}, we can account for the imperfect syndrome measurements by modifying the effective Lindblad operator to be
\begin{align} \label{eq: measurement error 3}
     \sum_{\sum \mathbf{n} \neq 0}  \Lambda_\mathbf{n} \,\mathcal{D}\left[U_x(\mathbf{n}) \Pi_x(\mathbf{n})\right] &\to \sum_{\mathbf{n}', \sum \mathbf{n} \neq 0}  \Lambda_\mathbf{n} \,p(\mathbf{n}'|\mathbf{n}) \, \mathcal{D}\left[U_x(\mathbf{n}) \Pi_x(\mathbf{n}')\right] \notag \\
     &\to \sum_{\sum \mathbf{n} \neq 0}  \Lambda_\mathbf{n}'  \, \mathcal{D}\left[U_x(\mathbf{n}) \Pi_x(\mathbf{n})\right] ,
\end{align}
where $p(\mathbf{n}'|\mathbf{n}) = q^{\Delta \mathbf{n}}\,(1-q)^{4-\Delta \mathbf{n}}$, $q$ is the error rate when measuring a single stabilizer and $\Delta \mathbf{n}$ is the number of measurement results that differ between $\mathbf{n}$ and $\mathbf{n}'$. The second line of Eq.~\eqref{eq: measurement error 3} comes after the cancellation of dynamics described in Sec.~\ref{section: measurement error}. In order to keep $\sigma$ stationary, we need to modify $\Lambda_\mathbf{n}$ to make the coefficient $\Lambda_\mathbf{n}'$ match Eq.~\eqref{eq: lambda_n}.
The modified coefficients $\Lambda_\mathbf{n}$, which represent the rate we apply feedback after measurements in the experiments, can be easily calculated numerically. 
For example, if we choose $\mu = \frac{1}{4} \log{2}$, demand that $\Lambda_\mathbf{n} = \Lambda(|\mathbf{n}|)$, and keep $U_x(\mathbf{n})$ fixed, $\Lambda(|\mathbf{n}|)$ as functions of $q$ is shown in Fig.~\ref{fig:measurement error}(a), from which we observe that we need $q\lesssim 0.3$ to be able to fix the errors, otherwise some of the $\Lambda$ must be negative to maintain stationarity.  

Note that from Fig.~\ref{fig:measurement error}(a), we have $\max_\mathbf{n}{\Lambda_\mathbf{n}} >1$ for nonzero $q$. Since the interpretation of $\Lambda$ is the probability of applying feedback, we can modify Eq.~\eqref{eq: example 2d} by replacing $\Lambda_\mathbf{n}$ with $\Lambda_\mathbf{n}/\max_\mathbf{n} \Lambda_\mathbf{n}$ and correspondingly rescale the measurement rate.

Similarly, when we correct incoherent Pauli errors with Eq.~\eqref{eq: correct nonunitary error}, if we include the measurement errors, we can follow the same analysis as Eq.~\eqref{eq: measurement error 3} to modify $\Lambda_\mathbf{n}$. If we simply require that $\Lambda_\mathbf{n}'$ to be the same as Eq.~\eqref{eq: correct nonunitary error}, we find that $\Lambda\left(|\mathbf{n}|>2\right)$ we need would be the same as $\Lambda\left(|\mathbf{n}|>2\right)$ in Fig.~\ref{fig:measurement error}(a). It seems that we again need $q\lesssim 0.3$ to fix the error for $\mu = \frac{1}{4}\log{2}$. However, since we can always add T-even dynamics \eqref{eqn:rep-code-diag-dynamics} without changing the stationary state, it turns out that we can choose some nonzero $\gamma_4$ and $\gamma_2$ to increase the threshold of $q$. One example is shown in Fig.~\ref{fig:measurement error}(b), from which we can see that the critical value of $q$ can be slightly greater than $0.3$.

\subsection{Biased quantum walks}
\label{subsec:Q break T}

The error correction scheme for the Hamiltonian errors breaks time-reversal symmetry, as one can explicitly check.  However, it is also of interest to break time-reversal symmetry in more ``intuitive ways'': for example, breaking T but preserving PT, where P is spatial inversion symmetry.   A simple motivating example already at the classical level is the addition of bias to the motion of a collection of particles (whose number is conserved) in one dimension: in this setting, the classical effective theory for particle density $\rho$ is modified from $\partial_t \rho = D\partial_x^2 \rho + \cdots $ to $\partial_t \rho = \partial_x \left(a \rho + b\rho^2 + \cdots + D\partial_x \rho + \cdots \right) $. Here the coefficients $a$ and $b$ break T, but preserve PT \cite{ACM}. To achieve this type of qualitative correction to the long wavelength dynamics in a quantum setting, we must consider  additional T-breaking modifications to the Lindblad operators of the previous subsection.   

In the discussion that follows, we focus on one-dimensional spin-1/2 chains of length $L$ satisfying periodic boundary conditions. The stationary state $\sigma = \ee^{-\Phi}$ with $\Phi = -\mu \sum_x Z_x Z_{x+1}$. We use configurations $\{s^{\,}_1, \dots, s^{\,}_{L}\}$ to represent eigenstates of $\sigma$, where $s^{\,}_x \in \lbrace \circ ,\bullet\rbrace$, which correspond to the eigenvalues $\{+1, -1\}$ of the stabilizer $Z_x Z_{x+1}$, respectively. Each stabilizer configuration corresponds to two spin configurations related by the Ising symmetry $\prod_x X_x$.  For simplicity, we project onto the subspace in which $\prod_x X_x=+1$ and consider dynamics that remains in this subspace. We remark that the only states in this ``Hilbert space'' are those where the total number of $\bullet$ is even.

\subsubsection{Classical T-odd dynamics}
We first consider effectively classical T-odd dynamics that protect our desired steady state.  Intuitively, one of the simplest T-odd classical dynamics is to transform states $|\cdots  \circ \bullet \cdots\rangle \to|\cdots  \bullet \circ \cdots\rangle$ in a translation-invariant way. Such dynamics will produce a biased drift of domain walls, and is exactly analogous to the classical biased random walk described in the introduction to this subsection.  The density matrix $\sigma$ is still stationary because for each specific stabilizer configuration $\{s^{\,}_1, \dots, s^{\,}_{L}\}$, the number of  $ \circ \bullet$ motifs is always the same as the number of $ \bullet \circ$ \cite{guo22}, so the rate of probability outflow from a particular configuration is always equal to the rate of probability inflow. In Appendix~\ref{sec:T-odd}, we show how to construct all translational invariant classical T-odd dynamics for 1D systems systematically.

The Lindbladian of such dynamics realized exclusively by measurements and feedback can be
\begin{align} \label{eq:classical T-odd L 2}
    \Liouvillian  = \gamma_{0} \sum_x\  &\mathcal{D}\left(\ket{ \bullet \circ_x}\bra{ \circ \bullet_x} \right) + \mathcal{D}\left(\ket{ \bullet \circ_x}\bra{ \bullet \circ_x} \right) + \mathcal{D}\left(\ket{\bullet \bullet_x}\bra{ \bullet \bullet_x} \right) 
    +
    \mathcal{D}\left(\ket{ \circ \circ_x}\bra{ \circ \circ_x} \right),
\end{align}
where 
\begin{subequations}
\begin{align}
\BKop{ \bullet \circ_{x}}{ \circ \bullet_{x}} &= \frac{1}{4} X_{x} \left(\ident + Z_{x-1} Z_{x}\right) \left(\ident - Z_{x} Z_{x+1}\right),  \\
\BKop{ \bullet \circ_{x}}{ \bullet \circ_{x}} &= \frac{1}{4}  \left(\ident - Z_{x-1} Z_{x}\right) \left(\ident + Z_{x} Z_{x+1}\right), \\
\BKop{ \circ \circ_{x}}{ \circ \circ_{x}} &= \frac{1}{4}  \left(\ident + Z_{x-1} Z_{x}\right) \left(\ident + Z_{x} Z_{x+1}\right), \\
\BKop{\bullet \bullet_x}{\bullet \bullet_x} &= \frac{1}{4}  \left(\ident - Z_{x-1} Z_{x}\right) \left(\ident - Z_{x} Z_{x+1}\right)
\, .
\end{align}
\end{subequations}
Note that this dynamics does protect quantum coherence between $\ket{000}$ and $\ket{111}$, since the measurements and feedback only detect and correct \emph{relative} bit flips between sites.  However, because the quantum dynamics looks strictly classical in the basis of stabilizer eigenstates, we refer to it as \emph{classical T-odd dynamics}. We identify any dynamics of stabilizer eigenvalues that cannot be mapped to classical Markov chains as quantum dynamics.

\subsubsection{Quantum T-odd dynamics}
\label{sec:quantumTodd}

Next, we 
consider quantum T-breaking dynamics that can produce a biased drift of domain walls across the system in a purely quantum manner; we also show that there is no classical model that captures the resulting drift of domain walls. For simplicity, we use $\ket{\alpha_x}$ to represent states $\ket{\cdots  \alpha \cdots }$, where the ``motif'' $\alpha$ is the local configuration at position $x$. The dynamics produced by local jump operators with coefficient $\gamma_{\alpha \beta}^{\alpha' \beta'}$ is represented by $\ket{\alpha}\bra{\beta} \to \BKop{\alpha'}{\beta'}$, where the order of indices of $\gamma$ is the same as Eq.~\eqref{eq:F} and the index of position is neglected for now. We say state $\BKop{\alpha}{\beta}$ and $\BKop{\alpha'}{\beta'}$ are coupled if both $\gamma_{\alpha \beta}^{\alpha' \beta'}$ and $\gamma_{\alpha' \beta'}^{\alpha \beta}$ are nonzero, which is represented by $\BKop{\alpha}{\beta} \rightleftharpoons \BKop{\alpha'}{\beta'}$. The rough picture of the dynamics is that terms of the form $\ket{ \circ \circ \bullet}\bra{ \circ \bullet \circ} \to \ket{ \circ \bullet \circ}\bra{ \bullet \circ \circ}$ can move domain walls to the left, thus producing the biased drift of domain walls. 

\begin{figure}[t]
\centering
\includegraphics[scale=0.52]{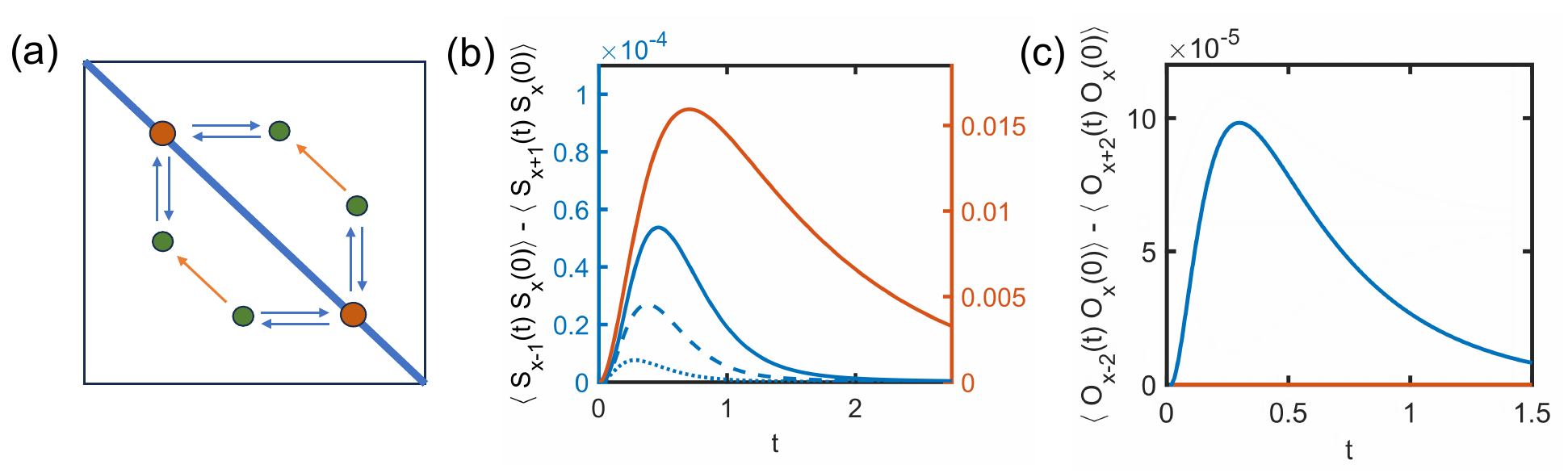}
\caption{ (a) The dynamics of the biased quantum walks in the density matrix: the blue arrows represent the coupling between diagonal elements with off-diagonal elements; the yellow arrows represent the T-breaking dynamics between off-diagonal elements. (b) Plot of the correlation function $\langle S_{x-1}(t) S_x(0)\rangle-\langle S_{x+1}(t) S_x(0)\rangle$ for biased quantum walks \eqref{eq:quantum flocking 1} (solid blue line), where $S_x = Z_x Z_{x+1}$ denotes the stabilizer at position $x$. We plot the same correlation function for the corresponding classical dynamics \eqref{eq:classical flocking 2} (orange line). We also plot the correlation function for dynamics with more phase damping \eqref{eq:quantum zeno}. The correlation functions for quantum dynamics with $\alpha = 2$ (dashed blue line) and $\alpha = 4$ (dotted blue line) are reduced in magnitude and the classical dynamics (orange line) are not affected. (c) We plot a different correlation function $\langle O_{x-2}(t) O_x(0)\rangle-\langle O_{x+2}(t) O_x(0)\rangle$ for the biased quantum walks \eqref{eq:quantum flocking 1} (blue line) and the corresponding classical dynamics \eqref{eq:classical flocking 2} (orange line), where $O_x = \BKop{ \bullet \circ_{x}}{\circ \bullet_{x}} +\BKop{\circ \bullet_{x} }{\bullet \circ_{x}} $. The function $\langle O(t)O(0)\rangle$ captures the drift in domain wall superposition. Therefore, it vanishes identically for the classical dynamics. }
\label{fig:qbrw}
\end{figure}

A schematic diagram of the dynamics is shown in Fig.~\ref{fig:qbrw}: we first couple diagonal states with some off-diagonal states $\ket{ \circ \circ \bullet}\bra{ \circ \bullet \circ} \rightleftharpoons \ket{ \circ \circ \bullet}\bra{ \circ \circ \bullet} \rightleftharpoons \ket{ \circ \bullet \circ}\bra{ \circ \circ \bullet}$ and $\ket{ \circ \bullet \circ}\bra{ \bullet \circ \circ} \rightleftharpoons \ket{ \bullet \circ \circ}\bra{ \bullet \circ \circ} \rightleftharpoons \ket{ \bullet \circ \circ}\bra{ \circ \bullet \circ}$. We then add the dynamics $\ket{ \circ \circ \bullet}\bra{ \circ \bullet \circ} \to \ket{ \circ \bullet \circ}\bra{ \bullet \circ \circ}$ and $\ket{ \circ \bullet \circ}\bra{ \circ \circ \bullet} \to \ket{ \bullet \circ \circ}\bra{ \circ \bullet \circ}$ so that the domain walls are moving to the left. We also need to add some additional jump operators (diagonal terms to $\gamma_{ij}$) to keep the dynamics completely positive. The result is translation-invariant dynamics, all of which is T even, with the crucial exception of the superposition drift. The transitions between the off-diagonal states effectively produce the classical dynamics $\ket{ \circ \circ \bullet}\bra{ \circ \circ \bullet} \to \ket{ \bullet \circ \circ}\bra{ \bullet \circ \circ}$ in a purely quantum way.

In order to keep $\sigma$ stationary, we can use the formalism developed in Appendix~\ref{sec:T-odd} to see that $\gamma_{\beta \beta}^{\alpha\alpha} = \gamma_{\alpha\alpha}^{\beta\beta}$ and $\gamma_{\alpha \alpha}^{\alpha\beta} = \gamma_{\beta\alpha }^{\alpha\alpha}$ should hold for any $\alpha$ and $\beta$. One example is: 
\begin{align} \label{eq:quantum flocking 1}
\mathcal{L}^\text{Q}(\rho)  =  \sum_x  \mathcal{L}_x^{\mathrm{Q}}(\rho) 
  &= \sum_x \ \mathcal{D}\left[\ket{ \circ \circ \bullet_x}\left(\bra{ \circ \circ \bullet_x}+\bra{ \circ \bullet \circ_x}\right)\right](\rho) + \mathcal{D}\left[\left(\ket{ \circ \circ \bullet_x}+\ket{ \circ \bullet \circ_x}\right)\bra{ \circ \circ \bullet_x}\right](\rho) \notag \\
&+\mathcal{D}\left[\ket{ \bullet \circ \circ_x}\left(\bra{ \bullet \circ \circ_x}+\bra{ \circ \bullet \circ_x}\right)\right](\rho) +\mathcal{D}\left[\left(\ket{ \bullet \circ \circ_x}+\ket{ \circ \bullet \circ_x}\right)\bra{ \bullet \circ \circ_x}\right] (\rho) \notag \\\
    &+\mathcal{D}\left[\ket{ \bullet \circ \circ_x}\bra{ \bullet \circ \circ_x}\right](\rho) +\mathcal{D}\left[\ket{ \circ \circ \bullet_x}\bra{ \circ \circ \bullet_x}\right](\rho) \notag \\
  &+\frac{1}{2}\, \left(\ket{ \circ \bullet \circ_x}\bra{ \circ \circ \bullet_x}\,\rho\,\ket{ \circ \bullet \circ_x}\bra{ \bullet \circ \circ_x} + \ket{ \bullet \circ \circ_x}\bra{ \circ \bullet \circ_x}\,\rho\,\ket{ \circ \circ \bullet_x}\bra{ \circ \bullet \circ_x}\right) ,
\end{align}
where we can express the domain-wall basis states above via
\begin{subequations}
\begin{align}
\BKop{ \circ \bullet \circ_{x}}{ \circ \circ \bullet_{x}} &= \frac{1}{8} X_{x+2} \left(\ident + Z_x Z_{x+1}\right) \left(\ident + Z_{x+1} Z_{x+2}\right) \left(\ident - Z_{x+2} Z_{x+3}\right) \\
\BKop{ \bullet \circ \circ_{x}}{ \circ \bullet \circ_{x}} &= \frac{1}{8} X_{x+1} \left(\ident + Z_x Z_{x+1}\right) \left(\ident - Z_{x+1}Z_{x+2}\right) \left(\ident + Z_{x+2} Z_{x+3}\right) \\
\BKop{ \circ \circ \bullet_{x}}{ \circ \circ \bullet_{x}} &= \frac{1}{8} \left(\ident + Z_x Z_{x+1}\right) \left(\ident + Z_{x+1} Z_{x+2}\right) \left(\ident - Z_{x+2} Z_{x+3}\right) \\
\BKop{ \bullet \circ \circ_{x}}{ \bullet \circ \circ_{x}} &= \frac{1}{8} \left(\ident - Z_x Z_{x+1}\right) \left(\ident + Z_{x+1}Z_{x+2}\right) \left(\ident + Z_{x+2} Z_{x+3}\right) \, .
\end{align}
\end{subequations}
Note that, in Eq.~\eqref{eq:quantum flocking 1}, the first two lines produce the couplings between diagonal states and off-diagonal states, while the last line produces the dynamics $\ket{ \circ \circ \bullet}\bra{ \circ \bullet \circ} \to \ket{ \circ \bullet \circ}\bra{ \bullet \circ \circ}$ and $\ket{ \circ \bullet \circ}\bra{ \circ \circ \bullet} \to \ket{ \bullet \circ \circ}\bra{ \circ \bullet \circ}$, which is the only T-breaking part. 

The effects of the quantum biased drift are shown in Fig.~\ref{fig:qbrw} (solid blue line): we numerically simulated this model in small one-dimensional systems with $\mu = \frac{1}{4}\log{2}$, and calculated correlation functions $\langle S_{x-1}(t) S_x(0)\rangle-\langle S_{x+1}(t) S_x(0)\rangle$ and $\langle O_{x-2}(t) O_x(0)\rangle-\langle O_{x+2}(t) O_x(0)\rangle$ for the dynamics~\eqref{eq:quantum flocking 1}, where $S_x = Z_x Z_{x+1}$ is the stabilizer, $O_x=\BKop{ \bullet \circ_{x}}{\circ \bullet_{x}} +\BKop{\circ \bullet_{x} }{\bullet \circ_{x}} $ and $\langle A(t) B(0) \rangle \equiv \tr\left[\ee^{\Liouvillian^\dagger t} (A)  B \sigma \right]$.\footnote{Note that both operators $S_x$ and $O_x$ commute with $\sigma$, so subtleties about the precise time-dependent correlation function of interest are unimportant here.} Both correlation functions are nonzero for the quantum dynamics. 

We now argue that the dynamical system \eqref{eq:quantum flocking 1} is intrinsically quantum: namely, the T-odd part of this dynamics cannot be mapped to an effective classical dynamics even if we change the basis of the system.\footnote{Since $\Phi$ is proportional to the number of domain walls in the system, it is hugely degenerate; thus, basis changes that preserve $\sigma$ do exist.} We can show, at a minimum, that if such a basis change exists, the basis change is nonlocal. This is because the T-odd Lindbladian~\eqref{eq:quantum flocking 1} \emph{locally} protects $\sigma$: dynamics at any position $x$.
Translation invariance is not needed for stationarity; indeed, \begin{equation}
     \mathcal{L}_x^{\mathrm{Q}} (\sigma) = 0
 \end{equation} for \emph{arbitrary} position $x$.  For classical dynamics, T-odd terms cannot be added locally in this manner while preserving stationarity, since T-odd classical dynamics only respect stationarity when they lead to biased flows around closed cycles in state space.  More physically, if domain walls drift to the right only locally in region $R$, then they must ``pile up'' on the right-hand side of region $R$, in contradiction with the assertion that the same translation-invariant $\sigma$ exists for the locally modified chain. Therefore, if there exists a change of basis that would make the dynamics classical, $\mathcal{L}$ cannot be transformed to any local dynamics, and the classical dynamics would be highly nonlocal.

The difference between quantum and classical dynamics is also reflected in certain correlation functions. In Fig.~\ref{fig:qbrw}, we depict the correlation function for classical dynamics (orange line) generated by
\begin{equation} \label{eq:classical flocking 2}
\mathcal{L}^{\text{c}}(\rho)  = \sum_x \ \mathcal{D}\left[\ket{\bullet \circ \circ_x}\bra{ \circ \circ \bullet_x}\right] (\rho) + \mathcal{D}\left[\ket{ \bullet \bullet \circ_x}\bra{ \circ \bullet \bullet_x}\right](\rho) \  , 
\end{equation}
which also has the effect of moving domain walls to the left. The last term of the classical dynamics is needed to keep $\sigma$ stationary. From Fig.~\ref{fig:qbrw}(b), the drifting for classical dynamics is much stronger than the quantum dynamics. However, in Fig.~\ref{fig:qbrw}(c), as a correlation function that captures quantum effect, the correlation function $\langle O_{x-2}(t)O_x(0) - O_{x+2}(t)O_x(0)\rangle$, 
which captures the drift in domain wall superpositions, and hence vanishes identically for the classical dynamics. This constitutes additional evidence that the dynamics \eqref{eq:quantum flocking 1} is intrinsically quantum.

We can also demonstrate a quantum Zeno effect \cite{zeno}, in which rapidly increasing the rate of measurement freezes out the bias in the quantum random-walk dynamics.  This is an intrinsically quantum phenomenon, and thus only exists for the quantum dynamics \eqref{eq:quantum flocking 1}.  We model this Zeno effect by increasing the coefficients of terms in the third line of 
Eq.~\eqref{eq:quantum flocking 1} by introducing the parameter $\alpha$, leading to
\begin{align} \label{eq:quantum zeno}
\mathcal{L}  \to \mathcal{L} + \alpha \left(\mathcal{D}\left[\ket{ \bullet \circ \circ_x}\bra{ \bullet \circ \circ_x}\right] +\mathcal{D}\left[\ket{ \circ \circ \bullet_x}\bra{ \circ \circ \bullet_x}\right] \right)
\, .
\end{align}
For the quantum dynamics, since the biased motion only comes from the last line of Eq.~\eqref{eq:quantum flocking 1}, which only couples off-diagonal terms, the motion of domain walls can be reduced in magnitude by the additional phase damping caused by Eq.~\eqref{eq:quantum zeno}. In Fig.~\ref{fig:qbrw}(b), we show the results of numerical simulation of Eq.~\eqref{eq:quantum zeno} with $\alpha = 2$ and $\alpha=4$, and calculate the correlation function $\langle S_{x-1}(t) S_x(0)\rangle-\langle S_{x+1}(t) S_x(0)\rangle$ for both classical (orange line) and quantum dynamics (dashed blue line). The correlation function for quantum dynamics decreases with increasing $\alpha$, while for classical dynamics, the correlation function remains the same. This is further evidence that the $\alpha$-dependent term serves to suppress a fundamentally quantum-mechanical, T-odd effect.

\section{Conclusion}
We have described a systematic classification of the possible open quantum systems that protect a target stationary state $\sigma = \mathrm{e}^{-\Phi}$.  When $\Phi$ can be expressed as a sum over commuting operators, we can further classify all possible \emph{few-body} (i.e., local) Lindbladians that protect a many-body state $\sigma$, up to a handful of exceptions related to the classification of classical Markov chains with known steady states~\cite{ClassicalMarkovChain}.

At the most mathematical level, our construction constitutes a significant generalization of the Davies' Lindbladian~\cite{Davies1974_I, Davies1976_II}, which drives a quantum system towards a Gibbs state with $\Phi=\beta H$.  Unlike Davies' Lindbladian, however, our protocols do \emph{not} require that the generator $H$ of unitary time evolution be the same as the generator $\Phi$ of the stationary state. This allows for the description and characterization of uniquely quantum mechanical phenomena, e.g., in which the ``drive'' is counteracted by measurement and feedback.   It is quite likely that, as in classical systems \cite{Bernard2009,Suwa2010}, the non-Davies quantum Lindbladians discussed herein have a larger spectral gap, and thus prepare the desired state more quickly -- an idea utilized in several recent works~\cite{Chen:2023zpu, Rouze:2024ufx,Ding:2024mxo}.

At a slightly more practical level, our framework is naturally suited towards the problem of designing passive error-correcting quantum codes; preliminary work along these lines appears in Ref.~\citenum{Hong:2024vlr}.  As we have illustrated at length, our framework can describe the feedback schemes needed to stabilize a target state in the presence of generic errors.  It is therefore possible to design targeted error-correcting protocols that are tuned towards hardware-specific error models and rates, including the highly biased noise \cite{Puri_2020} that characterizes superconducting qubits, among other platforms.  Using this framework, it may be possible to optimize more systematically over the space of possible code modifications to protect against biased error models, following Ref.~\citenum{Bonilla_Ataides_2021}. Our approach naturally handles coherent noise as well, and may provide more optimized error-correcting protocols than a general-purpose decoder.  Such coherent noise is possible in both transmon qubits \cite{abdo22} as well as neutral-atom qubits \cite{deist, Singh_2023}.

At a more physical level, the methods we present herein represent a promising route towards engineering experimentally detectable measurement-induced phase transitions, by designing particular feedback schemes that precisely compensate for known Hamiltonian ``errors.'' We present one such example in Sec.~\ref{sec:examples}, though we suspect that many more exist. Moreover, the formalism we develop may aid the systematic classification of universal quantum dynamics that can realize in open systems, and particularly, the possible phases of ``active'' quantum systems. For example, Ref.~\citenum{FisherSymmetryBreaking} argued that using measurement and feedback, it was possible to break a continuous symmetry \emph{spontaneously} in one dimension, but the protocol utilized showed extreme sensitivity to noise.  Intuitively, that protocol can be understood in our framework as the $\beta \to \infty$ limit of dynamics that protects the steady state
\begin{equation}
    \Phi \sim -\beta \sum_i \mathbf{S}_i \cdot \mathbf{S}_{i+1}.
\end{equation}
Since this Heisenberg model does not have long-range order in one spatial dimension, the conclusions of Ref.~\citenum{FisherSymmetryBreaking} are consistent with our framework's expectations, in which adding a small amount of noise (taking $\beta$ finite) destabilizes the long-range order.

At the same time, we stress that, while our methods are quite powerful at preparing desired quantum states, they are designed for relatively simple generators $\Phi$ of the stationary state. However, certain simple models of local  dynamics -- both classical and quantum -- are believed to have quite complicated generators $\Phi$ \cite{HiddenTRS,bertini_rmp}.  Hence, we cannot rule out the possibility that some quantum dynamical universality classes lie beyond the purview of our methods, precisely because they admit local Lindbladians with highly nonlocal $\Phi$. It remains an interesting open problem to determine whether such models nonetheless admit some reasonable notion of a stable ``phase of matter'' \cite{Rakovszky:2023btf}, and whether or not models with nonlocal $\Phi$ realizing novel universality classes can be described in our framework using a new set of degrees of freedom.

Lastly, we note that when $\Phi=\beta H$, a very powerful effective field theory based on the Schwinger-Keldysh path integral \cite{haehl2016fluid,eft1,eft2,jensen2018dissipative}  has been developed for studying \emph{quantum} dissipative effective field theories.  It would be fascinating if these methods can be generalized to the nonthermal dynamics described herein.  Recent progress along these lines was described in Ref.~\citenum{Huang:2024rml}.   One possible difficulty in achieving this goal is that, in any saddle point limit where such path integrals can be readily analyzed, nonthermal dynamics may already be largely, if not entirely, captured by the \emph{classical} effective field theory developed in Ref.~\citenum{ACM}.  Such classical methods seem unlikely to capture all of the quantum coherent phenomena described in this work. 

\section*{Acknowledgements}
We thank Marvin Qi and Charles Stahl for many helpful discussions. This work was supported by the Alfred P. Sloan Foundation under Grant FG-2020-13795 (AL), the Gordon and Betty Moore Foundation under Grant GBMF10279 (JG, AL), the Department of Energy under Quantum Pathfinder Grant DE-SC0024324 (AJF, AL), and the Air Force Office of Scientific Research under Grant FA9550-20-1-0222 (OH) and Grant FA9550-24-1-0120 (JG, AL).

\appendix

\section{Generating T-odd \texorpdfstring{${\gamma_{ab}}$}{} with \texorpdfstring{${c_a = c_b}$}{}} \label{sec:T-odd}

Here we show how to generate T-odd dynamics of ${\gamma_{ab}}$ with ${c_a = c_b}$. For a general open quantum system, one can choose a basis $\set{\ket{a}}$ that diagonalizes the stationary state ${\sigma = \sum_a \sigma_a \,\BKop{a}{a}}$. Using jump operators ${F_{ab} = \BKop{a}{b}}$, the Lindbladian can be written as 
\begin{align} \label{L_B1}
    \Liouvillian(\rho) = &- \sum_{ab} \ii\, h_{ab} \comm{F_{ab} }{\rho} + \sum_{abcd}  \gamma_{bd}^{ac} \left( F^{\vpd}_{ab} \rho F_{cd}^{\dagger}  - \frac{1}{2}\acomm{F_{cd}^{\dagger} F^{\vpd}_{ab}}{\rho} \right) .
\end{align}
In this language, the constraints imposed by stationarity can be separated into two parts,
\begin{subequations}\label{eq:128}
\begin{align}
    \matel{a}{\Liouvillian[\sigma]}{a} &= \sum_c \left( \sigma_c \gamma_{cc}^{aa} -\sigma_a \gamma_{aa}^{cc} \right) = 0, \label{cons_classic}\\
    \matel{a}{\Liouvillian[\sigma]}{b} &= -\ii (\sigma_b - \sigma_a) h_{ab} + \sum_c \left[\sigma_c \gamma_{cc}^{ab} - \frac{1}{2} (\sigma_a + \sigma_b) \gamma_{ba}^{cc} \right] = 0, \; a\neq b \label{cons_quantum}.
\end{align}
\end{subequations}
In Eq.~\eqref{cons_classic}, the diagonal part of the ${\gamma}$ matrix, ${\gamma_{cc}^{aa}}$ represents the transition rate of ${\BKop{c}{c} \to \BKop{a}{a}}$, which include all the dynamics between the diagonal part of ${\sigma}$. This part of dynamics can be mapped to classical Markov chains, so we refer to them as the ``classical part'' of dynamics. The off-diagonal part of the ${\gamma}$ matrix that appears in Eq.~\eqref{cons_quantum}, ${\gamma_{cc}^{ab}}$ and ${\gamma_{ba}^{cc}}$ can be roughly understood as the transition rate of ${\BKop{c}{c} \to \BKop{a}{b}}$ and ${\BKop{b}{a} \to \BKop{c}{c}}$, which we refer to as the ``quantum part'' of dynamics. Note that the coefficients $\gamma^{ab}_{cd}$, where all the indices are different, are also in the quantum part. These coefficients are not constrained by the stationarity of $\sigma$. The only constraint for them is the positivity of $\gamma$.

The constraints of the classical dynamics~\eqref{cons_classic} seem hard to solve in complete generality.  Each variable ${\gamma_{cc}^{aa}}$ appears in two different equations that cannot be neatly decoupled from the rest. It can formally be solved by finding all cycles in the state space and attempting to add nonzero transition rates that cause the system to flow around each cycle in a biased way.  However, such biased ``random walks'' are directly in the many-body state space, so they do not generically correspond to local Lindbladians. However, as we will show shortly, we know how to generate one-dimensional, translation-invariant, local classical dynamics for systems with $\sigma = \ee^{-\Phi}$ in the form of Eq.~\eqref{eqn: Phi} \cite{ClassicalMarkovChain}. 

The quantum part of the constraints is easier: each variable only appears once. For many-body systems, even with the constraints of locality, we only need to solve a finite number of equations for each variable, so a complete solution can be found, including when restricting to local dynamics. 

In summary, therefore, we will systematically show in this appendix how to classify all one-dimensional, translation-invariant classical T-odd dynamics, and all quantum dynamics.

We now explicitly show how to generate these dynamics, starting with a simple example before moving on to the general case. Consider a 1D spin-1/2 chain with length $L$ and periodic boundary conditions (PBCs) -- i.e.,  $x+L \cong x$ for any site $x$. We take the stationary state  $\sigma \propto \exp\left(\,\mu \sum_x Z_x\right)$, and use $\ket{s}$ to represent a configuration $\{s^{\,}_1, \dots, s^{\,}_{L}\}$, where $s^{\,}_x \in \lbrace-1, 1\rbrace$ is the eigenvalue of the stabilizer $Z_x$. $\sigma = \sigma_{ss'}\ket{s}\bra{s'}$ is diagonal with the basis $\set{\ket{s}}$, so (as above) the transitions between density matrices $|s\rangle \langle s|$ and $|s^\prime\rangle\langle s^\prime|$ is defined to be the classical dynamics. Since we want local dynamics, we only consider $q$-body spatially local jump operators. Namely, we consider transitions that take $|s\rangle \langle s|$ to $|s^\prime\rangle\langle s^\prime|$ when $s$ and $s^\prime$ differ on at most $q$ adjacent sites -- e.g., $s = s_{1\cdots x-1} \otimes \alpha_{x \cdots x+q-1} \otimes s_{x+q\cdots L}$ and $s^\prime = s_{1\cdots x-1} \otimes \alpha^\prime_{x \cdots x+q-1} \otimes s_{x+q\cdots L}$, where $\alpha$, $\alpha'$ denote two ``motifs'' (of length $\le q$ sites) by which the microstates differ. Similar to Eq.~\eqref{eqn:stabilizer-eigenvector}, we choose the jump operators to be 
\begin{equation} \label{eq:jump operators}
    A_{\beta\alpha,x} = \ident_{1\cdots x-1}\otimes |\beta\rangle\langle \alpha|_{x\cdots x+q-1} \otimes \ident_{x+q\cdots L}.
\end{equation}
where $\alpha$ and $\beta$ denote motifs of $q$ adjacent stabilizer eigenvalues in the local 1D chain. The $\alpha\beta$ notation will prove more convenient than that of Eq.~\eqref{eqn:stabilizer-eigenvector}.  We define 
\begin{equation}
    c_{\beta\alpha,x} = \exp\left[\mu \sum_{y=x}^{x+q-1} \frac{\beta_y-\alpha_y}{2}\right]
\end{equation}
for use in what follows.
With these jump operators, a general $q$-local Lindbladian can be written as

\begin{align} \label{L_B2}
    \Liouvillian(\rho) = &- \ii \comm{H}{\rho} + \sum_{\alpha \beta \alpha' \beta',x}  \gamma_{ \beta \beta',x}^{\alpha \alpha' } \left( \ket{\alpha^x}\bra{\beta^x} \,\rho\, \ket{\beta'^x}\bra{\alpha'^x}  - \frac{1}{2}\acomm{ \langle \alpha'^x| \alpha^x\rangle \, \ket{\beta'^x}\bra{\beta^x}}{\rho} \right) .
\end{align}

\noindent We now discuss the ``classical part'' of the dynamics, which was defined to depend only on the coefficients $\gamma_{\beta\beta,x}^{\alpha\alpha}$, and realize maps on the space of diagonal density matrices. There can also be ``quantum dynamics,'' which realize nontrivial maps on the \emph{off-diagonal} components of the density matrix -- e.g., terms in $\rho$ of the form $|s\rangle\langle r|$ are endowed with dynamics, when $r = r_{1\cdots x-1} \otimes \alpha_{x \cdots x+q-1} \otimes r_{x+q\cdots L}$ and $s$ share the motif $\alpha$, but are otherwise distinct. Because of locality, the classical dynamics is always  accompanied by some quantum dynamics. However, since these quantum dynamics does not affect the stationarity criterion, we do not discuss it further, and focus on the truly classical dynamics, which is constrained by stationarity.

For convenience, we assume translation invariance, so that $\gamma_{\beta\beta,x}^{\alpha\alpha}=\gamma_{\beta\beta}^{\alpha\alpha}$ is independent of the position $x$. The utility of this assumption will soon become clear. Then the stationarity of ${\sigma}$ requires that, for each configuration $\ket{s}$,
\begin{align} \label{eq: dss}
    \frac{\text{d} \sigma_{ss}}{\text{d} t}  &= \sigma_{ss} \left( \sum_{\beta\alpha } N_{\beta}^s \, c_{\beta\alpha}^2 \gamma^{\beta\beta}_{\alpha\alpha} - \sum_{\beta\alpha} N_{\beta}^s \gamma_{\beta\beta}^{\alpha\alpha}\right) =  \sigma_{ss} \sum_{\alpha} N_{\beta}^s f_\beta = 0,
\end{align}
where $N_\beta^s$ is the number of motifs $\beta$ in the microstate $s$. In the second equation above, the first term denotes the number of classical configurations entering $|s\rangle\langle s|$, and the second term denotes the rate at which the system decays out of $|s\rangle\langle s|$. The function $f$ is defined as
\begin{align} \label{eq:f and gamma}
f_\beta = \sum_{\alpha } \left(c_{\beta\alpha}^2 \gamma^{\beta\beta}_{\alpha\alpha} - \gamma_{\beta\beta}^{\alpha\alpha}\right),
\end{align}
and depends only on the motif $\beta$.  By identifying all possible functions ${f}$ that satisfy
\begin{equation}
    \sum_{\alpha} N_{\alpha}^s f_\alpha = 0  \label{eq:fconstraint}
\end{equation}
for all configurations $s$, we can then solve for all possible $\gamma_{\alpha\alpha}^{\beta\beta}$ out of the function $f$, which include all classical dynamics.

One can prove \cite{ClassicalMarkovChain} that all solutions $f$ to Eq.~\eqref{eq:fconstraint} can be written as 
\begin{align} \label{f function}
    f_\alpha = g \left( \alpha_1, \dots, \alpha_{q-1} \right) - g \left( \alpha_{2}, \dots, \alpha_{q} \right) \, ,~~
\end{align}
where $g$ is an arbitrary function and ${\{\alpha_1,\cdots, \alpha_q\}}$ is the motif $\alpha$, and that there are $2^{q-1}-1$ nontrivial linearly independent choices of function $g$.  The latter claim follows transparently from counting all possible linearly independent functions on $\mathbb{Z}_2^{q-1}$ (noting that $g=1$ does not contribute to $f_\alpha$); the former claim follows, in part, from the observation that the linear relation \eqref{eq:fconstraint} fails in generality if $f_\alpha$ has any term proportional to $\alpha_1\alpha_q$. From Eq.~\eqref{f function}, we can get all possible $f$ functions and all possible classical dynamics. Note that this part of the dynamics can always be generated by measuring stabilizers and applying feedback, possibly after adding additional dissipative terms to the Lindbladian (that protect $\sigma$).

These dynamics can be generalized to systems with stationary states $\sigma \propto \exp\left( \sum_x \mu_x Z_x\right)$. In Eq.~\eqref{eq:f and gamma}, if we take into account the position of the coefficients $\gamma_{P\Bar{\alpha},x}^{P\Bar{\alpha}}$ and $\gamma_{P\alpha,x}^{P\alpha}$, we find that coefficients with different $x$ don't appear in the same equation. Therefore, we can keep function $f$ unchanged and demand that for every $x$,
\begin{align} 
f_\beta = \sum_{\alpha } \left(c_{\beta\alpha,x}^2 \gamma^{\beta\beta}_{\alpha\alpha,x} - \gamma_{\beta\beta,x}^{\alpha\alpha}\right).
\end{align}

\noindent Unfortunately, we have found it very challenging to generalize this method beyond one-dimensional lattices.  We remark, however, that the above method easily generalizes to models where the stabilizers act on multiple sites, as in Sec.~\ref{sec:quantumTodd}.

Now we discuss the quantum part of the constraints. Here, we will not need to assume that some aspect of the dynamics is translation invariant. Still, for simplicity, we will mainly focus on $\sigma \propto \exp\left(\mu \sum_x Z_x\right)$, as the generalization is direct. The constraints of quantum dynamics come from Eq.~\eqref{cons_quantum}, where we can regard $\ket{a}$ and $\ket{b}$ as specific configurations.  From Eq.~\eqref{eqn:stationary-general}, we can always find a Hamiltonian to counteract the effects of $\gamma_{ab}$ whenever $c_a \ne c_b$, and so the classification problem becomes trivial: $\langle a|H|b\rangle $ is whatever is needed to obey Eq.~\eqref{eqn:stationary-general}.  Notice that as discussed in the main text, locality will be respected. Hence, we only need to focus on the case where $c_a=c_b$. With $q$-local dynamics, only when $\ket{a}$ and $\ket{b}$ differ by up to $q$ adjacent sites,
\begin{subequations}\label{eq:microstatesA}
    \begin{align}
        |a\rangle &= |s\rangle_{1\cdots x-1}\otimes |\alpha\rangle_{x\cdots x+q-1-n}\otimes |s^\prime\rangle_{x+q-n\cdots L} \\
        |b\rangle &= |s\rangle_{1\cdots x-1}\otimes |\beta\rangle_{x\cdots x+q-1-n}\otimes |s^\prime\rangle_{x+q-n\cdots L} ,
    \end{align}
\end{subequations}
can the coefficients on the right-hand side of Eq.~\eqref{cons_quantum} can be nonzero.  Notice that we can take $0\le n \le q-1$.   For each pair of $n$, $x$, $\alpha$ and $\beta$, we obtain a decoupled set of $4^n$ equations to solve (this counting assumes the stabilizers take values $\pm 1$).   For the simplest case of $n=0$, we get a single equation to solve,
\begin{align} \label{eqn:cons quantum 1}
    \frac{\text{d} \sigma_{ab}}{\text{d} t} &= \sum_s \left(\sigma_{ss} \gamma_{ss}^{ab} - \frac{1}{2} (\sigma_{aa} + \sigma_{bb}) \gamma_{ba}^{ss} \right) = \sigma_{aa} \sum_{\gamma} \left( 
c_{\gamma\alpha}^2 \, \gamma^{\alpha\beta}_{\gamma\gamma,x} - \gamma^{\gamma\gamma}_{\beta\alpha,x} \right) = 0.
\end{align}
Note that $\sigma_{aa} = \sigma_{bb}$, because we only discuss dynamics of $\gamma_{ab}$ with $c_a = c_b$. Due to spatial locality, this constraint does not depend on $|s\rangle$ or $|s^\prime\rangle$ in Eq.~\eqref{eq:microstatesA}. Lastly, there is no other equation that constrains $\gamma^{\alpha\beta}_{\gamma\gamma,x}$ or $\gamma^{\gamma\gamma}_{\beta\alpha,x}$: the sole constraint on these matrix elements is Eq.~\eqref{eqn:cons quantum 1}.

If $n>0$, things are a little more complicated: the $q$-site jump operator can be of the form $A_{\gamma\delta, y}$ for $x-n \le y \le x$.  We can solve for the resulting constraints on $\gamma$s as follows. Consider the states
\begin{subequations}
    \begin{align}
        |\tilde a(r,r^\prime)\rangle &= |s\rangle_{1\cdots x-1-n}\otimes |r\rangle_{x-n\cdots x-1}\otimes |\alpha\rangle_{x\cdots x+q-1-n}\otimes |r^\prime\rangle_{x+q-n\cdots x+q-1} \otimes |s^\prime\rangle_{x+q\cdots L} \\
        |\tilde b(r,r^\prime)\rangle &= |s\rangle_{1\cdots x-1-n}\otimes |r\rangle_{x-n\cdots x-1}\otimes   |\beta\rangle_{x\cdots x+q-1-n}\otimes |r^\prime\rangle_{x+q-n\cdots x+q-1} \otimes |s^\prime\rangle_{x+q\cdots L} ,
    \end{align}
\end{subequations}
where $|r\rangle$ and $|r^\prime\rangle$ are again identical between $|\tilde a\rangle$ and $|\tilde b\rangle$.  We must now consider the $4^n$ equations that arise from evaluating $\langle \tilde a(r,r^\prime)|\mathcal{L}[\sigma]|\tilde b(r,r^\prime)\rangle$.  There is not an elegant notation to express the general form of these constraints, but we can illustrate their form with a simple example that straightforwardly generalizes.  Consider the simplest nontrivial case of $q=2$ and $n=1$, where $|\alpha\rangle = |0\rangle$ and $|\beta\rangle = |1\rangle$, and $x=2$. By analogy to Eq.~\eqref{eqn:cons quantum 1}, we find a set of four constraints,
\begin{equation}\label{eqn:cons quantum}
    \sum_{k,k^\prime = 0,1}\left[ c_{(kk^\prime)(r0),1}^2 \gamma_{(kk^\prime)(kk^\prime),1}^{(r0)(r1)} - \gamma^{(kk^\prime)(kk^\prime)}_{(r1)(r0),1}  + c_{(kk^\prime)(0r^\prime),2}^2 \gamma_{(kk^\prime)(kk^\prime),2}^{(0r^\prime)(1r^\prime)} - \gamma^{(kk^\prime)(kk^\prime)}_{(1r^\prime)(0r^\prime),2}  \right] = 0
\end{equation}
for $r,r^\prime = 0,1$.  Notice that these equations are not all independent since, e.g., the last term in the above equation is independent of the value of $r$.  It is tedious, but straightforward, to generalize this construction to general $n$ and $q$.

For systems with stabilizers that introduce degeneracy, so long as the operators that transition between such states (e.g., logical operators in an error-correcting code) are nonlocal, the existence of such degeneracy does not modify the discussion above. Therefore, for systems with stationary states $\sigma = \exp\left(-\Phi\right)$ in the form of Eq.~\eqref{eqn: Phi}, we can get all possible quantum dynamics of $\gamma_{ab}$ with $c_a = c_b$ by solving the generalizations of Eq.~\eqref{eqn:cons quantum 1} and Eq.~\eqref{eqn:cons quantum}. In contrast, for the dynamics of $\gamma_{ab}$ with $c_a = c_b$, we can only classify 1D classical T-odd dynamics with translation invariance.

\renewcommand{\thesection}{}
\bibliographystyle{unsrtnatcustom.bst}
\bibliography{aqm}

\end{document}

%% file: preamble.tex
\usepackage{amsmath,amssymb,amsfonts,mathtools,dsfont,relsize}
\usepackage{microtype} 
\usepackage{graphicx}

\usepackage{amsthm}

\usepackage{enumerate}
\usepackage{suffix} 
\usepackage{comment}
\usepackage{xpatch}
\makeatletter
\patchcmd{\@ssect@ltx}
    {\addcontentsline{toc}{#1}{\protect\numberline{}#8}}
    {}
    {}
    {}
\makeatother

\makeatletter
\g@addto@macro\bfseries{\boldmath}
\makeatother

\newcommand{\vpd}[0]{\vphantom{\dagger}}
\newcommand{\vps}[0]{\vphantom{*}}
\newcommand{\vpp}[0]{\vphantom{\prime}}

\newcommand{\Liouvillian}[0]{\mathcal{L}}

\makeatletter
\newcommand{\notpropto}{\propto\kern-1\@ptsize pt \diagup}
\makeatother

\DeclareMathOperator{\sgn}{sgn}
\DeclareMathOperator{\tr}{tr}

\DeclareMathOperator{\spec}{spec}

\DeclarePairedDelimiter{\set}{\lbrace}{\rbrace}

\DeclarePairedDelimiter{\abs}{\lvert}{\rvert}
\DeclarePairedDelimiter{\expval}{\langle}{\rangle}

\DeclarePairedDelimiter{\ket}{\lvert}{\rangle}
\DeclarePairedDelimiter{\bra}{\langle}{\rvert}
\DeclarePairedDelimiter{\braket}{\langle}{\rangle}

\newcommand{\betterbe}[0]{\overset{!}{=}}
\newcommand{\ii}[0]{\mathrm{i}}
\newcommand{\ee}[0]{\mathrm{e}}
\newcommand{\bvec}[1]{\boldsymbol{#1}}
\newcommand{\pd}[1]{\partial^{\vpd}_{#1}}\WithSuffix\newcommand\pd*[1]{\partial_{#1}}

\newcommand{\thed}[0]{\mathrm{d}}

\newcommand{\kron}[1]{\delta_{#1}}

\newcommand{\ident}[0]{\mathds{1}}

\newcommand{\Comps}{\mathbb{C}}
\newcommand{\Ints}{\mathbb{Z}}

\newcommand{\SU}[1]{\mathsf{SU} ( #1 )}

\newcommand{\BKop}[2]{\left| #1 \middle\rangle \hspace{-0.4mm} \middle\langle #2 \right|}\WithSuffix\newcommand\BKop*[2]{| #1 \rangle \hspace{-0.4mm} \langle #2 |}
\newcommand{\BKbkop}[4]{\left| #1 \middle\rangle \hspace{-0.4mm}  \middle\langle #2 \middle| #3 \middle\rangle \hspace{-0.4mm} \middle\langle #4 \right|}\WithSuffix\newcommand\BKbkop*[4]{| #1 \rangle \hspace{-0.4mm} \langle #2 | #3 \rangle \hspace{-0.4mm} \langle #4 |}

\newcommand{\inprod}[2]{ \left\langle #1 \middle| #2 \right\rangle}\WithSuffix\newcommand\inprod*[2]{\langle #1 \middle| #2 \rangle}

\newcommand{\matel}[3]{\left\langle #1 \middle| #2 \middle| #3 \right\rangle}
\WithSuffix\newcommand\matel*[3]{\langle #1 | #2 | #3 \rangle}

\newcommand{\comm}[2]{\left[ #1, \, #2 \right]}\WithSuffix\newcommand\comm*[2]{[ #1 , \, #2 ]}
\newcommand{\acomm}[2]{\left\{ #1, \, #2 \right\} }\WithSuffix\newcommand\acomm*[2]{\{ #1 , \, #2 \}}

\newcommand{\PX}[1]{X^{\vps}_{#1}}\WithSuffix\newcommand\PX*[1]{X_{#1}}
\newcommand{\PY}[1]{Y^{\vps}_{#1}}\WithSuffix\newcommand\PY*[1]{Y_{#1}}
\newcommand{\PZ}[1]{Z^{\vps}_{#1}}\WithSuffix\newcommand\PZ*[1]{Z_{#1}}

\usepackage{tikz}
\usetikzlibrary{quantikz2}